\documentclass{elsart}
\usepackage{graphicx}
\usepackage{amssymb}

\newcommand{\cc}{cosmological constant}
\newcommand{\del}{\partial}
\newcommand{\dphi}{\partial_i \phi \partial^i \phi}

\begin{document}

\begin{frontmatter}         %% this is the macro which includes title, author  etc.etc. incl abstract
\title{COSMOLOGICAL CONSTANT - THE WEIGHT OF THE VACUUM}
\author{T. Padmanabhan}
% \ead{nabhan@iucaa.ernet.in}
 %\ead[url]{http://www.iucaa.ernet.in/~paddy}
\address{IUCAA, Pune University Campus, Ganeshkhind, Pune 411 007, India. \\
email: nabhan@iucaa.ernet.in}

\begin{abstract}
Recent cosmological observations  suggest the existence of a positive \cc\ $\Lambda$
with the magnitude $\Lambda(G\hbar/c^3) \approx 10^{-123}$. This review 
 discusses several aspects of the \cc\ both from the cosmological  (sections \ref{intro}--\ref{cmbrani}) 
 and field theoretical  (sections \ref{interpretcc}--\ref{ccstring})  perspectives.  
   After a brief  introduction to the key issues related to 
   cosmological constant and a  historical overview,  
   a summary of 
    the kinematics and dynamics of the standard Friedmann model of the universe
   is provided. The observational evidence for cosmological constant,  especially from the supernova
   results, and the  constraints from the age of the universe, 
   structure formation,   Cosmic Microwave Background Radiation (CMBR)
    anisotropies and a few others  are described in detail, followed by a discussion of  the
   theoretical
   models (quintessence, tachyonic scalar field, ...)    from different perspectives. 
      The latter  part  of the review (sections \ref{interpretcc}--\ref{ccstring}) 
 concentrates on more conceptual and fundamental aspects of the \cc\  like
 some alternative interpretations of the \cc, 
  relaxation mechanisms  to reduce 
 the cosmological constant to the currently observed value,
 the geometrical structure of the de Sitter spacetime,
  thermodynamics of the de Sitter universe 
  and the role of string theory in the \cc\ problem.
\end{abstract}

\end{frontmatter}

\tableofcontents

\section{Introduction}\label{intro}
   
   This review discusses several aspects of the \cc\ both from the cosmological and field theoretical
   perspectives with the emphasis  on conceptual and fundamental
   issues rather than on observational details.   
   The plan of the review is as follows: This section introduces the key issues related to 
   cosmological constant and provides a brief historical overview. 
   (For previous reviews of this subject, from
   cosmological point of view, see \cite{jpbr,carolcomb1,carolcomb3,carolcomb2}.)
   Section \ref{framework}
   summarizes the kinematics and dynamics of the standard Friedmann model of the universe
   paying special attention to features involving the cosmological constant. Section \ref{evidencecc}
   reviews the observational evidence for cosmological constant, especially the supernova
   results, constraints from the age of the universe and a few others. We next study models with evolving
   cosmological `constant' from different perspectives. 
   (In this review, we shall use the term
  \cc\ in a generalized sense including the scenarios in which cosmological ``constant''
  is actually varying in time.)  A phenomenological parameterization is 
   introduced in section  \ref{paraeqn} to compare theory with observation and is followed
   up with explicit models involving scalar fields in section \ref{theorydark}. The emphasis
   is on quintessence and tachyonic scalar field models and the cosmic degeneracies 
   introduced by them. Section \ref{sfinuniv} discusses cosmological constant 
   and dark energy in the context of models for structure formation and section \ref{cmbrani}
 describes the constraints arising from CMBR anisotropies. 
 
 The latter part  of the review 
 concentrates on more conceptual and fundamental aspects of the \cc. 
 ( For previous reviews of this subject, from  a theoretical 
   physics perspective, 
   see \cite{swlambda,weincomb1,weincomb2}.)
 Section \ref{interpretcc} provides 
 some alternative interpretations of the \cc\ which might have a bearing on the possible solution to the
 problem. Several relaxation mechanisms have been suggested in the literature to reduce 
 the cosmological constant to the currently observed value and some of these
 attempts are described in section \ref{relaxcc}.
 Section \ref{desittergeom} gives a brief description of 
 the geometrical structure of the de Sitter spacetime and the 
 thermodynamics of the de Sitter universe is taken up in section \ref{horizons}. The relation between horizons,
 temperature and entropy are presented at one go in this section and the last section deals with the role of string theory in the \cc\ problem. 
 
  \subsection{The many faces of the cosmological constant}\label{facescc} 
  
  Einstein's equations, which determine the dynamics of the spacetime, can be derived
  from the action (see, eg. \cite{LL2}):
  \begin{equation}
  A = \frac{1}{16\pi G} \int R \sqrt{-g} d^4 x + \int L_{\rm matter}(\phi, \partial \phi) \sqrt{-g} d^4x
  \label{one}
  \end{equation}
  where $L_{\rm matter}$ is the Lagrangian for matter depending on some dynamical variables
  generically denoted as $\phi$. (We are using units with 
  $c=1$.) The variation of this action with respect to $\phi$ will lead to 
  the equation of motion for matter $ (\delta L_{\rm matter}/\delta \phi) =0 $,  in a given background geometry,
  while the variation of the action with respect to the metric tensor $g_{ik}$
  leads to the Einstein's equation 
  \begin{equation}
  R_{ik} - \frac{1}{2} g_{ik} R = 16 \pi G \frac{\delta L_{\rm matter}}{\delta g^{ik}} \equiv 8 \pi G T_{ik}
  \label{three}
  \end{equation}  
  where the last equation defines the energy momentum tensor of matter to be $T_{ik} \equiv 
  2(\delta L_{\rm matter}/\delta g^{ik})$. 
  
  Let us now consider a new matter action $L'_{\rm matter}=L_{\rm matter}
  -(\Lambda/8\pi G)$ where $\Lambda$ is a real constant. Equation of motion for the matter 
  $ (\delta L_{\rm matter}/\delta \phi) =0 $,  does not change
  under this transformation since $\Lambda $ is a constant; but the action now picks up an extra term proportional to $\Lambda$ 
  \begin{eqnarray}
  A  &=&  \frac{1}{16\pi G} \int R \sqrt{-g} d^4 x + \int \left(L_{\rm matter} -\frac{\Lambda}{8\pi G}\right) 
  \sqrt{-g} d^4x\nonumber \\
  &=&  \frac{1}{16\pi G} \int (R-2\Lambda) \sqrt{-g} d^4 x+ \int L_{\rm matter} \sqrt{-g} d^4x
   \label{four}
  \end{eqnarray}  
  and equation (\ref{three}) gets modified. 
  This  innocuous looking addition of a constant to the  matter Lagrangian leads to one of the most
  fundamental and fascinating problems of theoretical physics. The nature of this problem and its
  theoretical backdrop acquires  different shades of meaning depending which of the two forms of 
  equations in  (\ref{four}) is used. 
  
  The first
  interpretation, based on the first line of equation (\ref{four}), treats $\Lambda $  as the 
  shift in the matter Lagrangian which, in turn, will lead to a shift in the matter Hamiltonian.
  This could be thought of as a shift in the zero point energy of the matter system. Such a 
  constant shift in the energy does not affect the dynamics of matter 
   while gravity --- which couples to the total energy of the system --- picks up an extra
  contribution in the form of a new term $Q_{ik}$ in the energy-momentum tensor, leading to: 
   \begin{equation}
  R^i_k - \frac{1}{2} \delta^i_k R = 8\pi G (T^i_k + Q^i_k); \qquad Q^i_k \equiv \frac{\Lambda}{8 \pi G} 
  \delta^i_k \equiv \rho_\Lambda \delta^i_k
  \label{six}
  \end{equation}  
  
    The second line in equation (\ref{four}) can be interpreted 
  as gravitational field, described by the Lagrangian of the form $L_{\rm grav} \propto (1/G) (R-2\Lambda)$,
  interacting with matter described by the Lagrangian $L_{\rm matter}$. In this interpretation, gravity
  is described by two constants, the Newton's constant $G$ and the {\it \cc}\  $\Lambda$. 
  It is then natural to modify the {\it left hand side} of Einstein's equations and write (\ref{six}) as:
  \begin{equation}
  R^i_k - \frac{1}{2} \delta^i_k R - \delta^i_k \Lambda = 8\pi G T^i_k
  \label{five}
  \end{equation}  
  In this interpretation, 
  the spacetime is treated as curved even in the absence of matter ($T_{ik} =0$) since
  the equation $R_{ik} - (1/2) g_{ik}R -\Lambda g_{ik} =0$ does not admit flat spacetime as
  a solution. 
  (This situation is rather unusual and is related to the fact that symmetries of the theory
  with and without a \cc\ are drastically different; the original symmetry of general
  covariance cannot be naturally broken in such a way as to preserve the sub group
  of spacetime translations.)
  
  In fact, it is possible to consider a situation in which both effects can occur.
  If the gravitational interaction is actually described by the Lagrangian of the form
  $(R-2\Lambda)$, then there is an intrinsic cosmological constant in nature
  just as there is a Newtonian gravitational constant in nature. If the
  matter Lagrangian contains energy densities which change due to dynamics,
  then $L_{\rm matter}$ can pick up constant shifts during dynamical evolution.
  For example, consider a scalar field with the Lagrangian $L_{\rm matter} =(1/2) \partial_i\phi
  \partial^i\phi - V(\phi)$ which has the energy momentum tensor 
 \begin{equation}
  T^a_b = \partial^a\phi \partial_b \phi - \delta^a_b \left( \frac{1}{2} \partial^i\phi\partial_i\phi - V(\phi)\right)
  \label{scalartab}
  \end{equation}
  For field configurations which are constant [occurring, for example, at the minima of the
  potential $V(\phi)$], this contributes an energy momentum tensor $T^a_b = \delta^a_b V(\phi_{\rm min})$
  which has exactly the same form
  as a \cc. 
  As far as gravity is concerned, it is the combination of these two effects ---
  {\it of very different nature} --- which is relevant and the source will be $T_{ab}^{\rm eff} 
  = [V(\phi_{\rm min}) + (\Lambda/8\pi G)]g_{ab}$, corresponding to an effective 
  gravitational constant 
  \begin{equation}
  \Lambda_{\rm eff} = \Lambda + 8\pi G V(\phi_{\rm min})
  \label{nine}
  \end{equation} 
  If $\phi_{\rm min}$ and hence $V(\phi_{\rm min})$ changes during dynamical
  evolution, the value of $\Lambda_{\rm eff}$ can also change in course of time.
  More generally, any field configuration which is varying slowly in time will lead to
  a slowly varying $\Lambda_{\rm eff}$.

 The extra term $Q_{ik}$ in Einstein's equation  behaves in a manner which is 
  very peculiar compared to the energy momentum tensor of normal matter. The term $Q^i_k =
  \rho_\Lambda \delta^i_k$ is in the form of the energy momentum tensor of an ideal fluid
  with energy density $\rho_\Lambda$ and pressure $P_\Lambda = - \rho_\Lambda$; obviously,
  either the pressure or the energy density of this ``fluid'' must be negative, which is unlike 
  conventional laboratory systems. (See, however, reference \cite{volovik}.)
  
  Such an equation of state, $\rho = -P $ also has another important implication in general
  relativity. The spatial part  ${\bf g}$  of the geodesic acceleration (which measures the 
  relative acceleration of two geodesics in the spacetime) satisfies the following exact equation
  in general relativity (see e.g., page 332 of \cite{probbook}):
  \begin{equation}
  \nabla \cdot {\bf g} = - 4\pi G (\rho + 3P)
  \label{nextnine}
  \end{equation} 
  showing that the source of geodesic  acceleration is $(\rho + 3P)$ and not $\rho$.
  As long as $(\rho + 3P) > 0$, gravity remains attractive while $(\rho + 3P) <0$ can
  lead to repulsive gravitational effects. Since the \cc\ has $(\rho_\Lambda + 3P_\Lambda) =
  -2\rho_\Lambda$, a positive
  \cc\ (with $\Lambda >0$) can lead to {\it repulsive} gravity.  For example, if the energy density of 
  normal, non-relativistic  matter with zero pressure 
   is $\rho_{\rm NR}$, then  equation  (\ref{nextnine}) shows that the geodesics will accelerate
  away from each other due to the repulsion of \cc\ when $\rho_{\rm NR} < 2\rho_\Lambda$.  
  A related feature, which makes the above conclusion practically relevant is the fact that,
  in an expanding   universe,
  $\rho_\Lambda$ remains constant while $\rho_{\rm NR}$ decreases.
   (More formally, the equation of motion, $d(\rho_\Lambda V) = - P_\Lambda dV$
    for the cosmological constant, treated as an
  ideal fluid, is identically satisfied with constant
  $\rho_\Lambda, P_\Lambda$.) Therefore,
   $\rho_\Lambda $ will eventually dominate over $\rho_{\rm NR}$ if the universe
   expands sufficiently. 
   Since $|\Lambda|^{1/2}$ has the dimensions of inverse
  length, it will set the scale for the universe when cosmological
  constant dominates. 
  
  It follows that the most stringent bounds on $\Lambda$ will  arise
  from cosmology when the expansion of the universe has diluted the matter
  energy density sufficiently.
  The rate of expansion of the universe today is usually expressed in terms of the Hubble constant:
  $H_0 = 100 h $ km s$^{-1}$ Mpc$^{-1}$ where 1 Mpc  $ \approx 3\times 10^{24}$ cm
  and  $h$ is a dimensionless parameter in the range $0.62\lesssim h \lesssim 0.82$ (see section \ref{ageuniv}).  
From $ H_0$ we can form
the time scale $t_{\rm univ}\equiv H_0^{-1}\approx 10^{10} h^{-1}$ yr and the length scale $  cH_0^{-1} \approx 3000 h^{-1}$ Mpc; $t_{\rm univ}$  characterizes the evolutionary time scale of the universe and $H_0^{-1}$ 
is of the order of the largest length scales currently accessible in cosmological observations.
  From the observation 
  that the universe is at least of the size $H_0^{-1}$,
  we can set a bound on $\Lambda $ to be $|\Lambda| < 10^{-56}$ cm$^{-2}$. This 
  stringent bound  leads to several issues which have been debated
  for decades without satisfactory solutions. 
  \begin{itemize}
  \item
  In classical general relativity,
    based on the constants $G, c $ and $\Lambda$,  it
  is not possible to construct any dimensionless combination from these constants.
  Nevertheless, it is clear that $\Lambda$ is extremely tiny compared to any other
  physical scale in the universe,  suggesting that $\Lambda$ is probably
  zero. We, however, do not know of any symmetry mechanism or invariance
  principle which requires $\Lambda$ to vanish. Supersymmetry 
  does require
  the vanishing of the ground state energy; however, supersymmetry is so badly
  broken in nature that this is not of any practical use \cite{supersym1,supersym2}. 
  
  \item
  We mentioned above that observations actually constrain $\Lambda_{\rm eff}$
  in equation (\ref{nine}), 
  rather than $\Lambda$. This requires   $\Lambda$ and
  $V(\phi_{\rm min})$ to be fine tuned to an enormous accuracy
  for the bound, $|\Lambda_{\rm eff}| < 10^{-56} \ {\rm cm}^{-2}$,  to be satisfied. This becomes more mysterious when 
  we realize that $V(\phi_{\rm min})$ itself could change by several
  orders of magnitude during the evolution of the universe.
  
  \item
  When quantum fields  in a given curved spacetime are considered
  (even without invoking any quantum gravitational effects)
  one introduces the Planck constant, $\hbar$, in the 
  description of the physical system. It is then possible
  to form the dimensionless combination $\Lambda (G\hbar/c^3) \equiv \Lambda L_P^2$.
  (This equation also defines the quantity $L_P^2$; 
  throughout the review we use the symbol `$\equiv$' to define variables.)  
  The bound on $\Lambda$ translates into the condition $\Lambda L_P^2 \lesssim 10^{-123}$.
  As has been mentioned several times in literature, this will require enormous fine tuning.
  
  \item
  All the above questions could have been satisfactorily answered if we take $\Lambda_{\rm eff}$
  to be zero and assume that the correct theory of quantum gravity will provide an explanation
  for the vanishing of \cc.
  Such a view was held by several people (including the  author) until very recently.
  Current cosmological observations however suggests that $\Lambda_{\rm eff}$ is actually
  non zero and $\Lambda_{\rm eff} L_P^2$ is indeed of order ${\mathcal O}(10^{-123})$.
  In some sense, this is the cosmologist's worst nightmare come true. {\it  If the observations
  are correct, then $\Lambda_{\rm eff}$ is non zero, very tiny and its value is extremely
  fine tuned for no good reason}. This is a concrete statement
  of the first of the two `\cc\ problems'.
  
  \item
  The bound on $\Lambda L_P^2$ arises from the expansion rate of the 
  universe or --- equivalently --- from the energy density which is present in the universe
  today. The observations require the energy density of normal, non relativistic matter to be of the 
  same order of magnitude as the energy density contributed by the cosmological
  constant. But in the past, when the universe was smaller, the energy density of 
  normal matter would have been higher while the energy density of \cc\
  does not change.  Hence we need to adjust the energy densities
  of normal matter and \cc\ in the early epoch very carefully so that
  $\rho_\Lambda\gtrsim \rho_{\rm NR}$ around the current epoch.
  If this had happened very early in the evolution of the universe, then the 
  repulsive nature of a positive cosmological constant would have initiated
  a rapid expansion of the universe,  preventing the formation
  of galaxies, stars etc. If the epoch of $\rho_\Lambda \approx \rho_{\rm NR}$
  occurs much later  in the future, then the current observations
  would not have revealed the presence of non zero 
  \cc. This raises the second of the two \cc\ problems:
  Why is it that $(\rho_\Lambda/ \rho_{\rm NR}) = \mathcal{O} (1)$ at the 
  {\it current} phase of the universe ?
  
  \item 
  The sign of $\Lambda$ determines the nature of solutions to Einstein's 
  equations as well as the sign of $(\rho_\Lambda + 3P_\Lambda)$. Hence
  the spacetime geometry with $\Lambda L_P^2 = 10^{-123}$
  is very different from the one with $\Lambda L_P^2 = - 10^{-123}$.
  Any theoretical principle which explains the near zero value of 
  $\Lambda L_P^2$ must also explain why the observed value of $\Lambda$ is positive.
  
  \end{itemize}
  
  At present we have no clue as to what the above questions
  mean and how they need to be addressed. This review summarizes
  different attempts to understand the above questions from
  various perspectives.

  \subsection{A brief history of \cc} \label{history} 
    
   Originally, Einstein introduced the \cc\ $\Lambda$ in 
    the field equation for gravity (as in equation (\ref{five}))
      with the motivation that it allows for a finite, closed, static
   universe in which the energy density of matter determines the geometry. The spatial
   sections of such a universe are closed 3-spheres with radius $l = (8\pi G \rho_{\rm NR})^{-1/2}
   = \Lambda^{-1/2}$
   where $\rho_{\rm NR}$ is the energy density of pressureless matter (see section \ref{geometry}) 
    Einstein had hoped that   normal matter is {\it needed} to curve the 
   geometry;   a demand, which ---  to him ---  was closely related to the Mach's principle. 
   This hope, however, was soon shattered when de Sitter produced a solution to 
   Einstein's equations with \cc\ containing no matter \cite{desitter}. 
    However,  in spite of two fundamental papers by Friedmann 
  and one by Lemaitre \cite{F1,F2}, most workers did not catch on with the idea of
  an expanding universe. In fact, Einstein originally thought Friedmann's work
  was in error but later published a retraction of his comment; similarly, in the
  Solvay meeting in 1927, Einstein was arguing against the solutions describing 
  expanding universe. Nevertheless, the Einstein archives do contain a postcard
  from Einstein to Weyl in 1923 in which he says: ``If there is no quasi-static
  world, then away with the cosmological term''. 
  The early
   history following de Sitter's discovery (see, for example,
   \cite{north}) is clearly somewhat confused,  to say the least.
  
  It appears that the community
  accepted the concept of an expanding universe largely due to the work of Lemaitre.
  By 1931, Einstein himself had rejected the cosmological term as superflous and 
  unjustified (see reference \cite{einstein}, which is a single authored paper;
  this paper has been mis-cited in literature often, eventually converting
  part of the journal name ``preuss'' to a co-author ``Preuss, S. B''!; see
  \cite{history}). There is no direct record that Einstein ever called \cc\
  his biggest blunder. It is possible that this often repeated ``quote'' arises
  from Gamow's recollection \cite{gamow}: ``When I was discussing cosmological
  problems with Einstein, he remarked that the introduction of the cosmological
  term was the biggest blunder he ever made in his life.''
  By 1950's  the view was decidedly against $\Lambda$ and the authors of several  classic texts
  ( like
  Landau and Liftshitz\cite{LL2}, Pauli \cite{pauli} and Einstein \cite{meaning}) 
  argued against the \cc.
  
  In later years, \cc\ had a chequered history and was often accepted or 
  rejected for  wrong or insufficient reasons. For example, the original value of the
  Hubble constant was nearly an order of magnitude higher \cite{hubble29} 
   than the currently
  accepted value thereby reducing the age of the universe by a similar factor.
  At this stage, as well as on several later occasions (eg., \cite{sandage61,gunntinsley}), 
   cosmologists have
  invoked \cc\ to reconcile the age of the universe with observations (see section \ref{ageuniv}).
  Similar attempts have been made in the past when it was felt that
  counts of quasars peak at a given phase in the expansion of the
  universe \cite{petro,shklov,kardash}. 
  These reasons, for the introduction of something as fundamental
  as \cc, seem inadequate at present.

  However, these attempts clearly showed that sensible cosmology can only be
  obtained if the energy density contributed by \cc\ is comparable to
  the energy density of matter at the present epoch. This remarkable
  property was probably noticed first by Bondi \cite{bondi}
  and has been discussed by McCrea \cite{mccrea}. It is mentioned in \cite{jpbr}
  that such coincidences were discussed in Dicke's gravity research group
  in the sixties; it is almost certain that this must have been noticed by several other
  workers in the subject.
  
  The first cosmological model to make central use of the \cc was the steady state model
  \cite{steady1,steady2,steady3}. It made use of the fact that a universe with a cosmological constant has a time translational invariance in a particular coordinate system. The model also used a scalar field with negative energy field to continuously create matter while maintaining energy conservation. While modern approaches to cosmology invokes negative energies or pressure without hesitation, steady state cosmology was discarded by most workers after the discovery of CMBR.    

   The discussion so far has been purely classical.
  The introduction of quantum theory adds a new dimension to this problem.
  Much of the early work \cite{earlywork1a,earlywork1b} as well as the definitive work
  by Pauli \cite{earlywork2a,earlywork2b} involved evaluating the sum of the zero point
  energies of a quantum field (with some cut-off) in order to estimate the vacuum
  contribution to the \cc. Such an argument, however,  is hopelessly naive
  (inspite of the fact that it is often
  repeated even today). In fact, Pauli himself was aware of the fact that one
  must {\it exclude} the zero point contribution from such a calculation.
  The first paper to stress this clearly and carry out a {\it second order}
  calculation was probably the one by Zeldovich \cite{zeldo} though
  the connection between vacuum energy density and cosmological
  constant had been noted earlier by Gliner \cite{gliner}
  and even by Lemaitre \cite{earlywork3}. Zeldovich assumed
  that the lowest order zero point energy should be subtracted out
  in quantum field theory and went on to compute the gravitational
  force between particles in the vacuum fluctuations. If $E$ is an
  energy scale of a virtual process corresponding to a length scale $l=\hbar c/E$,
   then   $l^{-3}=(E/\hbar c)^3$ particles  per unit
  volume of energy $E$ will lead to the gravitational self energy density of the order of
  \begin{equation}
  \rho_\Lambda \approx \frac{G(E/c^2)^2}{l}l^{-3} = \frac{GE^6}{c^8\hbar^4}
   \end{equation}
   This will correspond to $\Lambda L_P^2\approx (E/E_P)^6$ where $E_P=(\hbar c^5/G)^{1/2}\approx 10^{19}$GeV
   is the Planck energy.
   Zeldovich took $E\approx 1$ GeV (without
  any clear reason) and obtained a $\rho_\Lambda$ which contradicted the observational bound  ``only'' by
  nine orders of magnitude.

  The first serious symmetry principle which had implications for \cc\
  was supersymmetry and it was realized early on \cite{supersym1,supersym2}
  that the contributions to vacuum energy from fermions and bosons
  will cancel in a supersymmetric theory. This, however, is not of much help
  since supersymmetry is badly broken in nature at sufficiently high energies (at $E_{\rm SS}
  > 10^2$ Gev).
  In general, one would expect the vacuum energy density to be comparable
  to the that corresponding to the supersymmetry braking scale, $E_{\rm SS}$.
  This will, again, lead to an unacceptably large value for $\rho_\Lambda$.
   In fact
  the situation is more complex and one has to take into account the coupling
  of matter sector and gravitation --- which invariably leads to a supergravity theory.
 The description of \cc\ in such models is more complex,
  though none of the attempts have  provided a clear direction of attack (see e.g, \cite{swlambda}
  for a review of early attempts).

  The situation becomes more complicated when the quantum field theory
  admits more than one ground state or even more than one local minima
  for the potentials. For example, the spontaneous symmetry breaking in the
  electro-weak theory arises from a potential of the form
  \begin{equation}
  V = V_0 - \mu^2  \phi^2 + g \phi^4 \qquad (\mu^2, g>0)
  \end{equation}
  At the minimum, this leads to an energy density $V_{\rm min} = V_0 - (\mu^4/4g)$.
  If we take $V_0 =0$ then $(V_{\rm min}/g) \approx -(300\ {\rm GeV})^4$.
  For an estimate, we will assume that the gauge coupling constant $g$
  is comparable to the electromagnetic coupling constant: 
   $g ={\mathcal O}(\alpha^2)$,
  where $\alpha \equiv (e^2/\hbar c)$ is the fine structure constant. Then,
   we get $|V_{\rm min}| \sim 10^6\ {\rm GeV}^4$ which misses
  the bound on $\Lambda$ by a factor of $10^{53}$. It is really   of no help to set $V_{\rm min}=0$
  by hand. At early epochs of the universe, the temperature dependent effective
  potential  \cite{thermo1,thermo2} will change minimum to $\phi =0$ with $V(\phi) = V_0$. In other words,
  the ground state energy changes by several orders of magnitude during the 
  electro-weak and other phase transitions. 
  
  Another facet is added to the discussion by the currently popular models of quantum gravity
  based on string theory \cite{stringtext1,stringtext2}. The currently accepted  paradigm of string theory
  encompasses several ground states of the same underlying theory 
  (in a manner which is as yet unknown). This could lead to the possibility that
  the final theory of quantum gravity might allow different ground states for nature
  and we may need an {\it extra} prescription to choose the actual  state in which we
  live in. The  different ground states can also have different values for \cc\ and
  we need to invoke a separate (again, as yet unknown) principle to choose the 
  ground state in which $\Lambda L_P^2 \approx 10^{-123}$ (see section \ref{ccstring}).
  
  \section{Framework of standard cosmology}\label{framework}

All the well developed models of standard cosmology start with two basic
assumptions: (i) The  distribution of matter in the
universe is homogeneous and isotropic at sufficiently large scales.
(ii) The large scale structure of the universe is essentially
determined by gravitational interactions and hence can be described
by Einstein's theory of gravity.  The geometry of the universe can then be determined via Einstein's
equations with the stress tensor of matter $T^i_k (t,{\bf x})$
acting as the source.
 (For a review of  cosmology, see  e.g. \cite{jvncosmobook,tpsfu,peacock,tapvol3,jvntparaa}).
 The first assumption determines the kinematics of the universe while the second one determines the dynamics.
 We shall  discuss the consequences of these two assumptions in the next two subsections.

  \subsection{Kinematics of the Friedmann model}\label{friedkine}

  The  assumption of isotropy and homogeneity implies   that the large scale
geometry can be described by a metric of the form
\begin{equation}
  d{\rm s}^2= dt^2 - a^2 (t) d{\bf x}^2 = dt^2-a^2(t)
   \left[{dr^2\over 1-kr^2}+r^2(d\theta^2+\sin^2\theta d\phi^2)\right]
  \label{frwmetric}
  \end{equation}
in a suitable set of coordinates called comoving coordinates.
 Here $a(t)$ is an arbitrary function of time (called {\it expansion factor}) and $k=0, \pm 1$. Defining a new coordinate
  $\chi$ through $\chi=(r,\sin^{-1}r,\sinh^{-1}r)$ for $k=(0,+1, -1)$ this line element becomes
  \begin{equation}
  ds^2 \equiv dt^2 -a^2d{\bf x}^2
  \equiv dt^2 - a^2(t) \left[ d\chi^2 + S_k^2(\chi) \left( d\theta^2 + \sin^2 \theta d\phi^2 \right) \right]
  \end{equation}
  where $S_k(\chi) = (\chi, \sin \chi, \sinh \chi)$ for $k=(0,+1, -1)$.
  In any range of time during which $a(t)$ is a monotonic function of $t$, one can use
  $a$ itself as a time coordinate. It is also convenient to define a quantity
  $z$, called the  redshift,  through the relation $a(t) = a_0 [1+z(t)]^{-1}$ where $a_0$ is
   the current value of the expansion factor. The line
  element  in terms of [$a,\chi,\theta,\phi$] or [$z,\chi,\theta,\phi$]  is:
    \begin{equation}
  ds^2 = H^{-2} (a) \left( \frac{da}{a}\right)^2 - a^2 d{\bf x}^2 = \frac{1}{(1+z)^2}\left[ H^{-2} (z)dz^2 - dx^2\right]
  \label{ten}
  \end{equation}
  where $H(a) = (\dot a/a) $, called the {\it Hubble parameter}, measures the rate of expansion of the
  universe.
  
   This equation allows us to draw an  important conclusion:
  The only non trivial metric function in a Friedmann universe is the function $H(a)$ (and the numerical 
  value of $k$ which is encoded in the spatial part of the line element.) Hence, any kind of
  observation based on geometry of the spacetime, however complex it may be, will not allow us to determine anything other than this
  single function $H(a)$. As we shall see, this alone is inadequate to describe the material
  content of the universe and any attempt to do so will require additional inputs.

  Since the geometrical observations often rely on photons received from
  distant sources, let us consider
   a photon traveling a distance $r_{\rm em}(z)$ from the time of emission (corresponding
  to the redshift $z$) till today.
 Using the fact that $ds=0$ for a light ray and  the
   second equality in  equation (\ref{ten}) we find that the distance traveled by light rays
  is related to the redshift by $dx = H^{-1}(z) dz$.
  Integrating this relation, we get
   \begin{equation}
   r_{\rm em}(z) = S_k (\alpha); \qquad \alpha \equiv
   \frac{1}{a_0} \int_0^z H^{-1}(z) dz
   \label{rem}
  \end{equation} 
  All other geometrical distances can be expressed in terms of $r_{\rm em}(z)$ (see
  eg., \cite{tpsfu}).
  For example, the flux of radiation $F$ received from a source of luminosity $L$ can
  be expressed in the form $F = L/(4\pi d_L^2)$ where
   \begin{equation}
  d_L(z)  = a_0 r_{\rm  em}(z) (1+z) = a_0 (1+z) S_k(\alpha)
  \label{dlz}
  \end{equation}
  is called the {\it luminosity distance}.
  Similarly, if $D$ is the physical size of an object which subtends an angle $\delta$ to the
  observer, then --- for small $\delta$ --- we can define an {\it angular diameter distance}
   $d_A$ through
  the relation $\delta = D/d_A$. The angular diameter distance is given by
  \begin{equation}
  d_A(z)=a_0 r_{\rm em}(z)(1+z)^{-1}
  \label{daz}
  \end{equation} 
  with $d_L = (1+z)^2 d_A$. 
  
  If we can identify some objects (or physical phenomena)
  at a redshift of $z$ having a characteristic transverse size $D$, then measuring the
  angle $\delta$ subtended by this object we can determine $d_A(z)$. Similarly,
  if we have a series of luminous sources at different redshifts having known
  luminosity $L$, then by observing the flux from these sources $L$, one can
  determine the luminosity distance $d_L(z)$.  
  Determining any of these functions
  will allow us to use the relations (\ref{dlz}) [or (\ref{daz})] and (\ref{rem}) to
   obtain $H^{-1}(z)$. For example, $H^{-1}(z)$ is related to $d_L(z)$ through
   \begin{equation}
  H^{-1}(z) = \left[ 1 -  \frac{k d^2_L(z)}{a_0^2(1+z)^2}\right]^{-1/2} \frac{d}{dz}
  \left[  \frac{d_L(z)}{1+z}\right] \to \frac{d}{dz} \left[\frac{d_L(z)}{1+z}\right]
  \label{hinvz}
  \end{equation} 
  where second equality holds  if the spatial sections of the universe are flat, corresponding to
  $k=0$;  then $d_L(z)$, $d_A(z), r_{\rm em}(z) $ and $H^{-1}(z)$ all contain
  the (same)  maximal amount of information about  the geometry.

  The  function $ r_{\rm em}(z)$ also determines the proper volume of the universe
  between the redshifts $z$ and $z+dz$ subtending a solid angle $d\Omega$ in the
  sky.
  The comoving volume element can  be
  expressed in the form 
  \begin{equation}
  \frac{dV}{dzd\Omega} \propto   r_{\rm em}^2 \frac{dr}{dz} \propto   \frac{d_L^3}{(1+z)^4} \left[ \frac{(1+z) d'_L}{d_L} - 1 \right]
  \end{equation}
  where the prime denotes derivative with respect to $z$. 
   Based on this, there has been a suggestion \cite{dlprime}
  that future observations of the number of dark matter halos
  as a function of redshift and circular velocities can be used to  determine
  the comoving volume element 
  to within a few percent accuracy. If this becomes possible, then
  it will provide an additional handle on constraining the cosmological parameters.

  The above discussion illustrates how cosmological observations
  can be used to determine the metric of the spacetime, encoded by the
  single function $H^{-1}(z)$. This issue is trivial in principle, though
  enormously complicated in practice because of observational uncertainties
  in the determination of $d_L(z), d_A(z)$ etc.
  We shall occasion to discuss these features
  in   detail  later on.

 \subsection{Dynamics of the Friedmann model}\label{frieddyn}

Let us now turn to the second assumption which determines the dynamics of the universe. 
   When several non interacting sources are present in the universe, the total
  energy momentum tensor which appear on the right hand side of the
  Einstein's equation will be the sum of the energy momentum tensor for
  each of these sources. Spatial homogeneity and isotropy imply that each
   $T^a_b $ is diagonal and has  the
   form $T^a_b = {\rm dia} \, [\rho_i(t), -P_i (t),  -P_i (t),-P_i (t)]$ where
   the index $i=1,2, ..., N$ denotes $N$ different kinds of sources (like
   radiation, matter, cosmological constant etc.).  Since the sources
   do not interact with each other, each  energy momentum tensor
   must satisfy the relation $T^a_{b;a}=0$ which translates to
   the condition $d(\rho_i a^3) = -P_ida^3$. It follows that the evolution of the energy densities of 
   each component is essentially dependent on the parameter 
   $w_i \equiv (P_i/\rho_i)$ which, in general, could be a function of time.
   Integrating  $d(\rho_i a^3) = - w_i \rho_i da^3$,  we get
    \begin{equation}
  \rho_i = \rho_i(a_0) \left( \frac{a_0}{a}\right)^3 \exp \left[ - 3 \int_{a_0}^a \frac{d\bar a}{\bar a} w_i(\bar a) 
  \right]
  \label{evolenergy}
  \end{equation} 
  which determines the evolution of the energy density of each of the species in terms  of the
  functions $w_i(a)$.

  This description  determines $\rho (a)$ for different sources but not $a(t)$. To determine the 
  latter we can use one of the Einstein's equations:
 \begin{equation}
 H^2(a) = \frac{\dot a^2}{a^2} = \frac{8\pi G}{3} \sum_i \rho_i(a) - \frac{k}{a^2}
 \label{frwone}
  \end{equation} 
  This equation shows that, once the evolution of the individual components of energy density
  $\rho_i(a) $ is known, the function $H(a)$ and thus the line element in equation (\ref{ten})
  is known. (Evaluating this equation at the present epoch one can determine the value of 
  $k$; hence it is not necessary to provide this information separately.) Given  $H_0$,
  the  current
  value of the Hubble parameter, one can construct a {\it critical density}, by the definition: 
 \begin{eqnarray}
 \rho_c  = \frac{3H_0^2}{8\pi G} & =& 1.88 h^2 \times 10^{-29}\ {\rm gm\ cm}^{-3}= 2.8 \times 10^{11} h^2 M_\odot \ {\rm Mpc}^{-3}\nonumber\\
& =& 1.1 \times 10^4 h^2 \ {\rm eV\ cm}^{-3} = 1.1 \times 10^{-5} h^2 \ {\rm protons \ cm}^{-3}
\end{eqnarray}
 and
  parameterize
  the energy density, $\rho_i(a_0)$,  of different components at the present epoch
   in terms of the critical density by $\rho_i(a_0) \equiv \Omega_i \rho_c$.
   [Observations  \cite{freedman,mould00}    
give  $h= 0.72 \pm 0.03$ (statistical) $\pm 0.07 $ (systematic)].
  It is obvious from equation (\ref{frwone}) that $k=0$ corresponds to 
  $\Omega_{\rm tot}=\sum_i \Omega_i =1$ while $\Omega_{\rm tot}>1$ and $\Omega_{\rm tot}<1$
  correspond to $k=\pm 1$.
  When $\Omega_{\rm tot}\ne 1$, equation (\ref{frwone}),
  evaluated at the current epoch, gives $(k/a_0^2)=H_0^2(\Omega_{\rm tot} -1)$, thereby fixing the value of
  $(k/a_0^2)$; when, $\Omega_{\rm tot}= 1$, it is conventional to take $a_0=1$ since its value can be rescaled.
   
  \subsection{Composition of the universe}\label{univcompo}

It is important to stress that {\it absolutely no progress in cosmology can 
be made until a relationship between $\rho$ and $P$ is provided, say, in the form of the functions $w_i(a)$s}.
This fact, in turn, brings to focus two 
issues which are not often adequately emphasized: 

(i) If we assume that
the source is made of normal laboratory matter, then the relationship
between $\rho$ and $P$  depends on our knowledge
of how the equation of state for matter behaves at different 
energy scales. This information needs to be provided by atomic physics, nuclear physics
and particle physics. Cosmological models can at best be only as accurate
as the input physics about $T^i_k$ is; any definitive assertion about the state of the universe is misplaced, if the  knowledge about $T^i_k$ which it is based on is itself speculative or non existent at the relevant
energy scales. At present we have laboratory results testing the behaviour of
matter up to about 100 GeV and hence we can, in principle, determine
the equation of state for matter up to 100 GeV. By and large, the equation of 
state for normal matter in this domain can be taken to be that of an
ideal fluid with $\rho $ giving the energy density and $P$
giving the pressure; the relation between the two is of the form
$P= w \rho$ with $w=0$ for non relativistic matter and $w=(1/3)$ for relativistic
matter and radiation.

(ii) The situation becomes more complicated when we realize that it
is entirely possible for the large scale universe to be dominated by
matter whose presence is undetectable at laboratory scales. For
example, large scale
scalar fields dominated either by kinetic energy or nearly constant potential
energy could exist in the universe and will not be easily detectable at
laboratory scales. 
We  see from (\ref{scalartab}) that such
 systems can have an equation of state of the form $P=w\rho$ with
$w=1$ (for kinetic energy dominated scalar field) or $w=-1$ (for potential
energy dominated scalar field). 
While the conservative procedure for doing cosmology would be 
to use only known forms of $T^i_k$ on the right hand side of Einstein's
equations, this has the drawback of preventing progress in our understanding
of nature, since cosmology could possibly be the {\it only} testing ground for the 
existence of forms of $T^i_k$ which are difficult to detect at 
laboratory scales.

%%%%%%%%%%%       FIGURE               %%%%%%%%%%%%%%%
     \begin{figure}[tp]    
   \begin{center}
   \includegraphics[angle=270,scale=0.5]{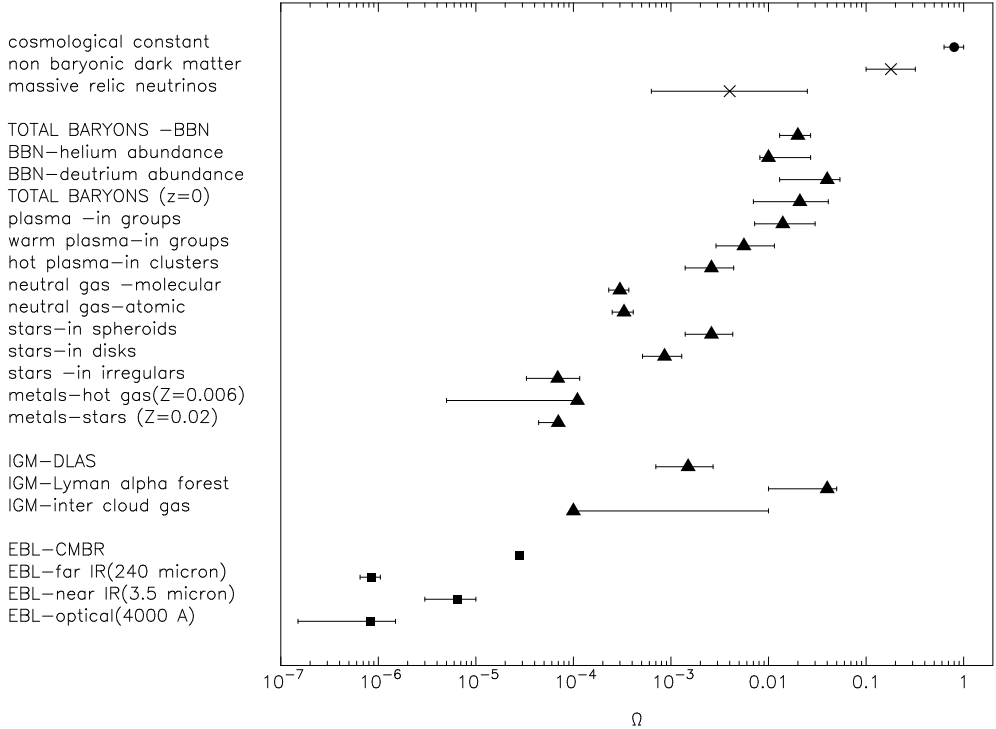}
   \end{center}
\caption{
Cosmic inventory of energy densities. See text for description. (Figure adapted
from \cite{tapvol3}.)
}
 \label{fig:baryons}
\end{figure}
%%%%%%%%%%%%%%%%%%%%%%%%%%%%%%%%%%%%%

One of the key issues in modern cosmology has to do with the conflict in 
principle between (i) and (ii) above. Suppose a model based on conventional
equation of state, adequately tested in the laboratory, fails to account
for a cosmological observation. Should one treat this as a failure of
the cosmological model or as a signal from  nature for the 
existence of a source $T^i_k$ not seen at laboratory scales? There
is no easy answer to this question and we will focus on many facets of
this issue in the coming sections.

 Figure \ref{fig:baryons} provides an inventory of the density
contributed by different forms of matter in the universe.
The x-axis is actually a combination $\Omega h^n$ of $\Omega$ and the Hubble parameter
$h$ since different components are measured by different techniques. (Usually
$n=1$ or 2; numerical values are for $h=0.7$.)
The density parameter contributed  today by visible, non relativistic, baryonic matter in the 
universe is about $\Omega_B \approx (0.01 - 0.2)$ (marked by triangles in the figure; different estimates 
are from different sources; see for a sample of references \cite{fuhopeeb,sarkar,burles1,burles2,thuan,bania,coc,burlnot,perc,wang}). 
 The
density parameter due to radiation is about $\Omega_R \approx 2\times 10^{-5}$
(marked by squares in the figure).
Unfortunately, models for the universe with just these two
constituents for the energy density are in violent disagreement with observations.
 It appears to be necessary
to postulate the existence of:

\begin{itemize}
\item Pressure-less $(w=0)$ {\it non} baryonic dark matter which
does not couple with radiation and having a density of  about  $\Omega_{\rm DM} \approx 0.3$.
Since it does not emit light, it is called {\it dark matter} (and marked by a cross in the figure).
Several independent techniques like cluster mass-to-light ratios \cite{carlberg} 
baryon densities in clusters \cite{mohr,grego} 
weak lensing of clusters \cite{mellier,wkl} 
and the existence of massive clusters at high redshift \cite{bahcfan} 
have been used to obtain a handle on $\Omega_{\rm DM}$. These observations are
all consistent with $\Omega_{\rm NR}= (\Omega_{\rm DM} + \Omega_B) \approx \Omega_{\rm DM}
 \approx (0.2 - 0.4)$.

\item  An exotic form of matter 
({\it cosmological constant} or something similar) with an equation of 
state $p\approx-\rho$ (that is, $ w \approx -1$) having a density parameter of about
$\Omega_\Lambda  \approx 0.7$ (marked by a filled circle  in the figure).
The evidence for $\Omega_\Lambda$ will be discussed in section \ref{evidencecc}.

\end{itemize}

So  in addition to $H_0$, at least  four more free parameters
are required to describe the background universe at low energies (say, below 50 GeV).
These are  $\Omega_{B},
\Omega_{R},\Omega_{\rm DM}$ and $\Omega_{\Lambda}$ describing the
fraction of the critical density contributed by baryonic matter, radiation 
 (including relativistic particles like e.g, massive neutrinos; marked by a cross
in the figure), dark matter 
and cosmological constant respectively. The first two certainly exist; the existence 
of last two is probably suggested by observations and is definitely  not contradicted
by any observations. Of these, only
$\Omega_{R}$ is well constrained and other quantities are plagued by
both statistical and systematic errors in their measurements. 
The top two positions in the contribution to $\Omega$ are from
cosmological constant and non baryonic dark matter.  
It is unfortunate that we do not have laboratory evidence for the
existence of the first two dominant contributions to the energy density
in the universe. (This  feature alone could make most of the 
cosmological paradigm described in this review irrelevant at a 
future date!)

 The simplest model for the universe is based on the assumption  that each of the sources which populate
  the universe has a constant $w_i$;  then equation (\ref{frwone}) becomes
  \begin{equation}
  \frac{\dot a^2}{a^2} = H_0^2 \sum_i \Omega_i \left(\frac{a_0}{a}\right) ^{3(1+w_i)}- \frac{k}{a^2}
  \label{qlsone}
  \end{equation} 
where each of these species is identified by density parameter
$\Omega_i$ and the equation of state characterized by
$w_i$. The most familiar form of energy densities
are those due to pressure-less matter with
$w_i = 0$
(that is, non relativistic matter with rest mass energy density $\rho c^2$
dominating over the kinetic energy density, $\rho v^2/2)$
 and radiation with $w_i=(1/3)$.
Whenever any {\em one} component of energy density dominates over others,
 $P\simeq w\rho$ and  it follows
from the equation (\ref{qlsone})  (taking $k=0$, for simplicity)
that
\begin{equation}
 \rho \propto a^{-3(1+w)}; \qquad a \propto t^{2/[3(1+w)]}
 \end{equation}
For example,  $\rho \propto a^{-4}, \ a\propto t^{1/2}$ if
the source is relativistic and $\rho \propto a^{-3}, \ a\propto t^{2/3}$ if
the source is non-relativistic.

This result shows that the  past evolution of the  universe  is characterized by two important
epochs (see eg. \cite{jvncosmobook,tpsfu}):                        
                  (i) The first is the radiation dominated epoch which occurs
at redshifts greater than $z_{\rm eq} \approx (\Omega_{\rm DM}/\Omega_{\rm R}) \approx 10^4$. For $z \gtrsim z_{\rm eq}$ the energy density is dominated by
hot relativistic matter and the universe is very well approximated as a
$k=0$ model with $a(t) \propto t^{1/2}$. (ii) The second phase occurs for
$z\ll z_{\rm eq}$ in which the universe is dominated by non relativistic matter and
--- in some cases --- the cosmological constant. The form of $a(t)$
in this phase depends on the relative values of $\Omega_{\rm DM}$ and $\Omega_\Lambda$.
In the simplest case, with $\Omega_{\rm DM} \approx 1$, $\Omega_\Lambda =0$,
$\Omega_{\rm B} \ll \Omega_{\rm DM}$ the expansion is a power law with $a(t) \propto t^{2/3}$.
(When cosmological constant dominates over matter, $a(t)$ grows exponentially.)

During all the epochs, the temperature of the radiation varies
as $T\propto a^{-1}$. When the temperature falls below $T\approx 10^3 $ K,
neutral atomic systems form in the universe and photons decouple from
matter. In this scenario, a relic background of such photons with Planckian
spectrum at some non-zero temperature will exist in the present day
universe. The present theory is, however, unable to predict the value of $T$ at $t=t_0$; it is therefore a free parameter  related $\Omega_R \propto T^4_0$.

\subsection{Geometrical features of a universe with a cosmological constant}\label{geometry}

   The evolution of the universe has different characteristic features
   if there exists sources in the universe for which $(1+3w) <0$. This is obvious
   from equation (\ref{nextnine}) which shows that if $(\rho+3P)=(1+3w)\rho$
   becomes negative, then the gravitational force of such a source (with $\rho>0$)
   will be repulsive. The simplest example of this kind of a source is the
   \cc\ with $w_\Lambda = -1$.

%%%%%%%%%%%%%%       	FIGURE   		%%%%%%%%%%%%%%%

\begin{figure}[ht]
\begin{center}
\includegraphics[scale=0.5]{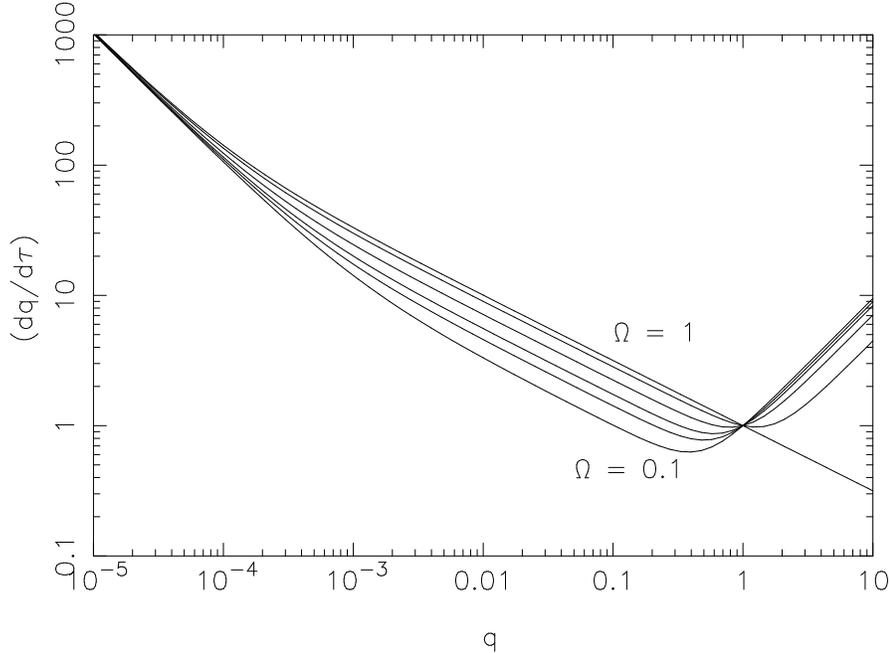}
\end{center}
\caption{The phase portrait of the universe, with the ``velocity'' of the universe $(dq/d\tau)$
plotted against the ``position'' $q = (1+z)^{-1}$ for
different  models with  $\Omega_R = 2.56 \times 10^{-5} h^{-2}, h=0.5,
\Omega_{\rm NR} + \Omega_\Lambda +\Omega_R = 1$. Curves are parameterized by the
value of $\Omega_{\rm NR} = 0.1, 0.2, 0.3, 0.5, 0.8, 1.0$ going
from bottom to top as indicated. (Figure adapted from \cite{tapvol3}.)}
\label{fig:veluniv}
\end{figure}

%%%%%%%%%%%%%%%%%%%%%%%%%%%%%%%%%%%%%%%%%%%

   To see the effect of  a \cc\ let us consider a
   universe with matter, radiation and a \cc.  Introducing a dimensionless time
coordinate $\tau = H_0t$ and writing $a= a_0 q(\tau)$  equation
   (\ref{frwone}) can  be cast in a more suggestive form describing the
   one dimensional motion of  a particle in a potential
\begin{equation}
\frac{1}{2} \left(\frac{dq}{d\tau}\right)^2 + V(q) = E
\label{qomeganr}
\end{equation} 
where
\begin{equation}
V(q) = - {1\over 2} \left[\frac{\Omega_R}{ q^2} + \frac{\Omega_{\rm
NR}}{ q} + \Omega_\Lambda  q^2\right]; \quad E=\frac{1}{ 2} \left( 1-\Omega\right).
\label{qveq}
\end{equation}  
This equation has the structure of the first integral for
motion of a particle with energy $E$ in a potential $V(q)$.
For models with $\Omega  = \Omega_R+\Omega_{\rm NR} + \Omega_\Lambda  =1$,
we can take $E=0$ so that $(dq/d\tau) = \sqrt{V(q)}$. Figure \ref{fig:veluniv}
is the phase portrait of the universe showing the velocity $(dq/d\tau)$ as a function of the position $q = (1+z)^{-1}$ for such models.
At high redshift (small $q$)
the universe is radiation dominated and $\dot q$ is independent
of the other cosmological parameters; hence all the curves
asymptotically approach each other at the left end of the figure.
At low redshifts, the presence of cosmological constant
makes a difference and -- in fact -- the velocity
$\dot q$ changes from being a decreasing function to an increasing function.
In other words, the presence of a cosmological constant leads to
an  {\it accelerating universe} at low redshifts.

  For a universe with non relativistic matter and \cc, the potential
  in (\ref{qveq}) has a simple form, varying as $(-a^{-1})$ for small $a$ 
  and $(-a^2)$ for large $a$ with a maximum in between at $q = q_{\rm max} = (\Omega_{\rm NR}/2 \Omega_\Lambda)^{1/3}$. 
 This system has been analyzed in detail in literature for both constant cosmological
 constant \cite{felten86}
 and for a time dependent \cc \ \cite{coop98}.
  A wide variety of explicit solutions 
  for $a(t)$ can be provided for these equations. We briefly summarize
  a few of them.
  \begin{itemize}
  \item If the ``particle'' is situated at the top of the potential, one
  can obtain a static solution with $\ddot a= \dot a =0$ by adjusting
  the cosmological constant and the dust energy density and taking 
  $k=1$. This solution, 
  \begin{equation}
  \Lambda_{crit} =4\pi G \rho_{\rm NR} = \frac{1}{a_0^2},
  \end{equation}
  was the one which originally prompted Einstein to introduce the cosmological
  constant (see section \ref{history}). 
  \item The above solution  is, obviously,  unstable and any small deviation
  from the equilibrium position will cause $a\to 0$ or $a\to \infty $. By
  fine tuning the values, it is possible to obtain a model for the 
  universe which ``loiters'' around $a=a_{\rm max}$ for a large period of time 
  \cite{edd30,lema27,rowan68,petro,shklov,kardash}.
    \item A subset of models  corresponds to those without matter and driven entirely
    by \cc\ $\Lambda$. These models have $k=(-1, 0, +1)$ and the corresponding expansion
    factors being proportional to  $[\sinh (Ht), \exp(Ht),$   $\cosh(Ht)]$ with
    $\Lambda^2 = 3H^2$. These line elements represent three different 
    characterizations of the de Sitter spacetime. The manifold is a four dimensional hyperboloid embedded
    in a flat, five dimensional space with signature $(+ - - -)$. We shall discuss this in greater detail in 
    section \ref{desittergeom}.
    \item It is also possible to obtain solutions in which the particle
    starts from $a=0$ with an energy which is insufficient for it 
    to overcome the potential barrier. These models
    lead to a universe which collapses back to a singularity. By arranging
    the parameters suitably, it is possible to make $a(t)$ move away or towards  the peak of the 
    potential (which corresponds to the static Einstein universe)
    asymptotically \cite{felten86}.
     \item In the case of $\Omega_{\rm NR} + \Omega_\Lambda =1$, the 
     explicit solution for $a(t)$ is given by 
     \begin{equation}
     a(t) \propto \left( \sinh \frac{3}{2} Ht\right)^{2/3};\qquad k=0; \qquad 3H^2 =\Lambda
     \label{univevol}
     \end{equation}
     This solution smoothly interpolates between a matter dominated universe
     $a(t) \propto t^{2/3}$ at early stages and a cosmological constant 
     dominated phase $a(t) \propto exp(Ht)$ at late stages. The transition 
     from deceleration to acceleration occurs at $z_{\rm acc}=
     (2\Omega_\Lambda/\Omega_{\rm NR})^{1/3} -1$, while the energy densities of 
     the \cc\ and the matter are equal at $z_{\Lambda m}=
     (\Omega_\Lambda/\Omega_{\rm NR})^{1/3} -1$. 
     \end{itemize}

\begin{figure}[ht]
\begin{center}
\includegraphics[scale=0.4]{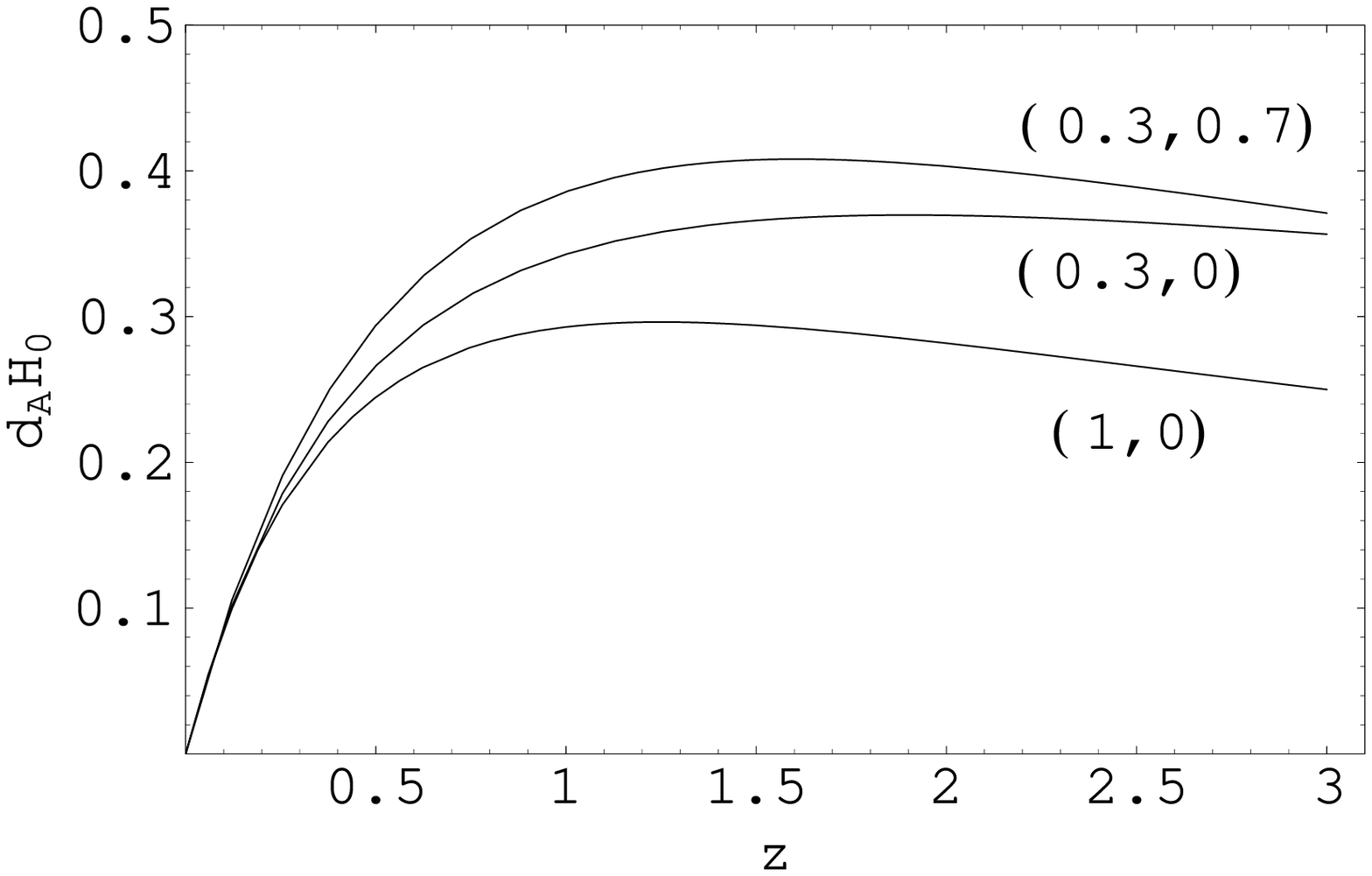}\ \includegraphics[scale=0.4]{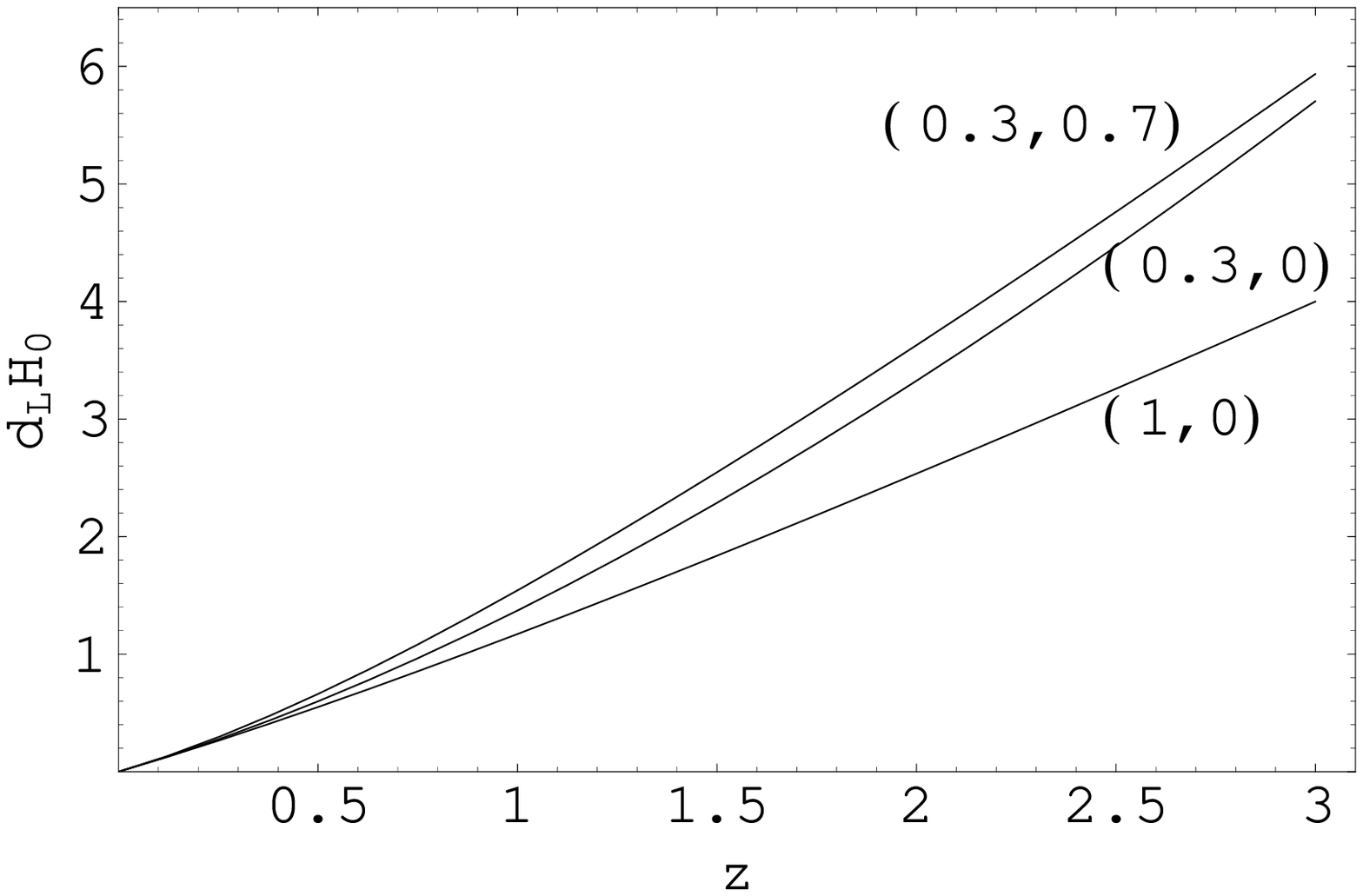}
\end{center}
\caption{The left panel gives the angular diameter distance in units of $cH_0^{-1}$ 
as a function of redshift.
The right panel gives the luminosity distance in units of $cH_0^{-1}$ 
as a function of redshift. Each curve is labelled by $(\Omega_{\rm NR}, \Omega_\Lambda)$.
 }
\label{fig:dazdlz}
\end{figure}

   The presence of a \cc\ also affects  other geometrical parameters in the universe.
   Figure \ref{fig:dazdlz} gives the plot of $d_A(z)$ and $d_L(z)$; (note that angular
diameter distance is not a monotonic function of z). Asymptotically,
for large z, these have the limiting forms,
\begin{equation}
d_A(z) \cong 2 ( H_0\Omega_{\rm NR})^{-1} z^{-1}; \quad d_L(z) \cong 2 ( H_0\Omega_{\rm NR})^{-1} z
\end{equation}

%%%%%%%%%%%%%%	FIGURE     %%%%%%%%%%%%%%%%

%%%%%%%%%%%%%%%%%%%%%%%%%%%%%%%%%%%%%

%%%%%%%%%%%%%%	  FIGURE       %%%%%%%%%%%%%%%

\begin{figure}[ht]
\begin{center}
\includegraphics[scale=0.6]{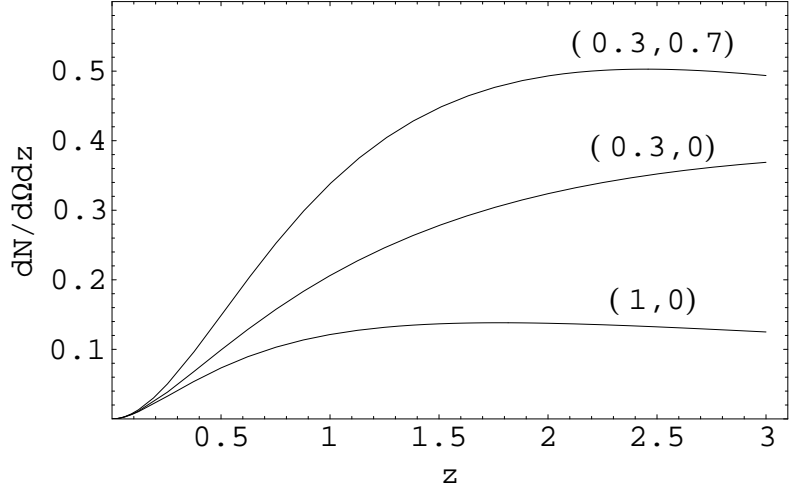}
\end{center}
\caption{The figure shows $(dN/d\Omega dz)$: 
  it is assumed that 
$n(z) = n_0(1+z)^3$. The y-axis is in units of $n_0H_0^{-3}$.
Each curve is labelled by $(\Omega_{\rm NR}, \Omega_\Lambda)$.
}
\label{fig:dnbydom}
\end{figure}

%%%%%%%%%%%%%%%%%%%%%%%%%%%%%%

The geometry of the spacetime also determines the proper volume of the universe
between the redshifts $z$ and $z+dz$ which subtends a solid angle $d\Omega$ in
the sky. If the number density of sources of a particular kind (say, galaxies, quasars, ...)
is given by $n(z)$, then the number count of sources
per unit solid angle per redshift interval should vary as
\begin{equation}
\frac{dN}{ d\Omega dz } = n(z) \frac{dV}{ d\Omega dz}  = \frac{n(z) a^2_0 r_{\rm em}^2(z) H^{-1} (z)}{ (1+z)^3}.  \label{qnc}
\end{equation} 
Figure \ref{fig:dnbydom}
shows $(dN/d\Omega dz)$;
 it is assumed that
$n(z) = n_0(1+z)^3$. The y-axis is in units of $n_0H_0^{-3}$.

\section{ Evidence for a non-zero cosmological constant}\label{evidencecc}

 There are several cosmological observations which suggests the  existence of 
  a non zero \cc\ with different levels of reliability.
   Most of these determine either the value of $\Omega_{\rm NR}$
  or some  combination of $\Omega_{\rm NR} $ and $\Omega_\Lambda$.
  When combined with the strong evidence from the CMBR observations that the $\Omega_{\rm tot}=1$
  (see section \ref{cmbrani}) or some other independent estimate of $\Omega_{\rm NR}$,
   one is led to a  non zero value for $\Omega_\Lambda$.
  The most reliable ones seem to be those based on high redshift supernova
   \cite{perlmutter98a,agr,sperl}
  and structure formation models \cite{cen98,kgb93,ostr95}.  We shall now discuss some of these
  evidence.

\subsection{Observational evidence for accelerating universe}\label{acceluniv}

\begin{figure}[tp]
\begin{center}
\includegraphics[scale=0.5,angle=-90]{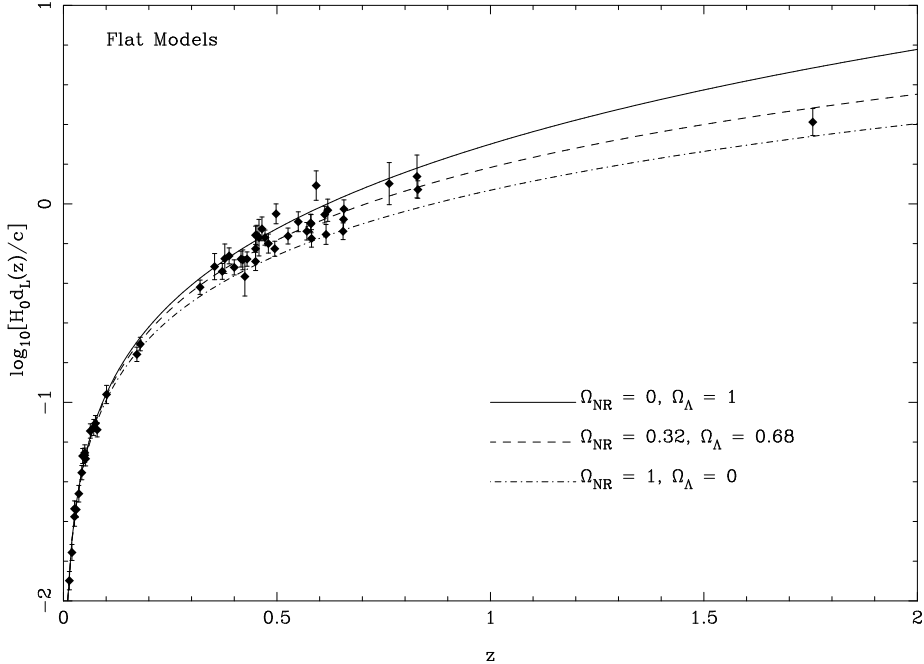}
\end{center}
\caption{The luminosity distance 
of a set of type Ia supernova at different redshifts and three theoretical models with $\Omega_{\rm NR} + \Omega_\Lambda =1$. The best fit curve has $\Omega_{\rm NR} =0.32, \Omega_\Lambda =0.68$.
}
\label{fig:datafit}
\end{figure}

\begin{figure}[htp]
\begin{center}
\includegraphics[scale=0.5,angle=-90]{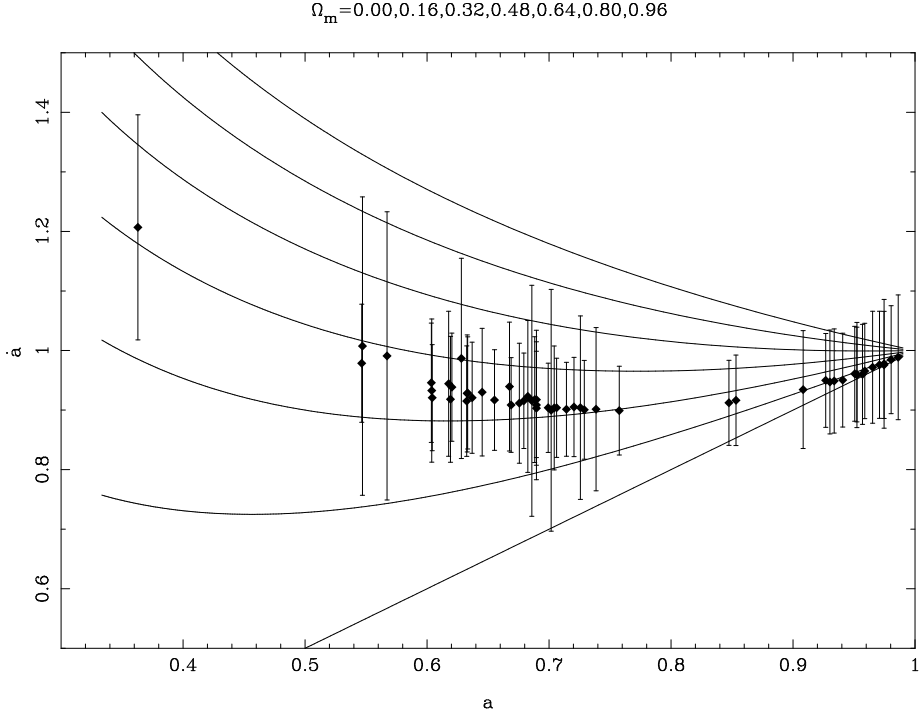}
\end{center}
\caption{Observations of SN are converted into the `velocity' $\dot a$ of the universe using a fitting function. The curves which are over-plotted corresponds to a cosmological model
with $\Omega_{\rm NR} + \Omega_\Lambda =1$. The best fit curve
has $\Omega_{\rm NR} = 0.32, \Omega_\Lambda=0.68$.
}
\label{fig:adotvsaomegam}
\end{figure}

Figure \ref{fig:veluniv} shows that the evolution of a universe with $\Omega_\Lambda \ne 0$
changes from a decelerating phase to an accelerating phase at late times.
If $H(a)$ can be observationally determined, then one can check
whether the real universe had undergone an accelerating phase
in the past. This, in turn, can be done if $d_L(z)$, say, can be observationally
determined for a class of sources. Such a study of several high redshift
supernova has led to the data which is shown in
 figures \ref{fig:datafit}, \ref{fig:adotvsaomegam}.        
 
 Bright supernova explosions are brief explosive stellar events which  are broadly classified as two types.
   Type-Ia supernova occurs when a degenerate dwarf star
  containing CNO enters a stage of rapid nuclear burning cooking
  iron group elements (see eg., chapter 7 of \cite{taptwo}). These are the brightest and most homogeneous
  class of supernova with hydrogen poor spectra. An empirical
  correlation has been observed between the sharply rising light curve
  in the initial phase of the supernova and its peak luminosity
  so that they can serve as standard candles. 
  These events typically last about a month and occurs approximately once in 300 years in
  our galaxy.
  (Type II supernova, which occur at the end of stellar evolution, are not useful as standard candles.)

  For any supernova, one can directly observe the apparent magnitude $m$
[which is essentially the logarithm of the flux $F$ observed]
and its redshift. The absolute magnitude $ M $ of the supernova is again related to the actual luminosity 
$L$ of the supernova
in a logarithmic fashion. Hence the relation $F=(L/4\pi d_L^2)$
can be written as
\begin{equation}
m- M  = 5 \log_{10}\left( \frac{d_L}{{\rm Mpc}} \right)+ 25
\end{equation}
The numerical factors arise from the astronomical conventions used in the definition of $m$ and $ M $. Very often,
one will use the dimensionless combination $ (H_0 d_L(z)/c)$ rather than $d_L(z)$ and the above equation
will change to $m(z)={\mathcal M} + 5 \log_{10}(H_0 d_L(z)/c)$ with  
 the quantity
${\mathcal M}$ being related to $M$ by
\begin{equation}
{\mathcal M}=M+25 +5 \log_{10}\left(\frac{c H_0^{-1}}{1 {\rm Mpc}}\right)
=M-5 \log_{10}h +42.38
\end{equation}
If the absolute magnitude of a class of Type I supernova can be determined
from the study of its light curve, then one can obtain the $d_L$ for
these supernova at different redshifts. 
(In fact, we only need the low-$z$ apparent magnitude at a given $z$ which is equivalent to
knowing $\mathcal{M}$.) 
Knowing $d_L$, one can determine
the geometry of the universe.

To understand this effect in a simple context, let us compare 
   the luminosity distance for a matter dominated model ($\Omega_{\rm NR} =1,
   \Omega_\Lambda =0)$ 
   \begin{equation}
d_L = 2H_0^{-1}\left[ (1 + z) - (1 + z)^{1/2}\right],
\label{dlone}
\end{equation}
   with that for a model driven purely by a  \cc\ ($\Omega_{\rm NR} =0,
   \Omega_\Lambda =1)$:
\begin{equation}
d_L = H_0^{-1} z(1 + z).
\label{dltwo}
\end{equation}
  It is clear that at a given $z$, the $d_L$ is larger for the \cc\ model.
   Hence, a given object, located at a fixed redshift, will appear brighter
   in the matter dominated model compared to the \cc\ dominated
   model. 
Though this sounds easy in principle, the statistical analysis turns out to be
quite complicated. 

The supernova cosmology project (SCP) has discovered \cite{sperl}
  42 supernova in the range $(0.18-0.83)$. 
The high-$z$ supernova search team (HSST) discovered 14 supernova
  in the redshift range $(0.16-0.62)$ and another 34 nearby supernova \cite{agr} and used
  two separate methods for data fitting. (They also included two previously published results
  from SCP.)
   Assuming $\Omega_{\rm NR}
  + \Omega_\Lambda =1$, the analysis of this data gives $\Omega_{\rm NR}= 0.28\pm 0.085
  $ (stat) $\pm 0.05$ (syst).

\begin{figure}[ht]
\begin{center}
\includegraphics[scale=0.5,angle=-90]{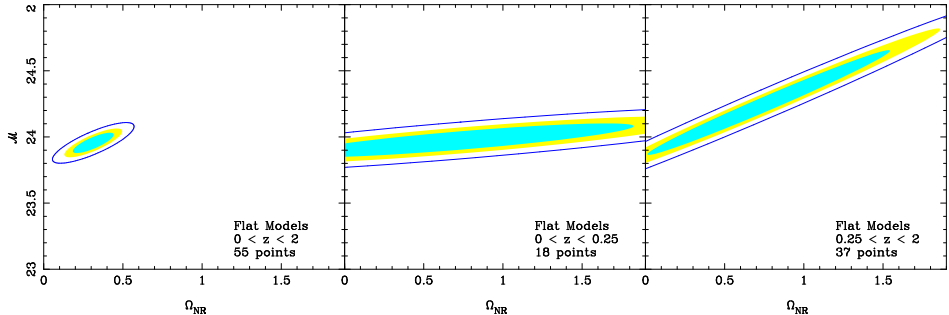}
\end{center}
\caption{Confidence contours corresponding to 68 \%, 90 \% and 99 \% based on SN data in the
$\Omega_{\rm NR} - {\mathcal M}$ plane for the flat models with $\Omega_{\rm NR} + \Omega_\Lambda=1$.
Frame (a) on the left uses all data while frame (b) in the middle uses low redshift data and the frame
(c) in the right uses high redshift data. While neither the low-z or high-z data alone can
exclude the $\Omega_{\rm NR}=1, \Omega_\Lambda=0$ model, the full data excludes
it to a high level of significance.
}
\label{fig:fitperlmutterflat}
\end{figure}

Figure  \ref{fig:datafit} shows the $d_L(z)$ obtained from the supernova data and three theoretical curves
all of which are for $k=0$ models containing non relativistic matter and cosmological constant. 
The data used here is based on the redshift magnitude relation of 54 supernova (excluding
6 outliers from a full sample of 60) and that of SN 1997ff at $z=1.755$; the magnitude used
for SN 1997ff has been corrected for lensing effects \cite{benitez}.
The best fit curve has $\Omega_{\rm NR}\approx 0.32,\Omega_\Lambda\approx 0.68$. In this analysis, 
one had treated $\Omega_{\rm NR}$ and the absolute magnitude $ M $ as free parameters
(with $\Omega_{\rm NR}+\Omega_\Lambda=1)$
and has done a simple best fit for both.  
The corresponding best fit value for ${\mathcal M}$ is ${\mathcal M} = 23.92 \pm 0.05$.
Frame (a) of figure \ref{fig:fitperlmutterflat} shows the confidence interval (for 68 \%, 90 \% and 99 \%)
in the $\Omega_{\rm NR}-{\mathcal M}$ for the flat models. It is obvious that
most of the probability is concentrated around the best fit value.
We shall discuss frame (b) and frame (c) later on.
(The discussion here is based on \cite{tptirthsn}.)

%%%%%%%%%%%%%%%

%%%%%%%%%%%%%%%%%%

%%%%%%%%%%%%%%%

%%%%%%%%%%%%%%%%%%

\begin{figure}[ht]
\begin{center}
\includegraphics[scale=0.7,angle=-90]{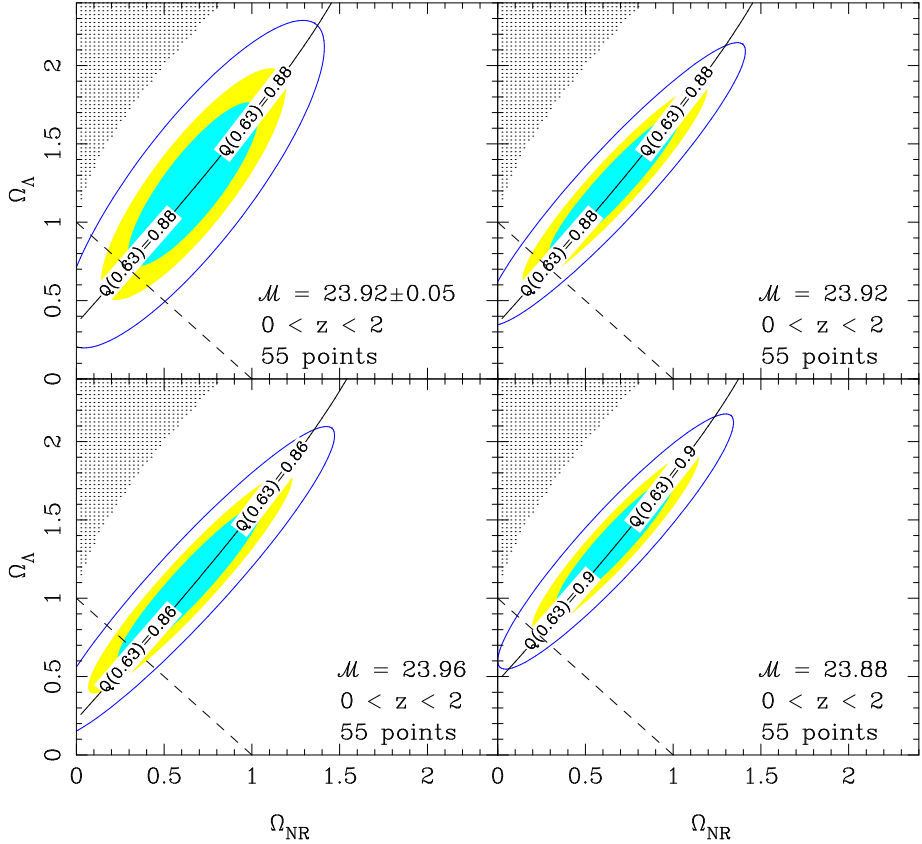}
\end{center}
\caption{Confidence contours corresponding to 68 \%, 90 \% and 99 \% based on SN data in the
$\Omega_{\rm NR} - \Omega_\Lambda$ plane. The top left frame is obtained after marginalizing over 
${\mathcal M}$ while the other three frames uses fixed values for ${\mathcal M}$. The values are chosen
to be the best-fit value for ${\mathcal M}$ and two others in the wings of $1\sigma $ limit.
The dashed line corresponds to the flat model. The unbroken slanted line
corresponds to $ H_0 d_L(z=0.63) =$ constant. It is clear that: (i) The data excludes
the $\Omega_{\rm NR}=1, \Omega_\Lambda=0$ model at a high significance level
irrespective of whether we marginalize over ${\mathcal M}$ or use an accepted
$1\sigma$ range of values for ${\mathcal M}$. (ii) The shape of the confidence contours
are essentially determined by the value of the luminosity distance at $z\approx 0.6$.
}
\label{fig:fitperlmutterallz}
\end{figure}

The confidence intervals in the $\Omega_\Lambda - \Omega_{\rm NR}$ plane are 
shown in figure \ref{fig:fitperlmutterallz} for the full data. The confidence regions
in the top left frame  are obtained
after marginalizing over ${\mathcal M}$. (The best fit value with $1\sigma$ error is 
indicated in each panel and the confidence contours correspond to 68 \%, 90 \% and 99 \%.)
The other three frames show the corresponding result with a constant value for
${\mathcal M}$ rather than by marginalizing over this parameter. The three frames
correspond to the mean value and two values in the wings of $1\sigma$ from the mean.
The dashed line connecting the points (0,1) and (1,0) correspond to a universe with
$\Omega_{\rm NR} + \Omega_\Lambda =1$.
From the figure we can conclude that: (i) The results do not change significantly whether we marginalize over
${\mathcal M}$ or whether we use the best fit value.  
This is a direct consequence of the result in frame (a) of figure (\ref{fig:fitperlmutterflat}) which
shows that the probability is sharply peaked. (ii) The results exclude the $\Omega_{\rm NR}=1,
\Omega_\Lambda=0$ model at a high level of significance  in spite of the uncertainty in
${\mathcal M}$.

%%%%%%%%%%%%%%%

%%%%%%%%%%%%%%%%%%

\begin{figure}[htp]
\begin{center}
\includegraphics[scale=0.5,angle=-90]{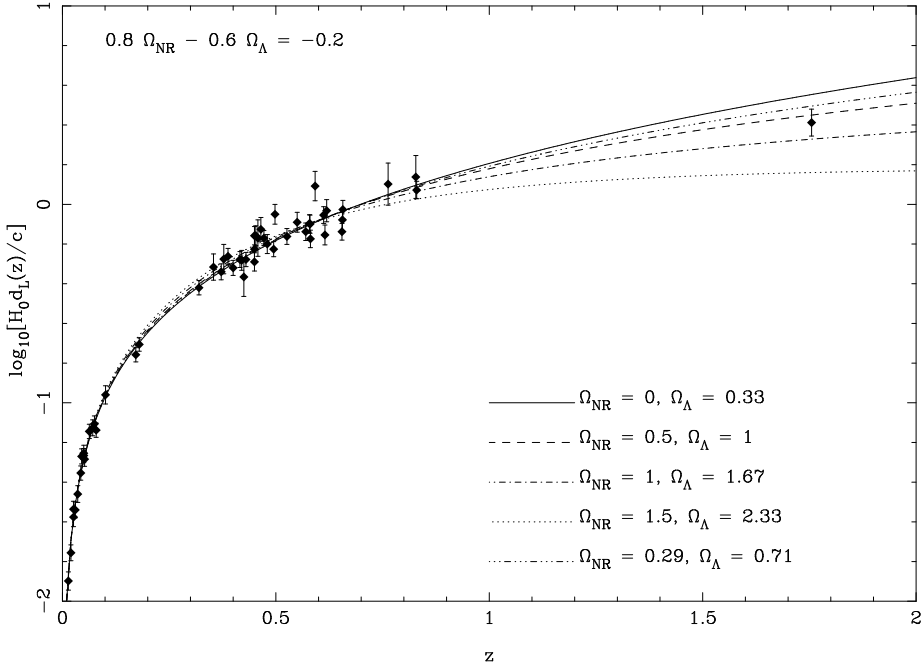}
\end{center}
\caption{The luminosity distance for a class of models with widely varying $\Omega_{\rm NR},\Omega_\Lambda$ but with a constant value for  $q \equiv (0.8 \Omega_{\rm NR} - 0.6 \Omega_\Lambda)$ are shown
in the figure. It is clear that when $q$ is fixed, low redshift observations cannot distinguish
between the different models even if $\Omega_{\rm NR}$ and $\Omega_\Lambda$ vary significantly.
}
\label{fig:datadeg1}
\end{figure}

The slanted shape of the probability ellipses shows that a particular linear combination of 
$\Omega_{\rm NR}$ and $\Omega_\Lambda$ is selected out by these observations \cite{goobarperl}.
This feature, of course, has nothing to do with supernova and arises purely
because the luminosity distance $d_L$ depends strongly on a particular 
linear combination of $\Omega_\Lambda$ and $\Omega_{\rm NR}$, as 
 illustrated in figure \ref{fig:datadeg1}.
In this figure, $\Omega_{\rm NR},\Omega_\Lambda$
 are treated as free parameters [without the $k=0$ constraint]
but a particular linear combination
$q\equiv (0.8\Omega_{\rm NR} - 0.6 \Omega_\Lambda)$ is held fixed. The  $d_L$ is not very sensitive to
individual values of $\Omega_{\rm NR},\Omega_\Lambda$  at low redshifts
 when $(0.8\Omega_{\rm NR} - 0.6 \Omega_\Lambda)$ is in the range of
$(-0.3,-0.1)$. Though some of
the models have unacceptable parameter values (for other reasons), supernova measurements alone cannot 
rule them out. Essentially the data at
$z<1$ is degenerate on the linear combination of parameters used to construct
the variable $q$.  The supernova data shows that most likely region is bounded by
$-0.3 \lesssim (0.8\Omega_{\rm NR} - 0.6 \Omega_\Lambda) \lesssim -0.1$.
In figure  \ref{fig:fitperlmutterallz}  we have also over plotted the line corresponding to $ H_0 d_L(z=0.63)=$ constant. The coincidence of this line (which roughly corresponds to $d_L$ at a redshift in the middle of the data) with the probability ellipses indicates that it is this quantity which essentially determines the nature of the result.

%%%%%%%%%%%%%%%

%%%%%%%%%%%%%%%%%%

We saw earlier that the presence of cosmological constant leads to an accelerating phase in the universe
which --- however ---is not obvious from the above figures. 
To see this explicitly one needs to display the data in the  $\dot a$
vs $a$ plane, which is done in
figure \ref{fig:adotvsaomegam}. 
Direct observations of supernova is  converted into $d_L(z)$ keeping 
 $M$ a free parameter. The $d_L$ is converted into  $d_H(z)$ 
 assuming $k=0$ and using (\ref{hinvz}). A best fit analysis, keeping $(M,\Omega_{\rm NR})$
 as free parameters now lead to the results shown in figure \ref{fig:adotvsaomegam}, which
confirms the
 accelerating phase in the evolution of the universe.
The curves which are over-plotted correspond to a cosmological model
with $\Omega_{\rm NR} + \Omega_\Lambda =1$. The best fit curve
has $\Omega_{\rm NR} = 0.32, \Omega_\Lambda=0.68$.

%%%%%%%%%%%%%%%

%%%%%%%%%%%%%%%%%%

  In the presence of the \cc, the universe accelerates at low redshifts while
  decelerating at high redshift. Hence, in principle, low redshift supernova data
  should indicate the evidence for acceleration. In practice, however, it is 
  impossible to rule out any of the cosmological models using low redshift
  ($z\lesssim 0.2$) data as is evident from figure \ref{fig:adotvsaomegam}.
  On the other hand, high redshift supernova data {\it alone} cannot be used to
  establish the existence of a \cc. The data for ($z\gtrsim 0.2$) in
  figure \ref{fig:adotvsaomegam} can be moved vertically up and made
  consistent with the decelerating $\Omega=1$ universe by choosing the
  absolute magnitude $M$ suitably. It is the interplay between both the high
  redshift and low redshift supernova which leads to the result quoted above.

\begin{figure}[ht]
\begin{center}
\includegraphics[scale=0.5,angle=-90]{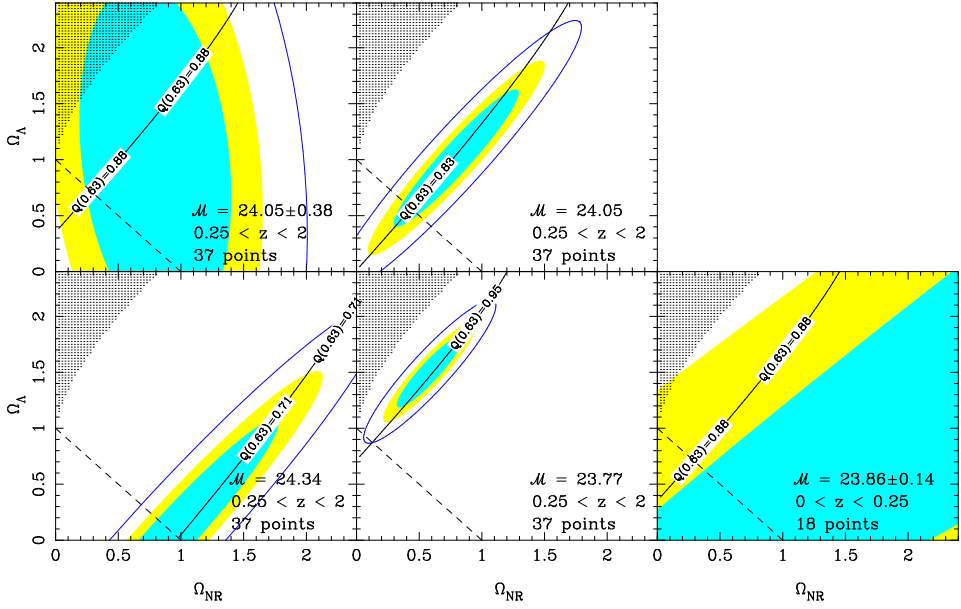}
\end{center}
\caption{Confidence contours corresponding to 68 \%, 90 \% and 99 \% based on SN data in the
$\Omega_{\rm NR} - \Omega_\Lambda$ plane using either the low-z data (bottom right
frame) or high-z data (the remaining four frames). The bottom right and the top left 
frames are obtained by marginalizing over 
  ${\mathcal M}$ while 
 the remaining three  uses fixed values for ${\mathcal M}$. The values are chosen
to be the best-fit value for ${\mathcal M}$ and two others in the wings of $1\sigma $ limit.
The dashed line corresponds to the flat model. The unbroken slanted line
corresponds to $ H_0 d_L(z=0.63) =$ constant. It is clear that:  (i)  The $1\sigma$ error  
in top left frame 
($0.38$) has
gone up by nearly eight times compared to the value ($0.05$) obtained using all data 
(see figure \ref{fig:fitperlmutterallz})
and the results are sensitive to the value of ${\mathcal M}$.
   (ii) The data can exclude the $\Omega_{\rm NR}=1,
\Omega_\Lambda =0$ model if the mean or low-end value of ${\mathcal M}$ is used
[see the two middle frames].
But, for the high-end of allowed $1\sigma$ range of ${\mathcal M}$, we cannot exclude
the 
$\Omega_{\rm NR}=1,
\Omega_\Lambda =0$ model [see the bottom left frame]. (iii) The low-z data [bottom right] 
cannot exclude any of the models.
}
\label{fig:fitperlmutterhighlowz}
\end{figure}

This important result can be brought out more quantitatively along the following lines.
 The data displayed in figure \ref{fig:adotvsaomegam}  divides 
the supernova into two natural classes: low redshift ones in the range $0<z\lesssim 0.25$ (corresponding to the accelerating phase of the universe)
and the high redshift ones in the range $0.25\lesssim z \lesssim 2$ (in the decelerating phase of the universe).
One can repeat all the statistical analysis for the full set as well as for the two
individual low redshift and high redshift data sets. Frame (b) and (c) of figure
\ref{fig:fitperlmutterflat} shows the confidence interval based on low redshift 
data and high redshift data separately. It is obvious that the $\Omega_{\rm NR}=1$ model
cannot be ruled out with either of the two data sets!
But, when the data sets are combined --- because of the angular slant of the 
ellipses --- they isolate a best fit region around $\Omega_{\rm NR} \approx 0.3$.
This is also seen in figure \ref{fig:fitperlmutterhighlowz} which plots the confidence intervals using just the high-z and low-z data separately. The right most frame in the bottom row is based on the low-z data alone (with marginalization over ${\mathcal M}$) and  this data cannot be used to  discriminate between cosmological models effectively. This is because the $d_L$ at low-z is only very weakly dependent on the cosmological parameters. So, even though the acceleration of the universe is a low-z phenomenon, we cannot reliably determine it using low-z data alone. The top left frame has the corresponding result with high-z data. As we stressed before, the $\Omega_{\rm NR}=1$ model cannot be excluded on the basis of the high-z data alone either. This is essentially because of the nature of probability contours seen earlier in frame (c) of 
figure \ref{fig:fitperlmutterflat}. The remaining 3 frames (top right, bottom left and bottom middle) show the corresponding results in which fixed values of ${\mathcal M}$ --- rather than by marginalizing over ${\mathcal M}$. Comparing these three figures with the corresponding three frames in \ref{fig:fitperlmutterallz} in which {\em all}
data was used, one can  draw the following conclusions: (i) The best fit value
for ${\mathcal M}$ is now ${\mathcal M}=24.05 \pm 0.38$; the $1\sigma$ error has now
gone up by nearly eight times compared to the result (0.05) obtained using all data.
Because of this spread, the results are sensitive to the value of ${\mathcal M}$ that one uses, 
unlike the situation in which all data was used.   (ii) Our conclusions will now depend on ${\mathcal M}$. 
For the mean value and lower end of ${\mathcal M}$, the data can exclude the $\Omega_{\rm NR}=1,
\Omega_\Lambda =0$ model
[see the two middle frames of figure \ref{fig:fitperlmutterhighlowz}].
But, for the high-end of allowed $1\sigma$ range of ${\mathcal M}$, we cannot exclude
the 
$\Omega_{\rm NR}=1,
\Omega_\Lambda =0$ model [see the bottom left frame of figure \ref{fig:fitperlmutterhighlowz}].

%%%%%%%%%%  FIGURE   %%%%%%%%%%%%%%%%%%%%%%%%%%%%%%%%

%%%%%%%%%%%%%%%%%%%%%%%%%%%%%%%%%%%%%%%%%%%%%%%%%%%

While these observations have enjoyed significant popularity, certain
key points which underly these analysis need to be stressed.
(For a sample of 
views which goes against the main stream, see  \cite{attila,mmr}).     
\begin{itemize}
\item
 The basic approach uses the supernova type I light curve
as a standard candle. While this is generally accepted, it must be remembered
that we do not have a sound theoretical understanding
of the underlying emission process.
\item
 The supernova data and fits are dominated by the region in the
parameter space around ($\Omega_{\rm NR}, \Omega_\Lambda) \approx (0.8, 1.5)$
which is strongly disfavoured by several other observations.
If this disparity is due to some other unknown systematic
effect, then this will have an effect on the estimate given above.
\item
 The statistical issues involved in the analysis of this
data to obtain best fit parameters are non trivial.
As an example of how the claims varied over time, let us note that
 the analysis of the first 7 high redshift SNe
Ia gave a value for $\Omega_{\rm NR} $ which is consistent with unity:
$\Omega_{\rm NR} = (0.94^{+0.34}_{-0.28})$. However, adding a single
$z=0.83$ supernova for which good HST data was available, lowered the
value to $\Omega_{\rm NR} = (0.6 \pm 0.2)$. More recently, the analysis
of a larger data set of 42 high redshift SNe Ia gives the results quoted above.
\end{itemize}
 
\subsection{Age of the universe and cosmological constant}\label{ageuniv}

\begin{figure}[ht]
\begin{center}
\includegraphics[scale=0.5]{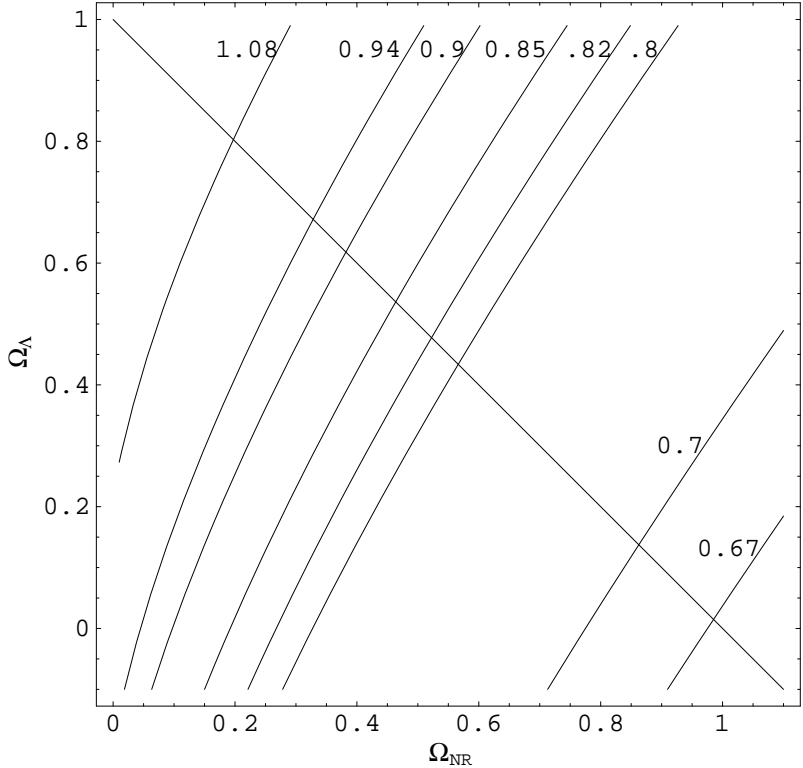}
\end{center}
\caption{Lines of constant $H_0t_0$  in the
 $\Omega_{\rm NR} - \Omega_\Lambda$ plane.
The eight lines are for $H_0t_0 = (1.08, 0.94, 0.9, 0.85, 0.82, 0.8, 0.7, 0.67) $  as shown.
 The diagonal line is the contour for models with $\Omega_{\rm NR} + \Omega_\Lambda  =1$. }
\label{fig:omnrh}
\end{figure}

From equation (\ref{qomeganr}) we can also determine the current age of the  universe
by the integral
\begin{equation}
H_0 t_0 = \int_0^1 \frac{dq}{\sqrt{2(E-V)}}
\label{currentunivage}
\end{equation}
Since most of the contribution to this integral comes from late times, we can ignore the
radiation term and set $\Omega_R \approx 0$.
When both $\Omega_{\rm NR}$ and $\Omega_\Lambda $ are present and are arbitrary, the age of the
universe is determined by the integral
\begin{eqnarray}
 H_0t_0 &=& \int_0^\infty \frac{dz}{(1+z) \sqrt{\Omega_{\rm NR}(1+z)^3 +\Omega_\Lambda }}
 \nonumber \\
& \approx & \frac{2}{3} \left(  0.7\, \Omega_{\rm NR} - 0.3\Omega_\Lambda  + 0.3\right)^{-0.3}
 \label{hoto}
 \end{eqnarray} 
The integral, which cannot be expressed in terms of elementary functions, is well approximated
by the numerical fit given in the second line.
Contours of constant $H_0t_0$  based on the (exact) integral are shown in figure \ref{fig:omnrh}.
It is obvious that, for a given $\Omega_{\rm NR}$, the age is higher
for models with $\Omega_\Lambda  \ne 0$.

%%%%%%%%%%%%%%%      FIGURE        %%%%%%%%%%%%%%%%%%

%%%%%%%%%%%%%%%%%%%%%%%%%%%%%%%%%%%%%%%%%%%%

 Observationally,
there is  a consensus \cite{freedman,mould00}
 that $h\approx 0.72 \pm 0.07$ and $t_0 \approx 13.5\pm 1.5$ Gyr
\cite{chaboyer}. 
 This will give $H_0t_0 = 0.94 \pm 0.14$.
Comparing this result with the fit in (\ref{hoto}), one can immediately draw several conclusions:
\begin{itemize}
\item
 If $\Omega_{\rm NR} >0.1$, then $\Omega_\Lambda  $ is non zero
if $H_0t_0 > 0.9$. A more reasonable assumption of $\Omega_{\rm NR} > 0.3$ we
will require non zero $\Omega_\Lambda$ if $H_0t_0 >0.82$.
\item
 If we take $\Omega_{\rm NR} =1, \Omega_\Lambda  =0$ and demand 
$t_0 > 12 $ Gyr (which is a conservative lower bound from stellar
ages) will require $h <0.54$.
Thus a purely matter dominated $\Omega=1$ universe would
require low Hubble constant which is contradicted by
most of the observations.
\item
 An open model with $\Omega_{\rm NR} \approx 0.2, \Omega_\Lambda =0$ will
require $H_0t_0 \approx 0.85$. This still requires
ages on the lower side but values like $h \approx 0.6, t_0 \approx 13.5 $ Gyr
are acceptable within error bars.
\item
 A straightforward interpretation of observations suggests maximum
likelihood for 
$H_0t_0 = 0.94$. This can be consistent with a $\Omega=1$ model only if
$\Omega_{\rm NR} \approx 0.3, \Omega_\Lambda  \approx 0.7$.
\end{itemize}

If the universe is populated by dust-like matter (with $w=0$) and another 
component with an equation of state parameter $w_X$, then the age
of the universe will again be given by an integral similar to the one in equation (\ref{hoto})
with $\Omega_\Lambda$ replaced by $\Omega_X(1+z)^{3(1+w_X)}$. This will give
\begin{eqnarray}
 H_0t_0 &=& \int_0^\infty \frac{dz}{(1+z) \sqrt{\Omega_{\rm NR}(1+z)^3 + \Omega_X(1+z)^{3(1+w_X)}}}
 \nonumber\\
 &=&\int_0^1 dq\, \left( \frac{q}{\Omega_{\rm NR} +\Omega_X q^{-3w_X}}\right)^{1/2}
 \end{eqnarray} 
 The integrand varies from 0 to $(\Omega_{\rm NR} + \Omega_X)^{-1/2}$ in the range of 
 integration for $w<0$ with the rapidity of variation decided by $w$. As a result, $H_0t_0$
 increases rapidly as $w$ changes from 0 to $-3$ or so and then saturates to a plateau.
 Even an absurdly negative value for $w$ like $w=-100$ gives $H_0t_0$ of only about
 1.48. This shows that even if some exotic dark energy is present in the universe with 
 a constant, negative $w$, it cannot increase the age of the universe beyond about
 $H_0t_0 \approx 1.48$.

The comments made above pertain to the current age of the  universe.
  It is also possible to obtain an expression similar to (\ref{currentunivage})  for the age
  of the universe at any given redshift $z$
  \begin{equation}
   H_0t(z) =\int_z^\infty \frac{dz'}{(1 + z')h(z')}; \qquad h(z) = \frac{H(z)}{H_0}
\label{ageatz}  
\end{equation}
and use it to constrain $\Omega_\Lambda$.
  For example, the existence of high redshift galaxies with evolved stellar
  population, high redshift radio galaxies and age dating of high redshift
  QSOs can all be used in conjunction with this equation to put constrains
  on $\Omega_\Lambda$ \cite{alcaniz99,dunlop96,dunlop98,krauss97,peacock98,ytk98}.
  Most of these observations require either $\Omega_\Lambda \ne 0$
  or $\Omega_{\rm tot} <1$ if they have to be consistent with $h \gtrsim 0.6$.
  Unfortunately, the interpretation of these observations at present requires
  fairly complex modeling and hence the results are not water tight.

  \subsection{Gravitational lensing and the \cc\ }\label{gravlcc}

   Consider a distant source at redshift $z$ which is lensed by some intervening 
  object. The lensing is most effective if the lens is located 
  midway between the source and the observer (see, eg., page  196 of \cite{tapvol3}). 
  This distance will be
   $(r_{\rm em}/2)$ 
  if the distance to the source is $r_{\rm em}$.
  (To be rigorous, one should be using angular diameter distances rather than
  $r_{\rm em}$ for this purpose but the essential conclusion does not change.)
    To see how this
   result depends on cosmology, let us consider a  source
 at redshift $z=2$, and a lens, located optimally, in:
(a) $\Omega=1$ matter dominated universe, (b) a very low density matter
dominated universe in the limit of $\Omega \to 0$, (c) vacuum dominated
universe with $\Omega_\Lambda =\Omega_{\rm tot}$.
In case (a), $d_H \equiv H^{-1}(z)  \propto (1+z)^{-3/2}$, so that 
\begin{equation}
 r_{\rm em}(z) \propto \int_0^z d_H(z) dz \propto \left( 1 - \frac{1}{\sqrt{1+z}}\right)
 \end{equation}
The lens redshift is determined by the equation 
 \begin{equation}
  \left( 1 - \frac{1}{\sqrt{1+z_L}}\right) = \frac{1}{2} \left( 1 - \frac{1}{\sqrt{1+z}}\right)
  \end{equation}
For $z=2$, this gives $z_L = 0.608$. 
   For case (b),  $a\propto t$ giving $d_H\propto (1+z)^{-1}$ and
$r_{\rm em}(z) \propto \ln (1+z)$. The equation to be solved is
$(1+z_L) = (1+z)^{1/2}$ which gives $z_L = 0.732$ for
$z=2$. 
Finally, in the case of (c),  $d_H$ is a constant giving $r_{\rm em}(z) \propto z$ and
$z_L = (1/2)z$. Clearly, the lens redshift
is larger for vacuum dominated universe compared to the
matter dominated universe of any $\Omega$.
When one considers a distribution of lenses, this will affect the probability for 
lensing in a manner which depends on $\Omega_\Lambda$. From the 
observed statistics of lensing, one can put a bound on $\Omega_\Lambda$.

More formally,  one can compute the probability 
 for
a source at redshift $z_s$ being lensed in a $\Omega_\Lambda + \Omega_{\rm NR} = 1$ universe (relative to the corresponding probability in a $\Omega_{\rm NR} =1, \Omega_\Lambda=0$ model). This relative probability is nearly five times larger at $z_s=1$ and about thirteen
  times larger for $z_s=2$  in a purely \cc\ dominated universe \cite{fuku91,fuku90,fuku92,cohn98,turner90,carolcomb1,carolcomb3}. This effect quantifies the fact that the physical 
volume associated with unit redshift interval is larger in models with \cc\ and
hence the probability that a light ray will encounter a galaxy is larger 
in these cases.
 
  Current analysis of lensing data gives somewhat differing constraints based on
  the assumptions which are made
   \cite{kochanek93,kochanek96,maoz93,falco98}; but typically 
  all these constraints are about $\Omega_\Lambda \lesssim 0.7$.  
  The  result \cite{class} from Cosmic Lens All Sky Survey (CLASS), for example,
  gives $\Omega_{\rm
       NR}=0.31^{+0.27}_{-0.14}$ (68\%) $^{+0.12}_{-0.10}$ (systematic)
       for a $k=0$ universe.

  \subsection{Other geometrical tests} \label{geotests}

  The existence of a maximum for $d_A(z)$ is a generic feature of cosmological
  models with $\Omega_{\rm NR}>0$. For a $k=0, \Omega_{\rm NR}=1$ model,
  the maximum occurs at $z_{\rm max} \approx 1.25$ and $z_{\rm max}$ increases 
   as $\Omega_\Lambda $ is increased. To use this as
  a cosmological test, we require a class of objects with known
  transverse dimension and redshift. The most reliable quantity
  used so far corresponds to the physical wavelength
  acoustic vibrations in the baryon-photon gas at $z\approx 10^3$.
  This length scale is imprinted in the temperature anisotropies
  of the CMBR and the angular size of these anisotropies will
  depend on $d_A$ and hence on the cosmological parameters;
  this is discussed in section \ref{cmbrani}. In principle,
  one could also use angular sizes of galaxies,
  clusters of galaxies, or 
  radio galaxies \cite{hoyle59,kapahi89,cham90}.
  Unfortunately, understanding of different physical effects and the redshift
  evolution of these sources make this a difficult test in practice.
  
   There is another geometrical feature of the universe in which angular
   diameter distance plays an interesting role. In a closed Friedmann model
   with $k=+1$, there is  possibility that an observer at $\chi =0$ will be able to 
   receive the light from the antipodal point $\chi =\pi$. In a purely
   matter dominated universe, it is easy to see that the light ray from the
   antipodal point $\chi =\pi$ reaches $\chi =0$ exactly at the time of 
   maximum expansion; therefore, in a closed, matter dominated universe,
   in the expanding phase, no observer can receive light from the antipodal
   point.
   The situation, however, is different in the presence of cosmological constant.
   In this case, $d_A(z) \propto (1+z)^{-1} \sin \mu$ where
   \begin{equation}
   \mu = |\Omega_{tot} - 1|^{\frac{1}{2}}\int_0^z {dz'\over
h(z')}, \quad h(z) = \frac{H(z)}{H_0}
   \end{equation}
   It follows that $d_A \to 0$ when $\mu \to \pi$. Therefore, the 
   angular size of an object near the antipodal point can diverge
   making the object extremely bright in such a universe. Assuming that
   this phenomena does not occur up to, say $z=6$, will imply that the 
   redshift of the antipodal point $z_a(\Omega_\Lambda, \Omega_{\rm NR})$
   is larger than 6. This result can be used to constrain the cosmological
   parameters \cite{durrer90,lahav91,coop98}
   though the limits obtained are not as tight as some of the other tests.
     
  Another test which can be used to obtain a handle on the geometry of the universe
  is usually called Alcock-Paczynski curvature test \cite{alpac}. The basic idea
  is to use the fact that when any spherically symmetric system at high redshift is observed,
  the cosmological parameters enter differently in the characterization of radial and transverse
  dimensions. Hence any system which can be approximated a priori to be intrinsically spherical
  will provide a way of determining cosmological parameters. The correlation function of SDSS
  luminous red galaxies seems to be promising in terms of both depth and density
  for applying this test (see for a detailed discussion, \cite{mat1,mat2}). 
  The main sources of error  arises from non linear clustering and the bias
  of the red galaxies, either of which can be a source of systematic error. A variant of 
  this method was proposed using observations of Lyman-$\alpha$ forest and 
  compare the correlation function along the line of sight and transverse to the 
  line of sight. In addition to the modeling uncertainties, successful application 
  of this test will also require about 30 close quasar pairs \cite{hui99,mcd99}.

\section{Models with evolving cosmological ``constant"}\label{ccmodels}

The observations which suggest the existence of  non-zero cosmological constant --- discussed in the last section ---
raises serious theoretical problems which we mentioned in section \ref{facescc}. These 
difficulties have led
people to consider the possibility that the dark energy in the universe is not just a cosmological constant but is of more complicated nature, evolving with time.
Its value today can then be more naturally set by the current expansion rate  rather than predetermined
earlier on --- thereby providing a solution to the cosmological constant problems.

Though a host of models have been constructed based on this hope,  none of them provides a satisfactory solution to the problems of fine-tuning.
Moreover, all of them involve an evolving equation of state parameter $w_X(a)$ for the unknown (``X") dark
energy component, thereby taking away all predictive power from cosmology \cite{tptachyon}. Ultimately, however, this
issue needs to settled observationally by checking whether $w_X(a)$ is a constant [equal to $-1$, 
for the cosmological constant] at all epochs
or whether it is indeed varying with $a$. We shall now discuss several observational and theoretical issues
connected with this theme.

 While the complete knowledge of the $T^a_b$ [that is, the knowledge of the right hand side
  of (\ref{frwone})] uniquely determines $H(a)$, the converse is not true. If we know only the function
  $H(a)$, it is not possible to determine the nature of the energy density
  which is present in the universe. We have already seen that  geometrical measurements
  can {\it only} provide, at best, the functional form of $H(a)$. It follows that purely geometrical
  measurements of the Friedmann universe will never allow us to determine the material
  content of the universe. 
  
  [The only exception to this rule is when we assume that
{\it each of the components in the universe has constant
$w_i$}. This is fairly strong assumption and, in fact,  will allow us to determine the
components of the universe from the knowledge of the function $H(a)$.
To see this, we first
 note that the term $(k/a^2)$ in equation (\ref{qlsone}) can be thought of
as contributed by a hypothetical species of matter with
$w=-(1/3)$. Hence equation (\ref{qlsone}) can be
written in the form
\begin{equation} \frac{\dot a^2}{a^2} = H_0^2 \sum_i \Omega_i\left( \frac{a_0}{a}\right)^{3(1+w_i)}
\label{thirty}
\end{equation}
with a term having $w_i = -(1/3)$ added to the sum.
Let
$\alpha \equiv
3(1+w)$
and $\Omega(\alpha)$ denote the fraction of the critical density contributed
by matter with $w=(\alpha/3)-1$. (For discrete values of $w_i$ and $\alpha_i$,
the function $\Omega(\alpha)$ will be a sum of Dirac delta functions.)
In the continuum
limit,  equation (\ref{thirty}) can be rewritten as
\begin{equation}
H^2 = H_0^2\int_{-\infty}^\infty d\alpha\ \Omega(\alpha)\  e^{-\alpha q}
\end{equation}
where $(a/a_0)= \exp(q)$. The function $\Omega(\alpha)$
is assumed to have finite support (or decrease fast enough) for the expression on the
right hand side to converge.
If the observations determine the function $H(a)$, then the
left hand side can be expressed as a function of $q$. An inverse Laplace
transform of this equation will then determine the form of $\Omega(\alpha)$
thereby determining the composition of the universe, as long as all
matter can be described by an equation of state of the form $p_i=w_i\rho_i$ with $w_i=$ constant 
for all $i=1,....,N$.]

More realistically one is interested in models which has a  complicated form of $w_X(a)$
for which  the above analysis is not applicable.
Let us divide the source energy density into two components: $\rho_{k}(a)$, which is known from
independent observations and a component $\rho_X(a)$ which is not known. From (\ref{frwone}), it follows that
\begin{equation}
\frac{8\pi G}{3}\rho_X(a)=H^2(a)(1-Q(a));\quad Q(a)\equiv\frac{8\pi G\rho_k(a)}{3 H^2(a)}
\label{findingrhox}
\end{equation}
Taking a  derivative of $\ln \rho_X(a)$  and using (\ref{evolenergy}), it is easy to obtain the relation
\begin{equation}
w_X(a)=-\frac{1}{3}\frac{d}{d\ln a}\ln[(1-Q(a))H^2(a)a^3]
\label{findingw}
\end{equation}
If geometrical observations of the universe give us $H(a)$ and other observations give us $\rho_k(a)$ then
one can determine $Q$ and thus $w_X(a)$. While this is possible, in principle the
uncertainties in measuring both $H$ and $Q$ makes this a nearly impossible route to follow
in practice. In particular, one is interested in knowing whether
$w$ evolves with time or a constant and this turns out to be a very difficult task observationally.
We shall now briefly discuss some of the issues.

\subsection{Parametrized equation of state and cosmological observations}\label{paraeqn}

  One simple, phenomenological,  procedure for comparing observations with theory is to parameterize
  the function $w(a)$ in some suitable form and determine a finite set of parameters
  in this function using the observations. Theoretical models  can then be reduced
  to a finite set of parameters which can be determined by this procedure.  To illustrate this
  approach, and the difficulties in determining the  equation of state of dark energy from the observations, we shall assume that $w(a)$ is given by the simple form: $w(a) =w_0+w_1(1-a)$;
  in the $k=0$ model (which we shall assume for simplicity), $w_0$ measures
  the current value of the parameter and $-w_1$ gives its rate of change at the present
  epoch. In addition to simplicity, this parameterization has the advantage of giving
  finite $w$ in the entire range $0<a<1$. 
  
  %%%%%%%%%%%%     FIGURE        %%%%%%%%%%%%%%%%%%%%

\begin{figure}[ht]
\begin{center}
\includegraphics[scale=0.6,angle=-90]{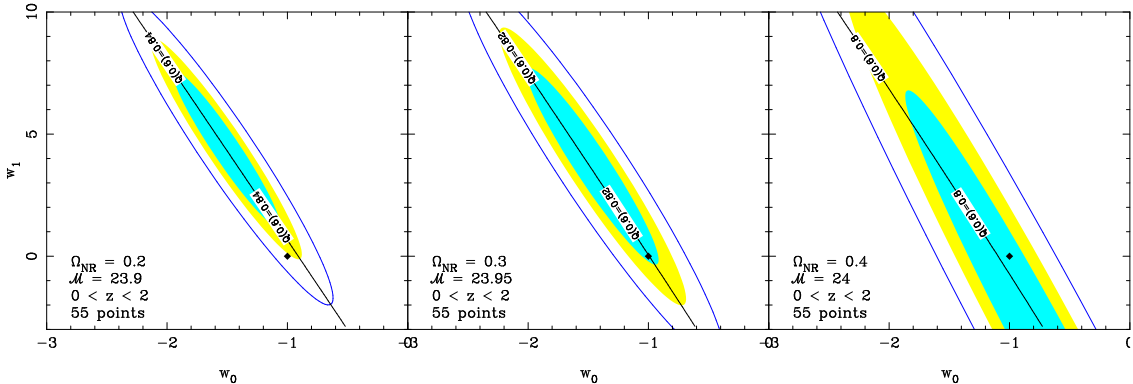}
\end{center}
\caption{Confidence interval contours in the $w_0-w_1$ plane arising from
  the full supernova data, for flat models with   $\Omega_{\rm NR} + \Omega_\Lambda
  =1$. The three frames are for $\Omega_{\rm NR} =(0.2,0.3,0.4)$. The data 
  cannot rule out cosmological constant with $w_0= -1, w_1=0$.
  The
  slanted line again corresponds to $H_0 d_L (z=0.63) = $ constant and shows that the shape
  of the probability ellipses arises essentially from this feature. 
}
\label{fig:permutterw0w1}
\end{figure}

  %%%%%%%%%%%%%%%%%%%%%%%%%%%%%%%%%%%%%%%

  Figure \ref{fig:permutterw0w1} shows confidence interval contours in the $w_0-w_1$ plane arising from
  the full supernova data,  obtained by assuming that $\Omega_{\rm NR} + \Omega_\Lambda
  =1$. The three frames are for $\Omega_{\rm NR} =(0.2,0.3,0.4)$. The following features are 
  obvious from the figure: (i) The cosmological constant corresponding to $w_0= -1, w_1=0$
  is a viable candidate and cannot be excluded. (In fact, different analysis of many observational
  results lead to this conclusion consistently; in other words, at present there is no  observational
  motivation to assume $w_1\ne 0$.) (ii) The result is sensitive to the value of $\Omega_{\rm NR}$ which
  is assumed. This is understandable from equation (\ref{findingw}) which shows that $w_X(a)$ depends
  on both $Q \propto \Omega_{\rm NR}$ and $H(a)$. (We shall  discuss this dependence
  of the results on $\Omega_{\rm NR}$ in greater detail below).  (iii) Note that the axes are not in equal units
  in figure \ref{fig:permutterw0w1}. The observations can determine $w_0$ with far greater accuracy than $w_1$. (iv) The
  slanted line again corresponds to $H_0 d_L (z=0.63) = $ constant and shows that the shape
  of the probability ellipses arises essentially from this feature. 
  
  In summary, the current data definitely supports a negative pressure component with $w_0 < -(1/3)$
  but is completely consistent with $w_1 =0$. If this is the case, then the cosmological constant is the 
  simplest candidate for this negative pressure component and there is very little observational motivation
  to study other models with varying $w(a)$. On the other hand, the cosmological constant has well
  known theoretical problems which could possibly be alleviated in more sophisticated
  models with varying $w(a)$. With this motivation, there has been extensive amount of work in the
  last few years investigating whether improvement in the observational scenario will allow us
  to determine whether $w_1$ is non zero or not. (For a sample of references, see
         \cite{SNDE1,SNDE2,SNDE3,SNDE4,SNDE5,SNDE6,SNDE7,SNDE8,SNDE9,SNDE10,SNDE11,SNDE12,SNDE13,SNDE14,SNDE15,SNDE16}.)
  In the context of supernova based determination of $d_L$, it is possible to analyze the situation
  along the following lines \cite{tptirthsn}.

\begin{figure}[ht]
\begin{center}
\includegraphics[scale=0.5,angle=-90]{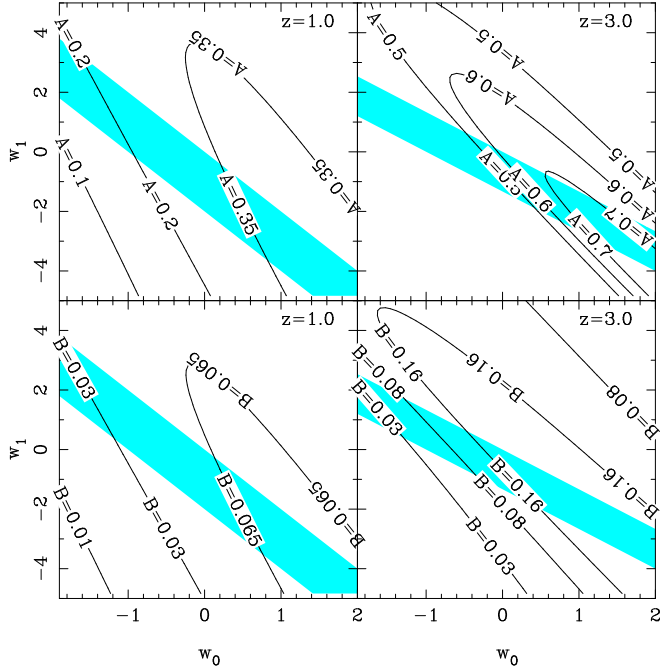}
\end{center}
\caption{Sensitivity of $d_L$ to the parameters $w_0,w_1$. The curves correspond to constant values for the percentage of change in $d_LH_0$ for unit change in $w_0$ (top frames), and $w_1$ (bottom frames). Comparison of the top and bottom frames shows that $d_LH_0$ varies by few tens of percent when $w_0$ is varied but changes by much lesser amount when$w_1$ is varied.
}
\label{fig:ab}
\end{figure}

  Since the supernova observations essentially measure $d_L(a)$, accuracy
  in the determination of
  $w_0$ and $w_1$ from  (both  the present and
  planned future \cite{snap}) supernova observations
  will crucially depend on how sensitive  $d_L$ is to the  changes in $w_0$ and
  $w_1$. A good measure of the sensitivity is provided by the two parameters
  \begin{eqnarray}
  A(z, w_0, w_1) &\equiv& \frac{d}{dw_0}\ln(d_L(z, w_0, w_1) H_0);\nonumber\\
  B(z, w_0, w_1) &\equiv& \frac{d}{dw_1}\ln(d_L(z, w_0, w_1) H_0)
  \label{abdef}
  \end{eqnarray}
  Since $d_L(z, w_0, w_1)$ can be obtained from theory, the
  parameters $A$ and $B$ can be computed form theory in a straight
  forward manner. At any given redshift $z$, we can plot contours of
  constant $A$ and $B$ in the $w_0 - w_1$ plane.
 Figure (\ref{fig:ab}) shows the result of such an analysis \cite{tptirthsn}.
 The two frames on the left are at $z=1$ and the two frames on
 the right are at $z=3$. The top frames give contours of constant
 $A$ and bottom frame give contours of constant $B$.
 From the definition in the  equation (\ref{abdef})  it is clear that $A$ and $B$
 can be interpreted as the fractional change in $d_L$ for unit change
 in $w_0,w_1$. For example, along the line marked $A=0.2$
 (in the top left frame) $d_L$ will change by 20 per cent for unit change in $w_0$. It is clear from the two top frames that for most of the 
 interesting region in the $w_0-w_1$ plane, changing $w_0$ by
 unity changes $d_L$ by about 10 per cent or more. Comparison of
 $z=1$ and $z=3$ (the two top frames) shows that
 the sensitivity is higher at high redshift, as to be expected.
 The shaded band across the picture corresponds to the region
 in which $-1 \le w(a) \le 0$ which is of primary
 interest in constraining dark energy with negative pressure.
 One concludes that determining $w_0$ from $d_L$ fairly accurately
 will not be too daunting a task.

%%%%%%%%%%%%%%     FIGURE      %%%%%%%%%%%%%%%%%

%%%%%%%%%%%%%%%%%%%%%%%%%%%%%%%%%%%%%%
 The situation, however, is quite different as regards $w_1$ as
 illustrated in the bottom two frames. For the same region of the
 $w_0 - w_1$ plane, $d_L$ changes only by a few percent
 when $w_1$ changes by unity. That is, $d_L$ is much less
 sensitive to $w_1$ than to $w_0$. It is going to be significantly
 more difficult to determine a value for $w_1$ from observations
 of $d_L$ in the near future. Comparison of $z=1$ and $z=3$ again
 shows that the sensitivity is somewhat better at high redshifts
 but only marginally.

%%%%%%%%%%%   FIGURE       %%%%%%%%%%%%%%%%%%%%

\begin{figure}[ht]
\begin{center}
\includegraphics[scale=0.5,angle=-90]{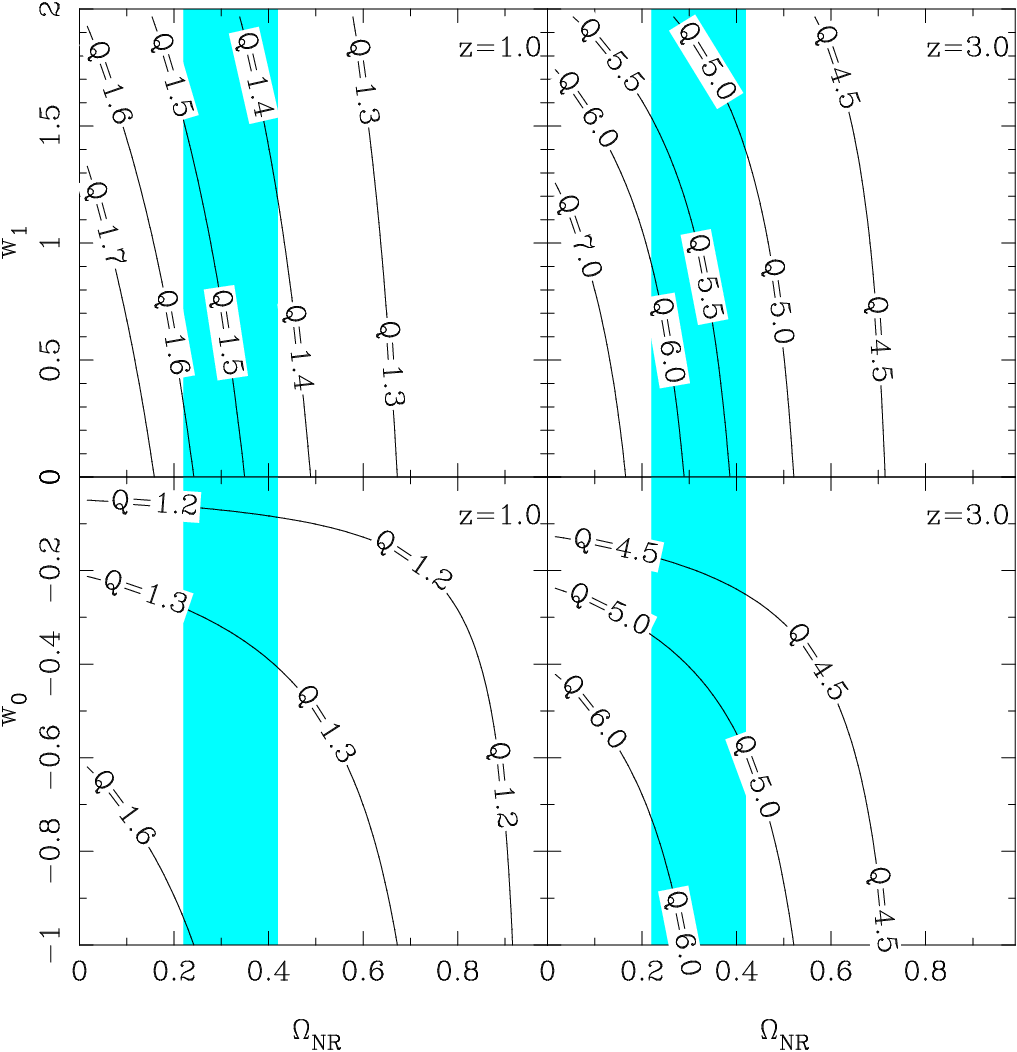}
\end{center}
\caption{Contours of constant $H_0 d_L$ in the 
 $\Omega_{\rm NR}-w_0$ and $\Omega_{\rm NR}-w_1$
 planes at two redshifts $z=1$ and $z=3$. The two top frames shows
 that a small variation of $\Omega_{\rm NR}$ in the
 allowed range of, say, ($0.2, 0.4$)corresponds to  fairly large
  variation in $w_1$  along
  the  curve of constant $d_L$.
}
\label{fig:qps}
\end{figure}

  %%%%%%%%%%%%%%%%%%%%%%%%%%%%%%%%%%%%%%%%

 The situation is made worse by the fact that $d_L$ also depends
 on the parameter $\Omega_{\rm NR}$. If varying $\Omega_{\rm NR}$
 mimics the variation of $w_1$ or $w_0$, then, one also needs
 to determine the sensitivity of $d_L$ to $\Omega_{\rm NR}$.
 Figure \ref{fig:qps} shows contours of constant $H_0 d_L$ in the 
 $\Omega_{\rm NR}-w_0$ and $\Omega_{\rm NR}-w_1$
 planes at two redshifts $z=1$ and $z=3$. The two top frames shows
 that if one varies the value of $\Omega_{\rm NR}$ in the
 allowed range of, say, ($0.2, 0.4$) one can move along 
  the  curve of constant $d_L$ and induce fairly large
  variation in $w_1$. In other words, large changes in $w_1$
  can be easily compensated by small changes in $\Omega_{\rm NR}$
  while maintaining the same value for $d_L$ at  a given redshift.
  This shows that the uncertainty in $\Omega_{\rm NR}$ introduces
  further difficulties in determining $w_1$ accurately from 
  measurements of $d_L$.  The two lower frames  show that
  the situation is better as regards $w_0$. The curves are much
  less steep and hence varying $\Omega_{\rm NR}$ does not
  induce large variations in $w_0$. We are once again 
  led to the conclusion that unambiguous determination of
  $w_1$ from data will be quite difficult. This is
  somewhat disturbing since $w_1\ne 0 $ is a clear
  indication of a dark energy component which is evolving.
   It appears that observations may not be of great help
  in ruling out cosmological constant as the major dark energy
  component. (The results given above are based on \cite{tptirthsn}; also see
  \cite{maor} and references cited therein.)

\subsection{Theoretical models with time dependent dark energy: cosmic degeneracy}\label{theorydark}
The approach in the last section was purely phenomenological and one might like to construct some physical model
which leads to varying $w(a)$. It turns out that this is fairly easy, and --- in fact ---- it is possible to construct
models which will accommodate virtually any form of evolution. We shall now discuss some examples.

  A simple form of the source with variable $w$ are   scalar fields with
  Lagrangians of different forms, of which we will discuss two possibilities:
    \begin{equation}
  L_{\rm quin} = \frac{1}{2} \partial_a \phi \partial^a \phi - V(\phi); \quad L_{\rm tach}
  = -V(\phi) [1-\partial_a\phi\partial^a\phi]^{1/2}
  \end{equation}
  Both these Lagrangians involve one arbitrary function $V(\phi)$. The first one,
  $L_{\rm quin}$,  which is a natural generalisation of the Lagrangian for
  a non-relativistic particle, $L=(1/2)\dot q^2 -V(q)$, is usually called quintessence (for
  a sample of models, see \cite{phiindustry1,phiindustry2,phiindustry3,phiindustry4,phiindustry5,phiindustry6,phiindustry7,phiindustry8,phiindustry9,phiindustry10,carolcomb2}).
    When it acts as a source in Friedman universe,
   it is characterized by a time dependent
  $w(t)$ with
    \begin{equation}
  \rho_q(t) = \frac{1}{2} \dot\phi^2 + V; \quad P_q(t) = \frac{1}{2} \dot\phi^2 - V; \quad w_q
  = \frac{1-(2V/\dot\phi^2)}{1+ (2V/\dot\phi^2)}
  \label{quintdetail}
  \end{equation}

The structure of the second scalar field  can be understood by a simple analogy from
special relativity. A relativistic particle with  (one dimensional) position
$q(t)$ and mass $m$ is described by the Lagrangian $L = -m \sqrt{1-\dot q^2}$.
It has the energy $E = m/  \sqrt{1-\dot q^2}$ and momentum $p = m \dot
q/\sqrt{1-\dot q^2} $ which are related by $E^2 = p^2 + m^2$.  As is well
known, this allows the possibility of having massless particles with finite
energy for which $E^2=p^2$. This is achieved by taking the limit of $m \to 0$
and $\dot q \to 1$, while keeping the ratio in $E = m/  \sqrt{1-\dot q^2}$
finite.  The momentum acquires a life of its own,  unconnected with the
velocity  $\dot q$, and the energy is expressed in terms of the  momentum
(rather than in terms of $\dot q$)  in the Hamiltonian formulation. We can now
construct a field theory by upgrading $q(t)$ to a field $\phi$. Relativistic
invariance now  requires $\phi $ to depend on both space and time [$\phi =
\phi(t, {\bf x})$] and $\dot q^2$ to be replaced by $\partial_i \phi \partial^i
\phi$. It is also possible now to treat the mass parameter $m$ as a function of
$\phi$, say, $V(\phi)$ thereby obtaining a field theoretic Lagrangian $L =-
V(\phi) \sqrt{1 - \del^i \phi \del_i \phi}$. The Hamiltonian  structure of this
theory is algebraically very similar to the special  relativistic example  we
started with. In particular, the theory allows solutions in which $V\to 0$,
$\dphi \to 1$ simultaneously, keeping the energy (density) finite.  Such
solutions will have finite momentum density (analogous to a massless particle
with finite  momentum $p$) and energy density. Since the solutions can now
depend on both space and time (unlike the special relativistic example in which
$q$ depended only on time), the momentum density can be an arbitrary function
of the spatial coordinate. This provides a rich gamut of possibilities in the
context of cosmology.
 \cite{cosmo1,cosmo2,cosmo3,cosmo5,cosmo6,cosmo7,cosmo8,cosmo9,cosmo10,cosmo12,cosmo13,cosmo15,cosmo16,cosmo17,cosmo18,cosmo19,cosmo20,cosmo21,cosmo22,cosmo23,armen,earl1,earl2,earl3,earl4,tpnew,tptachyon},
  This form of scalar field arises  in string theories \cite{asen} and --- for technical reasons ---
   is called a tachyonic scalar field.
   (The structure of this Lagrangian is similar to those analyzed in a wide class of models
   called {\it K-essence}; see for example, \cite{armen}. We will not discuss K-essence
   models in this review.)

   The stress tensor for the tachyonic scalar  field can be written in a
perfect fluid form
\begin{equation}
T^i_k = (\rho + p) u^i u_k - p \delta^i_k
\end{equation}
with
\begin{equation}
u_k = \frac{\del_k \phi}{\sqrt{\del^i \phi \del_i \phi}};\quad 
\rho = \frac{V(\phi)}{\sqrt{1 - \del^i \phi \del_i \phi}};\quad 
p = -V(\phi) \sqrt{1 - \del^i \phi \del_i \phi}
\end{equation}
The remarkable feature of this stress tensor is that it could be considered
as {\it the sum of a pressure less dust component and a cosmological constant} \cite{tpnew}
To show this explicitly,
we  break up the density $\rho$ and the pressure $p$
and write them in a more suggestive form as
\begin{equation}
\rho = \rho_\Lambda  + \rho_{\rm DM}; ~~
p = p_V  + p_{\rm DM}
\end{equation}
where
\begin{equation}
\rho_{\rm DM} = \frac{V(\phi) \del^i \phi \del_i \phi}
{\sqrt{1 - \del^i \phi \del_i \phi}};\ \  p_{\rm DM} = 0;\ \
\rho_\Lambda  = V(\phi) \sqrt{1 - \del^i \phi \del_i \phi};\ \
p_V  = -\rho_\Lambda
\end{equation}
This means that the stress tensor can be thought of as made up of two components
-- one behaving like a pressure-less fluid, while the other having a negative
pressure. In the cosmological context, the tachyonic field is described by:
    \begin{equation}
   \rho_t(t) = V[1-\ \dot\phi^2]^{-1/2}; \quad  P_t =-V[1-\ \dot\phi^2]^{1/2} ; \quad  w_t
  =\dot\phi^2-1
  \label{tachdetail}
  \end{equation}

   When $\dot\phi$ is small (compared to $V$ in the case of quintessence or
   compared to unity in the case of tachyonic field), both these sources have $w\to -1$ and
   mimic a cosmological constant. When $\dot \phi \gg V$, the quintessence has $w\approx 1$ leading to
   $\rho_q\propto (1+z)^6$; the tachyonic field, on the other hand, has $w\approx 0$ for $\dot\phi\to 1$
   and behaves like non-relativistic matter. In both the cases, $-1<w<1$, though it is possible to construct more complicated scalar field Lagrangians with even $w<-1$.
  (See for example, \cite{caldwell}; for some other alternatives to scalar field models, see
  for example, \cite{avelinoetal}.)
  
   Since  the quintessence field (or the tachyonic field)   has
   an undetermined free function $V(\phi)$, it is possible to choose this function
  in order to produce a given $H(a)$.
  To see this explicitly, let
   us assume that the universe has two forms of energy density with $\rho(a) =\rho_{\rm known}
  (a) + \rho_\phi(a)$ where $\rho_{\rm known}(a)$ arises from any known forms of source 
  (matter, radiation, ...) and
  $\rho_\phi(a) $ is due to a scalar field.  When $w(a)$
  is given, one can determine the $V(\phi)$ using either (\ref{quintdetail}) or (\ref{tachdetail}). For quintessence,
(\ref{quintdetail}) along with (\ref{findingrhox}) gives
\begin{eqnarray}
&&\dot\phi^2(a)=\rho(1+w)=\frac{3H^2(a)}{8\pi G}(1-Q)(1+w);\nonumber \\
&&2V(a)=\rho(1-w)=\frac{3H^2(a)}{8\pi G}(1-Q)(1-w)
\label{vfromwquint}
\end{eqnarray}
For tachyonic scalar field,
(\ref{tachdetail}) along with (\ref{findingrhox}) gives
\begin{equation}
\dot\phi^2(a)=(1+w);\quad
V(a)=\rho(-w)^{1/2}=\frac{3H^2(a)}{8\pi G}(1-Q)(-w)^{1/2}
\label{vfromwtach}
\end{equation}
Given $Q(a)$, $w(a)$ these equations implicitly determine $V(\phi)$.
We have already seen that, for any cosmological evolution specified
  by the functions $H(a)$ and $\rho_k(a)$, one can determine $w(a)$; see equation (\ref{findingw}).
  Combining (\ref{findingw}) with either (\ref{vfromwquint}) or (\ref{vfromwtach}), one can completely solve the problem.

  Let us first consider quintessence. Here, using (\ref{findingw}) to express $w$ in terms of $H$
  and $Q$,   the potential is given implicitly by the form
  \cite{ellis,tptachyon}
  \begin{equation}
  V(a) = \frac{1}{16\pi G} H (1-Q)\left[6H + 2aH' - \frac{aH Q'}{1-Q}\right]
  \label{voft}
   \end{equation} 
    \begin{equation}
    \phi (a) =  \left[ \frac{1}{8\pi G}\right]^{1/2} \int \frac{da}{a}
     \left[ aQ' - (1-Q)\frac{d \ln H^2}{d\ln a}\right]^{1/2}
    \label{phioft}
    \end{equation} 
   where $Q (a) \equiv [8\pi G \rho_{\rm known}(a) / 3H^2(a)]$.
   We shall now discuss some examples of this result:
   \begin{itemize}
   \item
  Consider a universe in which observations suggest that $H^2(a) = H_0^2 a^{-3}$.
  Such a universe could be populated by non relativistic matter with density parameter
  $\Omega_{\rm NR} =\Omega =1$. On the other hand, such a universe could be populated entirely by a scalar
  field with a potential $V(\phi) = V_0 \exp[-(16\pi G/3)^{1/2} \phi]$.
  One can also have a linear combination of non relativistic matter and scalar field with the
  potential having a generic form $V(\phi) = A \exp[-B\phi]$.
 \item
  Power law expansion of the universe can be generated by
  a quintessence model with $V(\phi)=\phi^{-\alpha}$. In this case, the energy
  density of the scalar field varies as $\rho_\phi \propto t^{-2\alpha/(2+\alpha)} $;
  if the background density $\rho_{\rm bg}$ varies as $\rho_{\rm bg} \propto t^{-2}$,
  the ratio of the two energy densities changes as $(\rho_\phi/\rho_{\rm bg} =
  t^{4/(2+\alpha)}$).
  Obviously, the scalar field density can dominate over the background at late
  times for $\alpha >0$.  
  \item
  A different class of models arise if the potential is taken
  to be exponential with, say, $V(\phi) \propto \exp(-\lambda \phi/M_{\rm Pl})$.
  When $k=0$, both $\rho_\phi$ and $\rho_{\rm bg}$ scale in the same manner
  leading to 
  \begin{equation}
  \frac{\rho_\phi}{\rho_{\rm bg} +\rho_\phi} = \frac{3(1+w_{\rm bg})}{\lambda^2}
  \end{equation}
  where $w_{\rm bg}$  refers to the background parameter value. In this 
  case, the dark energy density is said to ``track'' the background energy
  density. While this could be a model for dark matter, there are 
  strong constraints on the total energy density of the universe
  at the epoch of nucleosynthesis. This requires $\Omega_\phi \lesssim 0.2$
  requiring dark energy to be sub dominant at all epochs. 
  
\item
  Many other forms of $H(a)$ can be reproduced by a combination of non-relativistic matter and a suitable form of scalar field with a potential $V(\phi)$. As a final example \cite{coop98}, suppose $H^2(a)=H_0^2[ \Omega_{\rm NR} a^{-3}+(1-\Omega_{\rm NR})a^{-n}]$. This can arise, if the universe is populated with non-relativistic matter with density parameter $\Omega_{\rm NR}$ and a scalar field with the potential,  determined using equations (\ref{voft}), (\ref{phioft}). We get 
  \begin{equation}
  V(\phi)=V_0 \sinh^{2n/(n-3)}[\alpha(\phi -\psi)]
   \end{equation}
where
  \begin{equation}
  V_0={(6-n)H_0^2\over 16\pi G}
  \left[ \frac{\Omega_{\rm NR}^n}{(1- \Omega_{\rm NR})^3}\right]^{\frac{1}{n-3}};\quad
  \alpha=(3-n)(2\pi G/n)^{1/2}
  \end{equation}
and $\psi$ is a constant.
  \end{itemize}
  
  Similar results exists for the tachyonic scalar field as well \cite{tptachyon}. For example, given
  any $H(t)$, one can construct a tachyonic potential $V(\phi)$ so that the scalar field is the 
  source for the cosmology. The equations determining $V(\phi)$  are now given by:
  \begin{equation}
  \phi(a) = \int \frac{da}{aH} \left(\frac{aQ'}{3(1-Q)}
   -{2\over 3}{a H'\over H}\right)^{1/2}
  \label{finalone}
  \end{equation}
   \begin{equation}
   V = {3H^2 \over 8\pi G}(1-Q) \left( 1 + {2\over 3}{a H'\over H}-\frac{aQ'}{3(1-Q)}\right)^{1/2}
   \label{finaltwo}
   \end{equation}
   Equations (\ref{finalone}) and (\ref{finaltwo}) completely solve the problem. Given any
   $H(t)$, these equations determine $V(t)$ and $\phi(t)$ and thus the potential $V(\phi)$. 

  As an example, consider  a universe with power law expansion
   $a= t^n$. If it is populated only by  a tachyonic scalar field, then $Q=0$; further,
    $(a H'/H)$ in equation (\ref{finalone}) is a constant
     making $\dot \phi $  a constant. The complete solution
   is given by
   \begin{equation}
   \phi(t) = \left({2\over 3n}\right)^{1/2} t + \phi_0; \quad
   V(t) = {3n^2\over 8\pi G}\left( 1- {2\over 3n}\right)^{1/2} {1\over t^2}
   \end{equation}
   where $n>(2/3)$.
   Combining the two, we find the potential to be
   \begin{equation}
    V(\phi) = {n\over 4\pi G}\left( 1- {2\over 3n}\right)^{1/2}
   (\phi - \phi_0)^{-2}
   \label{tachpot}
   \end{equation}
   For such a potential, it is possible to have arbitrarily rapid expansion with large $n$.   
   (For the cosmological model, based on this potential, see \cite{cosmo23}.)
   
   A wide variety of phenomenological models with time dependent
  \cc\ have been considered in the literature. They involve
  power law decay of \cc\ like $\Lambda \propto t^{-\alpha}$ 
  \cite{bertolami86,cha77,bsom90,ef77,kwe95,lopna96,coop98}
  or $\Lambda \propto a^{-\alpha}$,
   \cite{ozer87,abdel92,chen90,gott87,kolb89,olson87,pavon91,maia94,matyj95,waga97,hoyle97,caldwell98a,caldwell98b,wangstein,hu99,garn98a},
  exponential decay $\Lambda \propto \exp(-\alpha a)$ \cite{rajeev}
  and  more complicated models (for  a summary, see
  \cite{coop98}).
  Virtually all these models can be reverse engineered and mapped to a 
  scalar field model with a suitable $V(\phi)$. Unfortunately, 
  all these models lack predictive power or clear particle physics
  motivation.

  This discussion also illustrates that even when $w(a)$ is known, it is not possible to proceed further and determine
  the nature of the source. The explicit examples given above shows that there
  are {\em at least} two different forms of scalar field Lagrangians (corresponding to
  the quintessence or the tachyonic field) which could lead to
  the same $w(a)$.  A  theoretical physicist, who would like to know which of these
  two scalar fields exist in the universe, may have to be  content with
  knowing $w(a)$. 
  The accuracy of the determination of $w(a)$  depends on the prior assumptions  made
  in determining $Q$, as well as on the observational accuracy with which the quantities $H(a)$ 
   can be measured. Direct observations usually give the luminosity
  distance $d_L$  or angular diameter
  distance $d_A$. To obtain $H(a)$ from either of these, one needs to calculate a derivative
  [see, for example, (\ref{hinvz})] which further limits the accuracy significantly. As we saw 
  in the last section, this is not easy.

  \section{Structure formation in the universe}\label{sfinuniv}

  The conventional paradigm for the formation of structures in the universe is based on the
  growth of small perturbations due to gravitational instabilities. In this picture, some
  mechanism is invoked to generate small perturbations in the energy density in the
  very early phase of the universe. These perturbations grow due to gravitational
  instability and eventually form the different structures which we see today.
  Such a scenario can be constrained most severely by observations of cosmic
  microwave background radiation (CMBR) at $z\approx 10^3$. Since the
  perturbations in CMBR are observed to be small ($10^{-5} - 10^{-4}$ depending
  on angular scales), it follows that the energy density perturbations were small compared
  to unity at the redshift of $z\approx 1000$.

  The central quantity one uses to describe
  the growth of structures during $0<z<10^3$ is 
   the {\it density contrast}
     defined as $\delta (t, {\bf x}) = [\rho(t,{\bf x}) - \rho_{\rm bg}(t)]/\rho_{\rm bg}(t)$ which characterizes 
     the fractional change in the energy density compared to the background. 
     (Here $\rho_{\rm bg}(t) $ is the mean background density of the smooth universe.)
     Since one is 
     often interested in the statistical behaviour of structures in the universe, it is conventional
     to assume that $\delta$ and other related quantities are  elements of an ensemble.
     Many popular models of structure formation suggest that the initial density perturbations
     in the early universe can be represented as a Gaussian random variable with zero mean
     (that is, $<\delta> =0$) and a given initial power spectrum. The latter quantity is defined through
     the relation $P(t, k) = <|\delta_k(t)|^2>$ where $\delta_{\bf k}$ is the Fourier transform
     of $\delta(t,{\bf x})$ and $< ... >$ indicates  averaging over the 
     ensemble. It is also conventional to define the two-point correlation function
     $\xi(t,x)$  as the Fourier transform of $P(t,{\bf k})$ over ${\bf k}$.
 Though gravitational clustering will make the density contrast non Gaussian at late times,
     the power spectrum and the correlation function continue to be of primary importance
     in the study of structure formation. 
     
     When the density contrast is small, its evolution can be studied by linear perturbation
     theory and 
     each of the spatial Fourier modes $\delta_{\bf k}(t)$ will grow independently. It 
     follows that
     $\delta (t,{\bf x})$ will have the form $\delta(t,{\bf x}) = D(t) f({\bf x})$ in the linear 
     regime where $D(t)$ is the growth factor and $f({\bf x}) $ depends on the initial
     configuration.
     When $\delta \approx 1$, linear perturbation theory breaks down and one needs
     to either use some analytical approximation or numerical simulations to study the non
     linear growth. A simple but effective approximation is based on spherical
     symmetry in which one studies the dynamics of a spherical region in the universe
     which has a constant over-density compared to the background. As the universe expands, the over-dense region will expand more slowly compared to the background, 
 will reach a maximum radius, contract and virialize to form a bound nonlinear system. 
    If the proper
     coordinates of the particles in a background Friedmann universe is given
     by ${\bf r} = a(t) {\bf x}$ we can take the proper coordinates of the particles in 
     the over-dense region to be ${\bf r} = R(t) {\bf x}$ where $R(t)$ is the expansion rate of the over-dense region. The relative acceleration of 
     two geodesics in the over-dense region will be
     ${\bf g} =\ddot R {\bf x} = (\ddot R/R) {\bf r}$. Using (\ref{nextnine}) and  $\nabla \cdot {\bf r} =3$, 
     we get 
     \begin{equation}
     \ddot R =- \frac{4\pi G}{3}(\rho +3P)R = - \frac{GM}{R^2} -  \frac{4\pi G}{3}(\rho +3P)_{\rm
     nondust} R
     \label{sphrevl}
     \end{equation}
     where the subscript `non-dust' refers to all components of matter other than the one
     with equation of state $P=0$; the dust component is taken into account by the first term
     on the right hand side with $M=(4\pi/3)\rho_{\rm NR}R^3$.
     The density contrast is related to $R$ by $(1+\delta) = (\rho/\rho_{\rm bg}) = (a/R)^3$.
     Given the  equation (\ref{sphrevl}) satisfied by $R$ and (\ref{frwone}), it is easy to 
     determine the equation satisfied by the density contrast. We get (see p. 404 of \cite{probbook}):
     \begin{equation}
\ddot\delta + 2 {\dot a \over a} \dot\delta = 4 \pi G \rho_b (1 + \delta) \delta + {4 \over 3}
 {\dot\delta^2 \over (1 + \delta)} 
 \label{dencont}
\end{equation} 
This is a fully nonlinear equation satisfied by the density contrast in a 
spherically symmetric over-dense region
in the universe.

\subsection{Linear evolution of perturbations}\label{linevolpert}
 
When the perturbations are small, one can ignore the second term in the right hand side
of (\ref{dencont}) 
and replace $(1+\delta)$ by unity in the first term on the right hand side. The resulting
equation is valid in the linear regime and hence will be satisfied by each of the 
Fourier modes $\delta_{\bf k}(t)$ obtained by Fourier transforming $\delta(t,{\bf x})$ with
respect to ${\bf x}$. Taking $\delta(t,{\bf x}) = D(t) f({\bf x})$, the $D(t)$ satisfies the equation
  \begin{equation}
\ddot D + 2 {\dot a \over a} \dot D = 4 \pi G \rho_b D
  \label{denconttwo}
\end{equation} 

 The power 
 spectra $P(k,t) = <|\delta_{\bf k} (t)|^2>$  at two different redshifts 
 in the linear regime are related by
\begin{equation}
 P(k,z_f)=T^2(k,z_f,z_i, {\rm bg})P(k,z_i) 
 \label{transfn}
 \end{equation}
where $T$  (called transfer function) depends only on the  parameters of the background universe (denoted by `bg') but
{\it not} on the initial power spectrum and can be computed by solving (\ref{denconttwo}). 
It is now clear that the only {\it new} input which structure formation scenarios
require is the specification of the initial perturbation at all relevant
scales, which requires  one arbitrary function of the wavenumber $k=2\pi/\lambda$.

Let us first consider the transfer function.
The rate of  growth of small  perturbations is essentially decided by two factors: (i)
The relative magnitudes of the proper wavelength of perturbation $\lambda_{\rm prop} (t) \propto a(t)$ and the Hubble radius $d_H(t) \equiv H^{-1}(t)= (\dot a/a)^{-1}$
and (ii) whether the universe is radiation dominated or matter dominated.
At sufficiently early epochs, the universe will be radiation dominated and $d_H(t) \propto t$ will be smaller than
the proper wavelength $\lambda_{\rm prop} (t) \propto t^{1/2}$. The density contrast of such modes, which are bigger than the 
Hubble radius, will grow as $ a^2$ until $\lambda_{\rm prop} = d_H(t)$. 
[When this occurs, the perturbation at a given wavelength is said to
enter the Hubble radius. One can use (\ref{denconttwo}) with the right hand side replaced by
$4\pi (1+w) (1+3w) G \rho$ in this case; this leads to $D\propto t \propto a^2$.]
When $\lambda_{\rm prop} < d_H$ and the universe is 
radiation dominated, the perturbation does not grow significantly and 
increases at best only logarithmically \cite{meszaros}. Later on, when the universe 
becomes matter dominated for $t>t_{\rm eq}$, the perturbations again
begin to grow. It follows from this result that modes with wavelengths greater
than $d_{\rm eq} \equiv d_H (t_{\rm eq})$ --- which enter the Hubble
radius only in the matter dominated epoch --- continue to grow at all 
times while modes with wavelengths smaller than $d_{\rm eq} $ suffer
lack of growth (in comparison with  longer wavelength modes) during the period
$t_{\rm enter} < t < t_{\rm eq}$.                        
This fact leads to a distortion of the shape of the 
primordial spectrum by suppressing the growth of small wavelength modes
in comparison with longer ones.
Very roughly, the shape of $T^2(k)$ can be characterized by the behaviour
$T^2(k) \propto k^{-4}$ for $k > k_{\rm eq}$ and $T^2\approx 1$ for
$k< k_{\rm eq}$. The wave number $k_{\rm eq} $ corresponds to the
length scale 
\begin{equation}
d_{\rm eq} = d_H(z_{\rm eq})=  (2\pi/k_{\rm eq}) \approx 13 (\Omega h^2)^{-1}  {\rm Mpc}
\label{shapeparam}
\end{equation}
(eg., \cite{tpsfu}, p.75).
The spectrum at wavelengths $\lambda \gg d_{\rm eq} $ is undistorted by
the evolution since $T^2$ is essentially unity at these scales.   
 Further evolution
can eventually lead to nonlinear structures seen today in the universe.

\begin{figure}[ht]
   \begin{center}
   \includegraphics[scale=0.5]{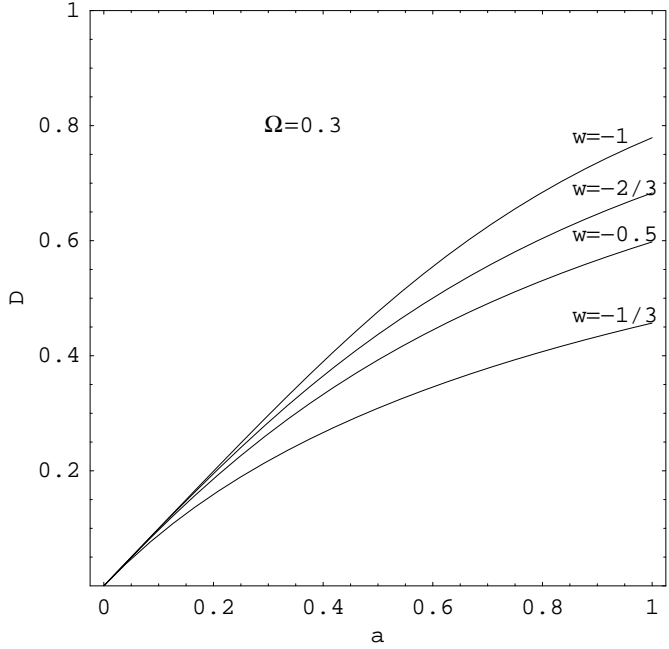}
   \end{center}
\caption{The growth factor for different values
  of $w$ including the one for \cc\ (corresponding to $w=-1$) and an open model (with $w=-1/3$).
 }
\label{fig:hypergeom}
\end{figure}

  At late times, we can ignore the effect of radiation in solving (\ref{denconttwo}).
  The linear perturbation equation (\ref{denconttwo}) has an exact solution (in 
  terms of hyper-geometric functions)  for cosmological models with non-relativistic matter and dark energy with a constant $w$. 
   It is given by 
  \begin{equation}
  \frac{D(a)}{a} = {}_2F_1 \left[ - \frac{1}{3w}, \frac{w-1}{2w}, 1 - \frac{5}{6w}, - a^{-3w} \frac{1- \Omega_{NR}}{\Omega_{NR}}\right] 
  \end{equation}
  [This result can be obtained by direct algebra. If the independent variable in equation (\ref{denconttwo})
  is changed from $t$ to $a^{-3w}$ and the dependent variable is changed from $D$ to $(D/a)$,
  the resulting equation has the standard form of hypergeometric equation for a universe
  with dark energy and non-relativistic matter as source.]
  Figure \ref{fig:hypergeom}   shows the growth factor for different values
  of $w$ including the one for \cc\ (corresponding to $w=-1$) and an open model (with $w=-1/3$.)

%%%%%%%%%%%  FIGURE  %%%%%%%%%%%%%%%%
%%%%%%%%%%%%%%%%%%%%%%%%%%%%%%%%
 For small values of $a$,
$D \approx a$ which is an exact result for $\Omega_\Lambda =0,\Omega_{\rm NR}=1$ model. The growth rate slows down in the cosmological constant dominated phase (in models with $\Omega_{\rm NR} + \Omega_\Lambda  =1$ with $w=-1$) or in the curvature dominated phase
(open models with $\Omega_{\rm NR} <1$ corresponding to $w=-1/3$).
 Between the two cases, there is
less growth in open models compared to models with cosmological
constant.

It is possible to rewrite equation (\ref{dencont})
 in a different form to find an approximate solution for even variable $w(a)$.
  Converting the time derivatives into derivatives with respect to
 $a$ (denoted by a prime) and using the Friedmann equations, we can write
 (\ref{dencont}) as 
 \begin{equation}
 a^2 \delta'' + \frac{3}{2} \left( 1 - \frac{p}{\rho} \right) a \delta' = \frac{3}{2} \frac{\rho_{\rm NR}}{\rho}
 \delta (1+ \delta) + \frac{4}{3} \frac{a^2\delta^{'2}}{1+\delta)}
 \end{equation}
 In a universe populated by only non relativistic matter and dark energy characterized by
 an equation of state function $w(a)$, this equation can be recast in a different manner
 by introducing a time dependent $\Omega$ [as in equation (\ref{findingrhox})]
  by the relation $Q(t) = (8\pi G /3) [\rho_{\rm NR}(t)/H^2(t)]$
  so that  $(dQ/d\ln a) = 3w Q (1-Q)$.
 Then equation (\ref{dencont}) becomes 
 in terms of the variable $f\equiv (d\ln \delta/d\ln a)$
 \begin{eqnarray}
 3wQ(1-Q) \frac{df}{dQ} &+& f^2 +f\left[\frac{1}{2} - \frac{3}{2} w ( 1 - Q)\right]\nonumber\\
 &=& \frac{3}{2} Q (1+ \delta ) + \frac{4}{3} \left( \frac{\delta }{1+\delta}\right) f^2
 \end{eqnarray}
 Unfortunately this equation is not closed in terms of $f$ and $Q$ since it also involves
 $\delta = \exp[\int (da/a) f]$. But in the linear regime, we can ignore the second term on the 
 right hand side and replace $(1+\delta)$ by unity in the first term thereby getting a closed
 equation:
 \begin{equation}
 3wQ(1-Q) \frac{df}{dQ} + f^2 +f\left[\frac{1}{2} - \frac{3}{2} w ( 1 - Q)\right]
 = \frac{3}{2} Q 
 \end{equation}
 This equation has approximate power law solutions \cite{wangstein}
 of the form $f = Q^n$ when $|dw/dQ|\ll
   1/(1-Q)$. Substituting this ansatz, we get 
  \begin{equation}
  n = \frac{3}{5-w/(1-w)}+\frac{3}{125} \frac{(1-w)(1-3w/2)}{(1-6w/5)^3}
  (1-Q) + \mathcal{O} [(1-Q)^2]
  \end{equation}
  [Note that $Q(t)\to 1$ at high redshifts, which is anyway the domain of validity of the linear perturbation
  theory].
  This result shows that $n$ is weakly dependent on $\Omega_{\rm NR}$; further, $n\approx (4/7)$
  for open Friedmann model with non relativistic matter and $n\approx (6/11)\approx 0.6$
  in a $k=0$ model with cosmological constant.

Let us next consider the initial power spectrum $P(k,z_i)$ in (\ref{transfn}). 
 The following points need to be emphasized regarding the initial
fluctuation spectrum.

(1) It can be {\it proved} that known local  physical phenomena, arising
from laws tested in
the laboratory in a medium with $(P / \rho) > 0$,  are incapable producing the initial
fluctuations of required magnitude and spectrum (eg., \cite{probbook}, p 458). The initial fluctuations,
  therefore,  must be treated as arising from  physics untested at the moment.

(2) Contrary to claims sometimes made in the literature, inflationary models
are  not capable of uniquely {\it predicting} the initial fluctuations. 
It is possible to come up with viable inflationary potentials 
(\cite{liddlelyth}, chapter 3)                  
which are capable
of producing any reasonable initial fluctuation. 

A  \emph{prediction} of the initial fluctuation spectrum was indeed made by
 Harrison\cite{harrison}   
and Zeldovich \cite{zeldovich72}, 
 who were years ahead of their times. They
 predicted --- based on very general arguments of scale invariance
--- that the initial fluctuations will have a power spectrum
$P=Ak^n$ with $n=1$. 
Considering the simplicity and importance of this result, we shall briefly
recall the arguments leading to the choice of $n=1$. 

If the power spectrum is $P\propto k^n$
at some early epoch, then the power per logarithmic band of wave numbers
is $\Delta^2\propto k^3P(k)\propto k^{(n+3)}$. Further, when the wavelength of the mode is larger than the Hubble radius, $d_H (t) = (\dot a/a)^{-1}$, during the radiation dominated phase,
the perturbation grows as $a^2$ making $\Delta^2\propto a^4 k^{(n+3)}$. We need to determine how $\Delta $ scales with $k$ when the mode enters the Hubble
radius $d_H(t)$. The epoch $a_{enter}$ at which this occurs 
is determined by the relation $2\pi a_{enter}/k=d_H$. Using $d_H\propto
t\propto a^2$ in the radiation dominated phase, we get $a_{enter}\propto k^{-1}$
so that
\begin{equation}
\Delta^2(k,a_{enter})\propto a_{enter}^4 k^{(n+3)}\propto k^{(n-1)}
\end{equation}
It follows that the amplitude of fluctuations is independent of scale $k$ at the
time of entering the Hubble radius, only if $n=1$. This is  the essence of
Harrison-Zeldovich and  which is independent of the inflationary paradigm. 
It follows  that  verification of $n=1$ by any observation is {\it not} a verification of inflation. At best it verifies a
far deeper principle of scale invariance. 
We also note that the power spectrum of
gravitational potential $P_\phi$ scales as $P_\phi\propto P/k^4
\propto k^{(n-4)}$. Hence
the fluctuation in the gravitational potential (per decade in $k$) $\Delta^2_\phi\propto k^3P_\phi$ is proportional to $\Delta^2_\phi\propto k^{(n-1)}$. This fluctuation in the gravitational potential is also independent of $k$ for $n=1$ clearly showing the special nature of this choice.[It is not possible to take $n$ strictly equal
to unity without specifying a detailed model; the reason has to do with 
the fact that scale invariance is always broken at some level and this
will lead to a small difference  between $n$ and unity].
Given the above description, the basic model of cosmology is based on seven parameters.
Of these 5 parameters $(H_0,\Omega_{\rm B}, \Omega_{\rm DM},\Omega_\Lambda,\Omega_{\rm R})$ determine the background universe and the two parameters
$(A,n)$ specify the initial fluctuation spectrum.

It is possible to provide simple analytic fitting functions for the transfer function,
incorporating all the above effects.
For models with a cosmological constant, the transfer function is well fitted by \cite{BBKS}
  \begin{equation}
  T^2_\Lambda(p) = \frac{\ln^2(1+2.34p)}{(2.34p)^2} [1+3.89p +(16.1p)^2 + (5.46p)^3 + (6.71p)^4]^{-1/2}
  \label{stdtrs}
  \end{equation}
  where $p = k/(\Gamma h\ {\rm Mpc}^{-1}) $ and $ \Gamma = \Omega_{\rm NR} h \exp[-\Omega_B(1+\sqrt{2h}/\Omega_{\rm NR})]$ is called the `shape factor'.
 The presence of dark energy, 
   with a constant $w$, 
   will also affect the transfer function and hence the final power spectrum. An approximate fitting formula
   can be given along the following lines \cite{ma}. 
    Let the power spectrum be written in the form 
   \begin{equation}
	P(k,a) = A_Q\, k^n\, T^2_Q(k) \,\left( 
	{a\,g_Q\over g_{Q,0}} \right)^2 
\label{pka}
\end{equation}
    where $A_Q$ is a normalization, $T_Q$ is the modified transfer function and $g_Q = (D/a)$ is
    the ratio between linear growth factor in the presence of dark energy compared to that in
    $\Omega=1 $ model. Writing $T_Q$ as the product $T_{Q\Lambda} T_\Lambda$ where
    $T_\Lambda$ is given by (\ref{stdtrs}), numerical work shows that 
    \begin{equation}
	T_{Q\Lambda}(k,a) \equiv {T_Q\over T_{\Lambda}} =
	{\alpha + \alpha\,q^2 \over 1 + \alpha\, q^2}\,, \qquad 
	q={k\over \Gamma_Q\,h}\,,
\label{tql}
\end{equation}
where $k$ is in Mpc$^{-1}$, and $\alpha$ is a scale-independent but
time-dependent coefficient  well approximated by $ \alpha  = (-w )^s$ with
\begin{eqnarray}
  	 s &=&  (0.012 - 0.036\,w  - 0.017/w ) [1-\Omega_{\rm NR}(a)]  \nonumber\\
	 &+&  (0.098 + 0.029\,w  - 0.085/w ) \ln\Omega_{\rm NR}(a) 
	\label{alpha}
\end{eqnarray}
where the matter density parameter is $\Omega_{\rm NR}(a) =
\Omega_{\rm NR}/[\Omega_{\rm NR} + (1 - \Omega_{\rm NR})\, a^{-3 w }]\,$. 
    Similarly, the relative growth factor can be expressed in the form
    $g_{Q\Lambda}  \equiv   (g_Q / g_\Lambda) = (-w )^t$ with
    \begin{eqnarray}
    t  &=& -(0.255 + 0.305\,w  + 0.0027/w ) [1-\Omega_{\rm NR}(a)]\nonumber\\
      &-& (0.366 + 0.266\,w  - 0.07/w ) \ln\Omega_{\rm NR}(a) 
\end{eqnarray}
    Finally the amplitude $A_Q$ can be expressed in the form 
     $A_Q=\delta_H^2 (c/H_0)^{n+3}/(4\pi)\,$,
where 
\begin{equation}
\delta_H =2 \times 10^{-5}\alpha_0^{-1}\,(\Omega_{\rm NR})^{c_1 + c_2
\ln\Omega_{\rm NR}}\exp{[c_3(n-1) + c_4(n-1)^2]}
\end{equation} 
and
$\alpha_0=\alpha(a=1)$ of
equation (\ref{alpha}), and 
\begin{eqnarray}
&&c_1= -0.789 |w |^{0.0754 - 0.211 \ln
|w |}, \quad c_2= -0.118 - 0.0727 w ,\nonumber \\
&&  c_3=-1.037,\quad 
c_4=-0.138
\end{eqnarray}
This fit is valid for 
  $-1\lesssim w  \lesssim -0.2$.

  \subsection{Nonlinear growth of perturbations}\label{nlpert}
  
\begin{figure}[ht]    
   \begin{center}
   \includegraphics[scale=0.5]{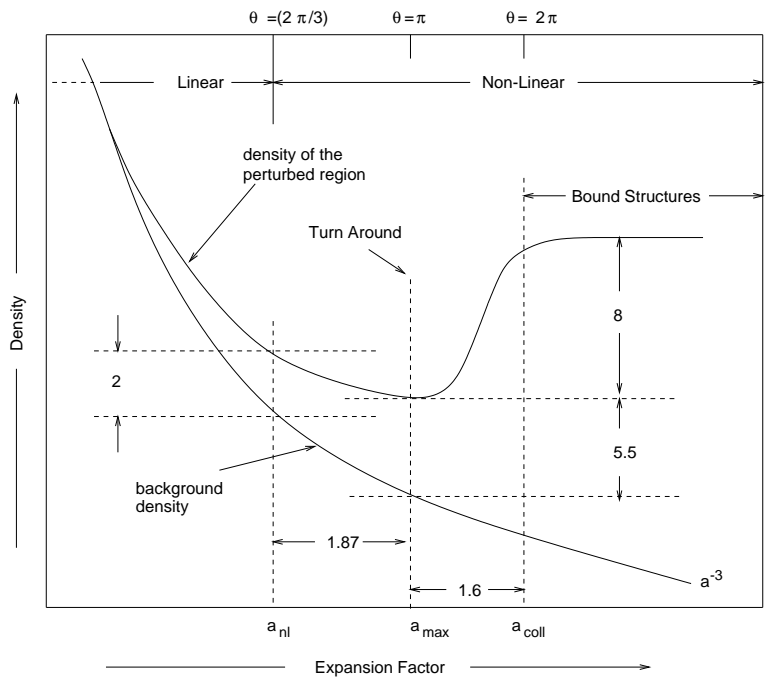}
   \end{center}
\caption{Evolution of an over dense region in spherical top-hat
approximation.}
 \label{fig:sphtophat}
\end{figure}

   In a purely matter dominated universe, equation (\ref{sphrevl}) reduces to $\ddot R=-GM/R^2$.
   Solving this equation one can obtain the non linear density contrast $\delta$ as a function 
   of the redshift $z$: 
  \begin{equation}
(1+z)=\left({4\over 3}\right)^{2/3}
{\delta_i(1+z_i) \over (\theta- \sin \theta)^{2/3}}
=\left({5 \over 3}\right)\left({4 \over 3}\right)^{2/3}
{\delta_0 \over (\theta - \sin \, \theta)^{2/3}}; 
\label{qredth}
\end{equation}
\begin{equation}
\delta = {9 \over 2} 
{(\theta 
- \sin \, \theta)^2 \over (1- \cos \, \theta)^3} - 1. 
\label{qdeuse}
\end{equation}
Here, $\delta_i>0$ is the initial density contrast at the redshift $z_i$
and $\delta_0$ is the density contrast at present if the initial density contrast was
evolved by linear approximation. In general, the linear density contrast
 $\delta_L$ is given by
\begin{equation}
\delta_L = {\overline \rho_L \over \rho_b}-1
={3 \over 5}
\left({3 \over 4}\right)^{2/3}
(\theta - \sin \theta)^{2/3}. 
\label{qrsle}
\end{equation}
 When
$\theta=(2 \pi/3), \delta_L = 0.568$
and $\delta =1.01 \simeq 1.$
If we interpret $\delta = 1$
as the transition point to nonlinearity, then such a
transition occurs at 
$\theta = (2\pi /3)$, 
$\delta_L \simeq 0.57$. 
From (\ref{qredth}),  we see that this occurs at the redshift
$(1+z_{\rm nl}) = (\delta_0/0.57).$
The spherical region reaches the maximum radius of
expansion  at $\theta = \pi$. 
This corresponds to
a density contrast of  $\delta_m\approx 4.6$
which is definitely in the nonlinear regime.
The linear evolution gives $\delta_L=1.063$
at $\theta= \pi$.
After the spherical over dense region turns around it will continue to
contract. Equation (\ref{qdeuse}) suggests that at
$\theta=2\pi$ all the mass will collapse to a point. 
A more detailed analysis of the spherical model \cite{modsph}, however,  shows that
the virialized systems formed at any given time have a mean density which is typically 200 times the
background density of the universe at that time in a $\Omega_{\rm NR}=1$. 
 This occurs at a redshift of about
$
(1+z_{\rm coll})=(\delta_0/
1.686).$
The  density of the virialized structure will be approximately
$
\rho_{\rm coll}\simeq 
170\rho_0(1+z_{\rm coll})^3 $
where $\rho_0$ is the present cosmological density.
The evolution is described schematically in figure \ref{fig:sphtophat}.

%%%%%%%%%      FIGURE     %%%%%%%%%%%%%%% 
%%%%%%%%%%%%%%%%%%%%%%%%%%%%%%%%

   In the presence of dark energy, one cannot ignore the second term
   in equation (\ref{sphrevl}). 
   In the case of a cosmological constant, $w=-1$ and $\rho =$ constant
   and this extra term is independent of time. This allows one
   to obtain the first integral to the equation (\ref{sphrevl}) and reduce the 
   problem to quadrature (see, for example \cite{lahavlambda1,lahavlambda2,lahavlambda3}). 
   For a more general case of constant $w$ with $w\ne -1$,
   the factor $(\rho +3P) = \rho(1+3w)$
   will be time dependent because $\rho$ will be time dependent \emph{even  for a 
   constant $w$}
    if $w\ne -1$.  
   In this case, one cannot obtain an energy integral for the equation
   (\ref{sphrevl}) and the dynamics has to be determined by actual
   numerical 
   integration.  
   Such an analysis leads to the following results \cite{wangstein},  
   \cite{kitayama}, \cite{mark}: 
   
   (i) In the case of matter dominated universe, it was
   found that the linear theory critical threshold for collapse,
   $\delta_c$, was about 1.69. This changes very little in the 
   presence of dark energy and an accurate fitting function
   is given by 
   \begin{eqnarray}
\delta_c & = & \frac{3(12 \pi)^{2/3}}{20} 
                       \left[ 1+\alpha \, \log_{10} \Omega_{\rm NR} \right],  \nonumber \\
         \alpha  & = &  0.353 w^4  +  1.044 w^3  +  1.128 w^2 + \,  0.555 w  +  0.131.
        \label{accuaratefit}
\end{eqnarray}

   (ii) The over density of a virialized structure as a function of the redshift 
   of virialization, however, depends more sensitively on the dark energy component.
   For $-1\le w \le -0.3$, this can be fitted by the function 
   \begin{equation}
\Delta_{\rm{vir}}(z) = 18 \pi^2 \left[ 1 + a \Theta^b(z) \right],
\end{equation}
where 
\begin{equation}
a  =  0.399 - 1.309 (|w|^{0.426} - 1);\quad
b  =  0.941 - 0.205 (|w|^{0.938} - 1),
\end{equation}
and $\Theta(z) = 1/\Omega_{\rm NR}(z)-1 = (1/\Omega_0-1)(1+z)^{3w}$.

   The importance of $\delta_c$ and $\Delta_{\rm vir}$ arises from the fact that these
   quantities can be used to study the abundance of non linear bound structures
   in the universe. The basic idea behind this calculation \cite{pressS} is as follows:
   Let us consider
a density field $\delta_R({\bf x})$ smoothed by a window function
$W_R$ of scale radius $R$. As a first approximation, we may assume
that the region with $\delta (R,t) >\delta_c$  (when smoothed on the scale
$R$ at time $t$) will form a gravitationally bound object with mass
$M\propto \overline\rho R^3$ by the time $t$.  The 
precise form of the  $M-R$ relation
depends on the window function used; for a step function $M=(4\pi/3)$
$\overline\rho R^3$, while for a Gaussian $M=(2\pi)^{3/2}\overline\rho R^3$. 
  Here $\delta_c$ is a critical value 
for the density contrast  given by (\ref{accuaratefit})
Since   $\delta(t) =D(t)$ for the growing mode, the probability for the region to form a bound structure at $t$ is the same as the probability $\delta > \delta_c [D(t_i) /D(t)]$ at some early epoch $t_i$. This probability can be easily estimated since  {\it at sufficiently early $t_i$}, the system is described by a Gaussian random field. 
  This fact can be used to calculate the number density of bound objects leading to the result 
\begin{equation}
N(M)dM=-{\overline\rho\over M}
\left({2\over\pi}\right)^{1/2}
{\delta_c\over \sigma^2}
\left({\partial\sigma\over\partial M}\right)\exp
\left(-{\delta^2_c\over 2\sigma^2}\right)dM, 
\label{numden}
\end{equation}
The quantity $\sigma$ here refers to the linearly extrapolated density contrast.   
We shall now describe the constraints on dark energy arising from structure formation.

  \subsection{Structure formation and constraints on dark energy }\label{clusterabund}

  Combining  the initial power spectrum,  $P(k) = Ak^n$, $n\approx 1$, 
  with the transfer function in (\ref{stdtrs}) we find that 
   the final spectrum has the approximate form 
  \begin{equation}
  \label{finalpower}
  P(k) \propto \left\{
  \begin{array}{ll}
  Ak^{-3} \ln^2 k& \quad\mbox{($k\gg k_{\rm eq}$)}\\
  Ak& \quad\mbox{($k\ll k_{\rm eq}$)}
  \end{array}
  \right.
  \end{equation}
  with $2\pi k_{\rm eq}^{-1} \approx d_H(z_{\rm eq}) \approx 13 (\Omega_{\rm NR} h^2)^{-1}$ Mpc =
  $ 13 (\Gamma h)^{-1} h^{-1}$ Mpc [see equation (\ref{shapeparam})] where $\Gamma \equiv \Omega_{\rm NR} h$
   is  the shape parameter (see equation (\ref{stdtrs}); we have assumed $\Omega_B\approx 0$ for 
   simplicity.) From equation   (\ref{finalpower}), it is clear
   that $P(k)$ changes from an increasing function to a decreasing function 
   at $k_{\rm eq}$ , the numerical value of which is decided by the shape parameter
   $\Gamma$. Smaller values of $\Omega_{\rm NR}$ and $\Gamma$ will lead
   to more power at longer wavelengths. 
   
   One of the earliest
   investigations which used power spectrum to determine $\Omega_\Lambda$ was based on the APM galaxy survey \cite{esm90}.
    This work showed that the existence of large scale power requires a non zero \cc.
   This result was  confirmed when the COBE observations fixed the amplitude
   of the power spectrum unequivocally (see section \ref{cmbrani}). It was pointed out in \cite{tpdna,efstath92}
   that the COBE  normalization  led to a wrong shape for  the power spectrum if we take $\Omega_{\rm NR}=1, \Omega_\Lambda=0$, 
   with  more power at small scales than observed. This problem could be
   solved by reducing $\Omega_{\rm NR}$ and changing the shape of the
   power spectrum. Current observations favour $\Gamma \approx 0.25$.
   In fact, an analysis of a host of observational data, including those mentioned above suggested 
   \cite{jsbtpjvn} that $\Omega_\Lambda\ne 0$ even before the SN data came up.

  Another useful constraint on the models for structure formation can be obtained
  from the abundance of rich clusters of galaxies with masses $M\approx 10^{15} M_\odot$.
  This mass scale corresponds to a length scale of about $8 h^{-1}$ Mpc and hence
  the abundance of rich clusters is sensitive to the root-mean-square fluctuation in the density contrast
  at $8 h^{-1}$ Mpc.  It is conventional to denote this quantity $<(\delta \rho/\rho)^2>^{1/2}$,
  evaluated at $8 h^{-1}$ Mpc, by $\sigma_8$. To be consistent with the observed abundance of rich clusters, equation (\ref{numden}) requires $\sigma_8\approx 0.5 \Omega_{\rm NR}^{-1/2}$. This is consistent
  with COBE normalization for $\Omega_{\rm NR} \approx 0.3, \Omega_\Lambda \approx 0.7$.  [Unfortunately, there is still some uncertainty about the $\sigma_8 -\Omega_{\rm NR}$ relation. There is a claim 
  \cite{sdsscluster} that
  recent analysis of SDSS data gives $\sigma_8\approx 0.33\pm 0.03\Omega_{\rm NR}^{-0.6}$.]

   The effect of dark energy component on the growth of 
  linear perturbations changes the value of  $\sigma_8$.
    The  results of section \ref{linevolpert} translate into the 
  fitting function \cite{wangstein}
  \begin{equation}
  \sigma_8= (0.50 - 0.1 \Theta)\Omega^{-\gamma(\Omega,\Theta)},
  \end{equation}
  where
  $\Theta = (n - 1 ) + (h - 0.65)$ and $\gamma(\Omega,\Theta)
  = 0.21- 0.22w + 0.33\Omega + 0.25\Theta$.
  For constant $w$ models with $w=-1,-2/3$ and $-1/3$, this gives $\sigma_8 = 0.96,
  0.80$ and 0.46 respectively. 
  Because of this effect, the abundance of clusters can be used to put stronger constraints on
  cosmology when the data for high redshift clusters improves. As mentioned
  before, linear perturbations grow more slowly in a universe with \cc\
  compared to the $\Omega_{\rm NR} =1$ universe. This means that 
  clusters will be comparatively rare at  high redshifts  in a $\Omega_{\rm NR} =1$
  universe compared to models with \cc. Only less than 10 per cent of massive clusters
  form at $z>0.5$ in a $\Omega_{\rm NR}=1$ universe whereas almost all
  massive clusters would have formed by $z\approx 0.5$ in a universe with \cc
  \cite{bahcall99,bahfan98,gioia97,vl98,cen98}.
  (A simple way of understanding this effect is by noting that if the clusters are
  not in place by $z\approx 0.5$, say, they could not have formed by today
  in models with \cc\ since there is very little growth of fluctuation
  between these two epochs.)
  Hence the evolution of cluster population as a function of redshift can be
  used to discriminate between these models. 
  
  An indirect way of measuring this
  abundance is through the lensing effect of a cluster of galaxy on extended background
  sources. Typically, the foreground clusters shears the light distribution of the
  background object and leads to giant arcs. Numerical simulations suggest \cite{bartel98}
  that a model with $\Omega_{\rm NR}=0.3, \Omega_\Lambda=0.7$ will produce
  about 280 arcs which is nearly an order of magnitude larger than the number
  of arcs produced in a $\Omega_{\rm NR}=1, \Omega_\Lambda=0$ model.
  (In fact, an open model with $\Omega_{\rm NR}=0.3, \Omega_\Lambda=0$
  will produce about 2400 arcs.) To use this effect, one needs a well defined
  data base of arcs and a controlled sample. At present it is not clear 
  which model is preferred though this is one test which seems to prefer
  open model rather than a $\Lambda$-CDM model.

    Given the solution to (\ref{sphrevl}) in the presence of dark energy, we can  repeat the above analysis and obtain the abundance of different
    kinds of structures in the universe   in the presence of dark energy. In particular
    this formalism can be used to study the abundance of weak gravitational lenses
    and virialized x-ray clusters which could act as gravitational lenses. The calculations
    again show \cite{mark} that the result is highly degenerate in $w$ and $\Omega_{\rm NR}$.
    If $\Omega_{\rm NR}$ is known, then the number count of weak lenses will be about a
    factor 2 smaller for $w=-2/3$ compared to the $\Lambda$CDM model with a 
    cosmological constant. However, if $\Omega_{\rm NR}$ and $w$ are allowed to 
    vary in such a way that the matter power spectrum matches with both COBE results
    and abundance of x-ray clusters, then the predicted abundance of lenses is 
    less than 25 per cent for $-1\le w \le -0.4$. It may be possible to constrain 
    the dark energy better by comparing relative abundance of virialized lensing
    clusters with the abundance of x-ray under luminous lensing halos. 
    For example, a survey covering about 50 square degrees of sky may be
    able to differentiate a $\Lambda$CDM model from $w=-0.6 $ model
    at a 3$\sigma$ level. 
  
  The value of $\sigma_8$ and cluster abundance can also be constrained via the
   Sunyaev-Zeldovich (S-Z)
  effect which is becoming a powerful probe of cosmological parameters \cite{komatsu}. The S-Z angular
  power spectrum scales as $\sigma_8^7 (\Omega_B h)^2$ and is almost independent 
  of other cosmological parameters. Recently the power spectrum of CMBR determined by
  CBI and BIMA experiments (see section \ref{cmbrani}) showed an excess at small scales which could be 
  interpreted as due to S-Z effect.  If this interpretation is correct, 
  then $\sigma_8 (\Omega_B h/0.035)^{0.29} =1.04 \pm 0.12$ at 95 per cent confidence level.
  This $\sigma_8$ is on the higher side and only future observations can decide whether 
  the interpretation is correct or not. The WMAP data, for example, leads to a more conventional
  value of $\sigma_8 = 0.84 \pm 0.04$.  
   
  Constraints on cosmological models can also arise from the modeling of 
  damped Lyman-$\alpha$ systems \cite{cen98,hui99,mcd99,dwein98,tirthtpsri1,tirthtpsri2}
  when the observational situation improves. At present these observations
  are consistent with $\Omega_{\rm NR}=0.3, \Omega_\Lambda=0.7$ model
  but do not exclude other models at a high significance level.  
    
  Finally, we comment on a direct relation between $\delta(a)$ and $H(a)$. 
  Expressing equation (\ref{dencont})  in terms of  $H(a)$  will lead to
  the form 
  \begin{equation}
  a^2 H^2 \delta'' + ( 3 H^2 + aHH') a \delta' = \frac{3}{2} \frac{H_0^2 \Omega_{\rm NR}}{a^3} \delta(1+\delta) + \frac{4}{3} \frac{a^2H^2}{(1+\delta)} \delta^{'2}.
  \end{equation}
  This can be used to determine $H^2(a)$ from $\delta(a)$ since  this 
  equation is linear and first order in $Q(a)\equiv H^2(a)$ (though it is second order in $\delta$).
  Rewriting it in the form 
  \begin{equation}
  A(a)Q' + B(a)Q = C(a)
  \end{equation}
  where 
  \begin{equation}
  A =\left( \frac{1}{2} a^2 \delta'\right) ; \ B = \left( 3a\delta' + a^2 \delta'' -
   \frac{4}{3} \frac{\delta' a^2}{1+\delta}\right) ; \ C = \frac{3}{2} \frac{H_0^2 \Omega_{\rm NR}}{a^3} \delta
   (1+\delta)
   \end{equation}
   We  can  integrate it  to give the solution 
   \begin{equation}
   H^2(a) = 3 H_0^2 \Omega_{\rm NR} \frac{(1+\delta)^{8/3}}{a^6 \delta^{'2}}\int da \frac{a\delta' \delta}{(1+\delta)^{5/3}}
\label{hfromdel}
   \end{equation}
   This shows that, given the non linear growth of perturbations $\delta(a)$ as a function of 
   redshift and the approximate validity of spherical model,
    one can determine $H(a)$ and thus $w(a)$ {\it even during the nonlinear phases of the evolution}. 
    [A similar analysis with the linear equation (\ref{denconttwo}) was done in \cite{star98}, leading
to the result which can be obtained by expanding (\ref{hfromdel}) to linear order in $\delta$.]
   Unfortunately, this is an impractical method from observational point of view at present.

   \section{CMBR anisotropies}\label{cmbrani}
   
     In the standard Friedmann model of the universe, neutral atomic systems form at a redshift of
   about $z\approx 10^3$ and the photons decouple from the matter at this redshift. These
   photons, propagating freely in spacetime since then, constitute the CMBR observed around us
   today. In an ideal Friedmann universe, for a comoving observer, this radiation will appear
   to be isotropic. But if physical process has led to inhomogeneities in  the $z=10^3$
   spatial surface, then these inhomogeneities will appear as angular anisotropies
   in the CMBR in the sky today. A physical process operating at a proper length scale
   $L$ on the $z=10^3$ surface will lead to an effect at an angle $\theta = L/d_A(z)$.
   Numerically, 
   \begin{equation}
   \theta(L)\cong \left({\Omega\over 2}\right)
\left({Lz\over H^{-1}_0}\right)=34.4^{''}(\Omega h)
\left({\lambda_0\over 1 {\rm Mpc}}\right).
\label{qthetal}
\end{equation}
  To relate the theoretical predictions to observations, it is usual to
expand the temperature anisotropies in the sky in terms of the 
spherical harmonics.
 The temperature anisotropy in the sky will provide
$\Delta =\Delta T/T$ as a function of two angles $\theta$ and $\psi$.
If we  expand the temperature 
anisotropy distribution on the sky in spherical harmonics: 
\begin{equation}
\Delta(\theta,\psi) \equiv {\Delta T\over T} (\theta,\psi) = \sum_{l,m}^\infty a_{lm} Y_{lm}(\theta, \psi).
\label{qdttwo}
\end{equation} 
all the information is now contained in the angular coefficients $a_{lm}$.

 If ${\bf n}$ and ${\bf m}$ are two directions in the sky with an angle $\alpha$ between them, the two-point correlation function of the temperature fluctuations in the sky can be defined as
\begin{equation}
{\mathcal C} (\alpha) = \langle S({\bf n}) S({\bf m})\rangle = \sum\sum \langle a_{lm}a_{l'm'}^*\rangle Y_{lm}({\bf  n})Y_{l'm'}^*({\bf m}).
\end{equation}
Since the sources of temperature fluctuations are related linearly to the density inhomogeneities, the coefficients $a_{lm}$ will be random fields with some power spectrum. In that case $<a_{lm}a^*_{l'm'}>$ will be nonzero only if $l=l'$ and $m=m'$. Writing 
\begin{equation}
 \langle a_{lm}a_{l'm'}^*\rangle = C_l \delta _{ll'} \delta_{mm'}
 \end{equation}
and using the addition theorem of spherical harmonics, we find that 
\begin{equation}
{\mathcal C} (\alpha) = \sum_l {(2l+1)\over 4\pi} C_l P_l(\cos \alpha)
\label{qcalpha}
\end{equation}  
with $C_l = <|a_{lm}|^2>$.
In this approach, the pattern of anisotropy is contained in the variation
of $C_l$ with $l$. Roughly speaking, $l\propto \theta^{-1}$ and we can
think of the $(\theta, l)$ pair as analogue of $({\bf x}, {\bf k})$ variables
in 3-D. The $C_l$ is similar to the power spectrum $P({\bf k})$.   
   
In the simplest scenario, 
the primary anisotropies of the CMBR arise from three different sources.
(i) The first is the gravitational potential fluctuations at the last
scattering surface (LSS) which will contribute an anisotropy $(\Delta T/T)_\phi^2
\propto k^3P_\phi(k)$ where $P_\phi(k)\propto P(k)/k^4$ is the 
power spectrum of gravitational potential $\phi$. This anisotropy arises
because photons climbing out of deeper gravitational wells lose
more energy on the average.  
(ii) The second source is the Doppler shift of the frequency of the photons
when they are last scattered by moving electrons on the LSS.
This is proportional to $(\Delta T/T)_D^2\propto k^3 P_v$ where 
$P_v(k)\propto P/k^2$ is the power spectrum of the velocity field.
(iii) Finally, we also need to take into account the intrinsic 
fluctuations of the radiation field on the LSS. In the case of adiabatic 
fluctuations, these will be proportional to the density fluctuations of
matter on the LSS and hence will vary as $(\Delta T/T)_{\rm int}^2\propto k^3P(k)$.
Of these, the velocity field and the density field (leading to the 
Doppler anisotropy and intrinsic anisotropy described in (ii) and (iii) above)
will oscillate at scales smaller than the Hubble radius at 
the time of decoupling
since pressure support will be effective at these scales.
At large scales, if $P(k)\propto k$, then 
\begin{equation}
\left({\Delta T\over T}\right)_\phi^2\propto \ {\rm constant};  
\left({\Delta T\over T}\right)_D^2\propto k^2\propto \theta^{-2};  
\left({\Delta T\over T}\right)_{\rm int}^2\propto k^4\propto \theta^{-4}
\end{equation}
where $\theta\propto \lambda \propto k^{-1}$ is the angular scale
over which the anisotropy is measured. Obviously, the fluctuations due
to gravitational potential dominate at large scales while 
 the sum of intrinsic and Doppler anisotropies
will dominate at small scales. Since the latter two 
are oscillatory, we sill expect an oscillatory behaviour in the 
temperature anisotropies at small angular scales. 

There is, however,
one more feature which we need to take into account. The above analysis 
is valid if recombination was instantaneous; but in reality the thickness
of the recombination epoch is about $\Delta z\simeq 80$ 
(\cite{joneswyse};\cite{tpsfu}, chapter 3). 
This implies that the anisotropies will
be  damped at scales smaller than the length scale corresponding to a 
redshift interval of $\Delta z = 80$. The typical value
for the peaks of the oscillation are at about 0.3 to 0.5 degrees depending
on the details of the model. At angular scales smaller than about
0.1 degree, the anisotropies are heavily damped by the thickness of the
LSS.

The fact that several different processes contribute to the 
 structure of angular anisotropies make CMBR  a valuable
tool for extracting cosmological information. To begin with, the 
anisotropy at very large scales directly probe modes which are 
bigger than the Hubble radius at the time of decoupling
and  allows us to directly determine the primordial spectrum.
Thus, in general, if the angular dependence of the spectrum at very large
scales is known, one can work backwards and determine 
the initial power spectrum.
If the initial power spectrum is assumed to be $P(k) = Ak$, then the observations
of large angle anisotropy  allows us to fix the amplitude $A$ of the power
spectrum \cite{tpdna,efstath92}. Based on the results of COBE satellite \cite{smot},
one finds that the amount of initial power per logarithmic band in
$k$ space is given by
\begin{equation}
\Delta^2(k)= \frac{k^3|\delta_k|^2}{2\pi^2} = \frac{Ak^4}{2\pi^2} 
\cong\left( \frac{k}{0.07 h {\rm Mpc}^{-1}}\right)^4
\end{equation}
(This corresponds to 
 $A\simeq (29 h^{-1}{\rm Mpc})^4$. 
 Since the actual $(\Delta T/T)$ is one realization of a Gaussian random process,
 the observed small-$l$ results are subject to unavoidable fluctuations called
 the `cosmic variance'.)
This result is powerful enough to rule out matter dominated, $\Omega=1$ models
 when combined with the data on the 
abundance of large clusters which   determines the
amplitude of the power spectrum at $R\approx 8h^{-1}$ Mpc. 
For example the parameter values $h=0.5,\Omega_0\approx
\Omega_{DM}=1, \Omega_\Lambda=0$, are ruled out by this observation when combined 
with COBE observations   \cite{tpdna,efstath92}.

%%%%%%%%%%%%  FIGURE    %%%%%%%%%%%%%%%%%%%%%

\begin{figure}[ht]
\begin{center}
\includegraphics[scale=0.5]{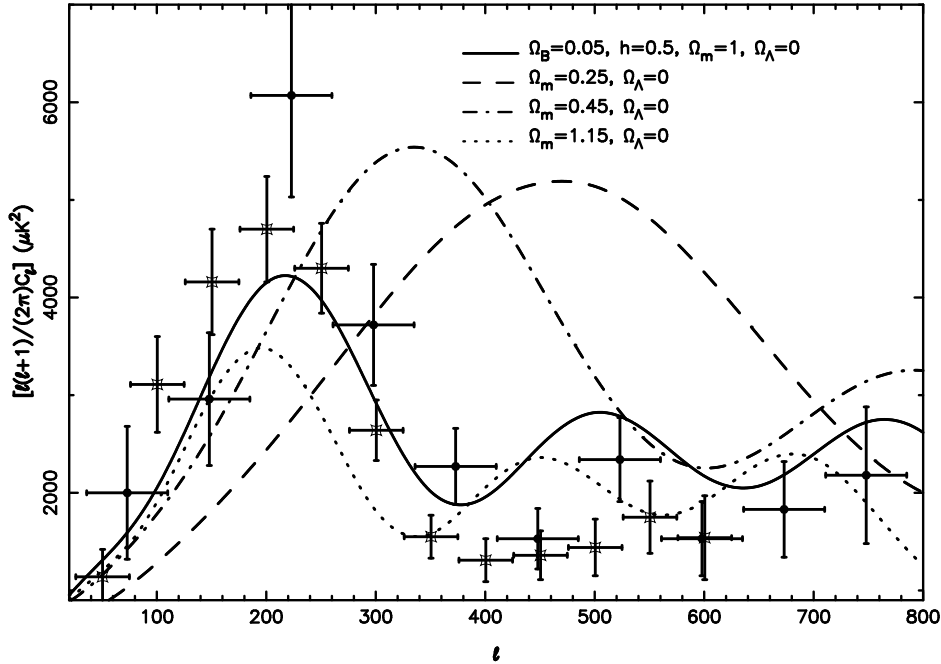}
\end{center}
\caption{The variation of the anisotropy pattern
 in universes with $\Omega_{\rm NR} = (0.25, 0.45, 1.0, 1.15), \Omega_\Lambda =0$
with the first acoustic peak moving from right to left. The y-axis
is essentially a measure of $(\Delta T/T)^2$ while the x-axis is a measure
of $1/\theta$.( Figure courtesy:
S.Sethi.)}
\label{fig:tpsethiacoustic}
\end{figure}

%%%%%%%%%%%%%%%%%%%%%%%%%%%%%%%%%%%%%%%%%%%%%

As we move to smaller scales we are probing the behaviour of baryonic gas coupled to the
photons. The pressure support of the gas leads to modulated acoustic oscillations
with a characteristic wavelength at the $z=10^3 $ surface. Regions of high and low
baryonic density contrast will lead to anisotropies in the temperature with the same
characteristic wavelength.
The physics of these oscillations has been studied in several papers in detail
\cite{peebles70,SZ70,wilson,vittorio,bondefstathiou,review,review2,review3,review5}.   
 The angle subtended by the wavelength of these
acoustic oscillations will lead to a series of peaks in the temperature anisotropy
which has been detected \cite{dber,ba-ha}. 
 The structure of acoustic peaks at small scales
 provides a  reliable procedure for estimating the 
cosmological parameters. 

To illustrate this point let us  consider the location of the 
first acoustic peak. 
Since all the Fourier components of the growing density perturbation
  start with zero amplitude at high redshift, the condition for a mode
  with a given wave vector $k$ to reach an extremum amplitude at
  $t=t_{\rm dec}$ is given by 
  \begin{equation}
   \int _0^{t_{\rm dec}}\frac{kc_s}{a}\, dt \simeq \frac{n   \pi }{2},
   \label{extremamp}
  \end{equation}
  where $c_s=(\partial P/\partial\rho)^{1/2} \approx (1/\sqrt{3})$ is the speed of sound in the baryon-photon fluid.
  At high redshifts, $t(z) \propto \Omega_{\rm NR}^{-1/2} (1+z)^{-3/2}$ and
  the proper wavelength of the first acoustic peak scales as 
  $\lambda_{\rm peak} \sim t_{\rm dec} \propto 
  h^{-1} \Omega_{\rm NR}^{-1/2}  $. The angle subtended by this scale in the sky
  depends on $d_A$. If $\Omega_{\rm NR} +\Omega_\Lambda =1$
  then the angular diameter distance varies as $\Omega_{\rm NR}^{-0.4}$
  while if $\Omega_\Lambda=0$, it varies as $\Omega_{\rm NR}^{-1}$.
  It follows that the angular size of the acoustic peak varies with the 
  matter density as 
  \begin{equation}
   \theta _{\rm peak}  \sim  \frac{z_{\rm dec}\lambda _{\rm peak}}{a_0 r }\propto
   \left\{
    \begin{array}{ll}
       \Omega_{\rm NR}^{1/2}&\qquad \mbox{(if $ \Omega_\Lambda =0$)}, \\
      \Omega_{\rm NR}^{-0.1}&\qquad \mbox{( if  $\Omega_\Lambda +\Omega_{\rm NR}=1$)}.
      \end{array}
      \right. 
     \label{flatpeak}
 \end{equation}
  Therefore, the angle subtended by acoustic peak is quite sensitive to $\Omega_{\rm NR}$ if
  $\Lambda =0$ but not if $\Omega_{\rm NR}+\Omega_\Lambda=1$.
  More detailed computations show that the multipole index corresponding to the acoustic
  peak scales as $l_p \approx 220 \Omega_{\rm NR}^{-1/2}$ if $\Lambda=0$
  and $l_p \approx 220$ if $\Omega_{\rm NR} + \Omega_\Lambda =1$ and
  $0.1 \lesssim \Omega_{\rm NR} \lesssim 1$. 
  This is illustrated in figure \ref{fig:tpsethiacoustic}
 which   shows the variation in the structure of 
acoustic peaks when $\Omega$ is changed keeping $\Omega_\Lambda  =0$.
The four curves are for $\Omega=\Omega_{\rm NR} = 0.25, 0.45, 1.0, 1.15$
with the first acoustic peak moving from right to left. 
The data points on the figures are from the first results of MAXIMA and BOOMERANG
experiments  and are included
to give a feel for the error bars in current observations.
It is obvious that the overall geometry of the 
universe can be easily fixed by the study of CMBR anisotropy.

The heights of acoustic peaks also contain important information.
In particular, the height of the first acoustic peak
relative to the second one depends sensitively on $\Omega_B$. 
However,  not all cosmological parameters
can be measured {\it independently}
using CMBR data alone. For example, different models with the same 
values for ($\Omega_{\rm DM} + \Omega_\Lambda $) and $\Omega_Bh^2$ will
give anisotropies which are fairly indistinguishable. The structure
of the peaks will be almost identical in these models. This shows
that while CMBR anisotropies can, for example, determine the total 
energy density  ($\Omega_{\rm DM} + \Omega_\Lambda $), we will need
some other independent cosmological observations to determine  
individual components.

At present there exists  several observations of the small scale anisotropies in 
  the CMBR from the balloon flights, BOOMERANG \cite{dber},
  MAXIMA \cite{ba-ha},
  and from radio telescopes CBI \cite{pe-ma},
   VSA \cite{scot},
    DASI \cite{halv,kova} and --- most recently --- from WMAP \cite{wmap}.
   These CMBR data has been extensively analyzed 
    \cite{wmap,perc,wang,smot,pe-ma,scot,leea,nett,coll,saun,crof,siev} in isolation as well as 
  in combination with other results.
  (The information about structure formation arises mainly from galaxy surveys
  like SSRS2, CfA2 \cite{daco},
    LCRS \cite{shec},
  Abell-ACO cluster survey \cite{retz},
  IRAS-PSC z \cite{saun},
   2-D survey \cite{peac,coll} and the Sloan survey \cite{sloansurvey}.)
  While there is some amount of variations in the results, by and large,
  they support the following conclusions. 
  \begin{itemize}
  \item The data strongly supports a $k=0$ model of the universe \cite{siev}
  with $\Omega_{\rm tot} = 1.00\pm^{0.03}_{0.02}$ from the pre-MAP data
  and $\Omega_{\rm tot} = 1.02\pm 0.02$ from the WMAP data.
  \item The CMBR data before WMAP, when 
  combined with large scale structure data,  suggest
  $\Omega_{\rm NR} = 0.29 \pm 0.05 \pm 0.04$ \cite{perc,wang,siev,roos}.
  The WMAP result \cite{wmap} is consistent with this giving
  $0.27 \pm 0.04$.
  The initial power spectrum is consistent with being scale invariant 
  and the pre-MAP value is 
   $n = 1.02 \pm 0.06 \pm 0.05$ \cite{perc,wang,siev}. 
   The WMAP gives the spectral index at $k=0.05 $ Mpc$^{-1}$ to be
   $0.93 \pm 0.03$.
  In fact, combining 2dF survey results with CMBR suggest \cite{george}
  $\Omega_\Lambda \approx 0.7$ independent of the supernova results.
  \item A similar analysis based on BOOMERANG data
leads to $\Omega_{\rm tot} =1.02\pm 0.06 $(see for example,
\cite{nett}). 
Combining this result with the HST constraint \cite{freedman}  
 on the 
Hubble constant $h=0.72 \pm 0.08 $, galaxy clustering data as well SN observations
one gets $\Omega_\Lambda = 0.62^{0.10}_{-0.18}, 
   \Omega_\Lambda = 0.55^{0.09}_{-0.09}$ and $\Omega_\Lambda = 0.73^{0.10}_{-0.07}$
   respectively    \cite{melchi}.   
   The WMAP data gives $h=0.71^{+0.04}_{-0.03}$.  
  \item The analysis also gives an independent handle on baryonic 
  density in the universe which is consistent with the BBN value: 
  The pre-MAP result was 
  $\Omega_B h^2 = 0.022 \pm 0.003$ \cite{perc,wang}. 
  (This is gratifying since the initial data had an error and gave too
  high a value \cite{tpshiv}.)
  The WMAP data gives $\Omega_B h^2 = 0.0224 \pm 0.0009$.
  \end{itemize}
  
  There has been some amount of work on the effect of dark energy on the CMBR anisotropy 
  \cite{brax00,bacci00,doran00,coraco,bacci02,wasser02,cora02,alessandro1,alessandro2}.
  The shape of the CMB spectrum is relatively insensitive to the dark energy 
  and the main effect is to alter the angular diameter distance to the last scattering
  surface and thus the position of the first acoustic peak. 
  Several studies have attempted to put 
  a bound on $w$ using the CMB observations. Depending on the assumptions which were invoked, 
  they  all lead to a bound  broadly in the range of $w\lesssim -0.6$. 
  (The preliminary analysis of WMAP data in combination with other astronomical data sets
  suggest $w < -0.78$ at 95 per cent confidence limit.)
  At present it is
  not clear  whether CMBR anisotropies
  can be of significant help in discriminating between different dark energy
  models.

   \section{Reinterpreting the cosmological constant}\label{interpretcc}
    
    It is possible to attack the \cc\ problem from various other directions
    in which the mathematical structure of equation (\ref{four}) is reinterpreted
    differently. Though none of these ideas have been developed into 
    a successful formal theory, they might contain ingredients which may eventually
    provide a solution to this problem. Based on this hope, we shall provide
    a brief description of some of these ideas.    (In addition to these ideas,
    there is extensive literature on several different paradigms for attacking the 
    \cc\ problem based on: (i)  Quantum field theory in curved spacetime
    \cite{odintsov,ilyashapiro1,ilyashapiro2},  (ii) quantum cosmological considerations
    \cite{mongan}, (iii)   models of inflation \cite{jekim}, (iv) string theory inspired
    ideas \cite{mersini}, and (v) effect of phase transitions \cite{mansouri}.)
   
   \subsection{Cosmological constant as a Lagrange multiplier}\label{cclagrange}
  
  The action principle for gravity in the presence of a cosmological constant 
   \begin{eqnarray}
  A  &=&  \frac{1}{16\pi G} \int (R-2\Lambda) \sqrt{-g} d^4 x \nonumber\\
  &=&
   \frac{1}{16\pi G} \int R \sqrt{-g} d^4 x
  - \frac{\Lambda}{8\pi G} \int  \sqrt{-g} d^4 x
   \label{lambdavol}
  \end{eqnarray}  
  can be thought of as a variational principle extremizing the integral over 
  $R$, subject to the condition that the 4-volume of the universe
  remains constant. To implement the constraint that the 4-volume
  is a constant, one will add a Lagrange multiplier term which is 
  identical in structure to the second term in the above equation.
  Hence, mathematically, one can think of the cosmological constant
  as a Lagrange multiplier ensuring the constancy of the 4-volume
  of the universe when the metric is varied.
  
  If we take this interpretation seriously, then it is necessary to 
  specify the 4-volume of the universe before the variation is
  performed and determine the cosmological constant so that
  the 4-volume has this specified volume. 
  A Friedmann model with positive cosmological
  constant in Minkowski space   will lead to a finite 3-volume
  proportional to $\Lambda^{-3/2}$ on spatial integration. (To achieve this,
  we should use the coordinates in which the spatial
  sections are closed 3-spheres.) The time integration, however,
  has an arbitrary range and one needs to restrict the integration
  to  part of this range by invoking some physical principle.
  If we take this to be typically the age of the universe, then
  we will obtain a 
  time dependent 
  cosmological constant $\Lambda(t)$ with $\Lambda(t) H(t)^{-2}$ remaining of order unity.
   While this appears to be a conceptually attractive idea, it is not easy
   to implement it in a theoretical model. 
   In particular, it is difficult to obtain this as a part of a generally covariant
   theory incorporating gravity. 
    
    \subsection{Cosmological constant as a constant of integration}\label{ccintc}
    
    Several people have suggested modifying the basic structure of 
    general relativity so that the  \cc\ will appear as a constant of integration.
    This does not solve the problem in the sense that it still leaves
    its value undetermined. But this changes the perspective
    and allows one to think of the cosmological constant as a 
    non dynamical entity 
    \cite{lambdaintconst1,lambdaintconst4}.
    
    One simple way of achieving this is to assume that the determinant $g$ of $g_{ab}$ is
    not dynamical and admit only those variations which obeys the condition $g^{ab}\delta g_{ab} =0$
    in the action principle. This is equivalent to eliminating the trace part  of Einstein's equations.
    Instead of the standard result, we will now be led to the equation
     \begin{equation}
  R^i_k - \frac{1}{4} \delta^i_k R = 8\pi G \left(T^i_k -\frac{1}{4} \delta^i_k T\right)
  \label{traceless}
  \end{equation}  
    which is just the traceless part of Einstein's equation. The general covariance of the 
    action, however, implies that $T^{ab}_{;b} =0$ and the Bianchi identities  
    $ (R^i_k - \frac{1}{2} \delta^i_k R)_{;i}  =0   $ continue to hold. These two conditions
    imply that $\partial_i R =-8\pi G \partial_i T$ requiring $R+8\pi G T$ to be a constant.
    Calling this constant $(-4\Lambda)$ and combining with equation (\ref{traceless}),
    we get
    \begin{equation}
  R^i_k - \frac{1}{2} \delta^i_k R - \delta^i_k \Lambda = 8\pi G T^i_k
  \end{equation} 
    which is precisely Einstein's equation in the presence of \cc.
    In this approach, the \cc\ has nothing to do with any term in the action or 
    vacuum fluctuations and is  merely an integration constant. Like any other
    integration constant its value can be fixed by invoking suitable boundary
    conditions for the solutions. 
    
    There are two key difficulties in this approach. The first, of course, is that
    it still does not give us any handle on the value of the cosmological constant
    and all the difficulties mentioned earlier still exists. This problem would have
    been somewhat less serious if the cosmological constant was strictly zero;
    the presence of a small positive \cc\ makes the choice of integration constant
    fairly arbitrary. The second problem is in interpreting the condition that 
    $g$ must remain constant when the variation is performed. It is not easy to
    incorporate this into the logical structure of the theory. (For  some
    attempts in this direction, see \cite{sorkin}.)
    
    \subsection{Cosmological constant as a stochastic variable}\label{ccstoc}
    
    Current cosmological observations can be interpreted as showing that 
 the {\it effective value} value of $\Lambda $ 
    (which will pick up contributions from all vacuum energy densities of matter fields) has been 
    reduced from the natural value of  $L_P^{-2}$ to $L_P^{-2}(L_PH_0)^2$ where $H_0$
    is the current value of the Hubble constant.
    One possible  way of thinking about this issue is the following \cite{tpcqgcc}: 
    Let us assume that the quantum
    micro structure of spacetime at Planck scale is capable of readjusting itself, soaking up any
    vacuum energy density which is introduced ---  like a sponge soaking up water.
    If this process is fully deterministic and exact, all vacuum energy densities will cease to have
    macroscopic gravitational effects. However, since this process is inherently quantum gravitational,
    it is subject to quantum fluctuations at Planck scales. 
    Hence, a tiny part of the vacuum energy will survive the 
    process and will lead to observable effects.      
    One may conjecture that the cosmological constant we measure corresponds to this  small 
    residual fluctuation which will depend on the volume of the spacetime region that is probed.  
    It is small, in the sense that it has been reduced from $L_P^{-2}$ to $L_P^{-2}(L_PH_0)^2$, 
    which indicates the fact that fluctuations --- when measured over a large volume --- is small
     compared to the bulk value. It is the wetness of the sponge we notice, not the water content inside.
     
     This is particularly relevant in the context of standard discussions of the contribution of 
 zero-point energies to cosmological constant. The correct theory is likely 
 to regularise the divergences and make the zero point energy finite and about $L_P^{-4}$. This contribution is most likely to modify the microscopic structure of spacetime (e.g if the spacetime is naively thought of as due to stacking of Planck scale volumes, this will modify the stacking or shapes of the volume elements) and will not affect the bulk gravitational field when measured at scales coarse grained over sizes much bigger than the Planck scales.
        
     Given a large 4-volume ${\mathcal V}$ of the  spacetime, we will divide it into 
     $M$  cubes of size $(\Delta x)^4$ and label the cubes by $n = 1, 2,  ....., M$.
     The contribution to the path integral amplitude ${\mathcal A}$, describing long wavelength
     limit of conventional Einstein gravity, can be expressed in the form     
     \begin{equation}
     {\mathcal A} = \prod _n \left[ \exp (c_1 (RL_P^2)+\cdots )\right]^{{i(\Delta x)^4\over L_P^4}} 
     \to 
     \exp {ic_1\over L_P^4} \int d^4x \sqrt{-g} (RL_P^2)\label{stdres}
     \end{equation}
     where  we have indicated the standard continuum limit. (In conventional units
      $c_1 = (16\pi)^{-1}$.) 
      Let us now ask how one could modify this result to 
      describe the ability of spacetime micro structure to readjust
     itself and absorb vacuum energy densities.     
     This would require some additional dynamical 
     degree of freedom that will appear in the path integral amplitude and survive in the 
     classical limit. It can be shown that \cite{tpcqgcc} the simplest implementation of this  feature is by modifying the standard path integral
     amplitude $[ \exp (c_1 (RL_P^2)+\cdots )]$
      by a factor $[\phi(x_n)/\phi_0]$ where $\phi(x) $ is a scalar
     degree of freedom and $\phi_0$ is a pure number introduced to keep this factor
     dimensionless. In other words, we modify the path integral  amplitude to the form:
     \begin{equation}
     {\mathcal A}_{\rm modify} = \prod_n \left[ {\phi(x_n)\over \phi_0} e^{[c_1RL_P^2+\cdots]}
     \right]^{{i(\Delta x)^4\over L_P^4}}\label{modres}
     \end{equation}
     
     In the long wavelength limit,
     the extra factor in (\ref{modres}) will lead to a term of the form 
      \begin{eqnarray}
     \prod_n \left({\phi\over \phi_0}\right)^{{i(\Delta x)^4\over L_P^4}}
     &=&\prod_n \exp \left[{{i(\Delta x)^4\over L_P^4}}
     \ln \left({\phi\over \phi_0}\right)\right]\nonumber\\
     & \to&  
     \exp {i\over L_P^4} \int d^4 x \sqrt{-g} \ln \left({\phi\over \phi_0}\right)
     \label{phifield}
     \end{eqnarray}
     Thus, the net effect of our assumption is to introduce a `scalar field  potential' $V(\phi) = 
     - L_P^{-4} \ln \left(\phi/ \phi_0\right)$ in the semi classical limit. It is obvious that
     the rescaling of such a scalar field by $\phi \to q \phi$ is equivalent to adding a cosmological
     constant with vacuum energy $-L_P^{-4} \ln q$. Alternatively, any vacuum energy
     can be re absorbed by such a rescaling.     
 The fact that the scalar degree of freedom occurs as a potential in (\ref{phifield})
     without a corresponding kinetic energy term shows that its dynamics is unconventional and non classical.

     The above description in terms of macroscopic scalar degree of freedom can, of course, be only approximate.
      Treated as a  vestige
     of a quantum gravitational degrees of freedom, the cancellation 
     cannot be
     precise because of fluctuations in the elementary spacetime volumes.
     These fluctuations will reappear as a ``small''  cosmological constant
     because of  two key ingredients:
 (i) discrete spacetime structure at Planck length and (ii) quantum gravitational
uncertainty principle. 

To show this, we use the fact noted earlier in section \ref{cclagrange} that the net
cosmological constant can be thought of as a Lagrange multiplier for proper volume
of spacetime in the action functional for gravity.
In any quantum cosmological models which leads to large volumes for the universe, phase of
the wave function will pick up a factor of the form
\begin{equation}
\Psi\propto \exp(-iA_0) \propto 
 \exp\left[ -i\left({\Lambda_{\rm eff}{\mathcal V}\over 8 \pi  L_P^2}\right)\right]
 \end{equation}
 from  (\ref{lambdavol}), where ${\mathcal V}$ is the four volume. 
Treating $(\Lambda_{\rm eff}/8 \pi  L_P^2,{\mathcal V})$ as conjugate variables $(q,p)$, we can invoke the standard uncertainty principle to predict
 $\Delta\Lambda\approx
8 \pi  L_P^2/\Delta{\mathcal V}$. Now we use the earlier  assumption regarding the microscopic structure of the spacetime: Assume that there is a zero point length  of the order of $L_P$
so that the volume of the universe is made of a large number ($N$) of cells, each of volume $(\alpha L_P)^4$ where 
   $\alpha$ is a numerical constant. Then  ${\mathcal V}=N(\alpha L_P)^4$, implying a Poisson fluctuation $\Delta{\mathcal V}\approx
\sqrt{{\mathcal V}}(\alpha L_P)^2$  and leading to 
\begin{equation}
\Delta\Lambda={8 \pi L_P^2\over \Delta{\mathcal V}}= \left({8\pi \over \alpha^2}\right){1\over\sqrt{{\mathcal V}}}\approx  {8\pi \over \alpha^2} H_0^2
\label{dellamb}
\end{equation}
This will give $\Omega_\Lambda= (8\pi/3\alpha^2)$ which will --- for example --- lead to $\Omega_\Lambda =(2/3)$ if $\alpha = 2 \sqrt{\pi}$. Thus
Planck length cutoff (UV limit) and volume of the universe (IR limit) combine to give the correct $\Delta\Lambda$. 
 (A similar result was obtained earlier in \cite{sorkin} based on a different model.)   The  key idea, in this approach,  is that $\Lambda$ is a stochastic variable with a zero mean and fluctuations. It is the rms fluctuation which is being observed in the cosmological context.
 
 This has 
 three implications: First, FRW equations now need to be solved with a stochastic term on the right hand side and one should check whether the observations can still be explained.
 The second  feature is that stochastic properties of $\Lambda$ need to be described by a quantum cosmological model. If the quantum state of the universe is expanded in terms of the eigenstates of some suitable operator (which does not commute the total four volume operator), then one should be able to characterize the fluctuations in each of these states.  Third, and most important, the idea of a 
 cosmological constant arising as 
 a {\it fluctuation} makes sense only if the bulk value is rescaled away.

 The non triviality of this result becomes clear when we compare it with few other alternative 
 ways of estimating the fluctuations --- none of which gives the correct result. The first
 alternative approach is based on the assumption
 that one can associate an entropy $S=(A_H/4L_P^2)$ with compact space
time horizons of area $A_H$ (We will discuss this idea in detail in section \ref{horizons}).  A popular interpretation of this result is that horizon areas are quantized
in units of $L_P^2$ so that $S$ is proportional to the number of bits of information contained in the horizon area.
In this approach, horizon areas can be expressed in the form $A_H=A_P N$ where $A_P\propto L_P^2$
is a quantum of area and $N$ is an integer. Then the {\it fluctuations} in the area will be 
$\Delta A_H=A_P\sqrt{N}=\sqrt{A_PA_H}$. Taking $A_H\propto \Lambda^{-1}$ 
for the de Sitter horizon, we find that $\Delta\Lambda\propto H^2(HL_P)$ which is a 
lot smaller than what one needs.
Further, taking $A_H\propto r_H^2$, we find that $\Delta r_H\propto L_P$; that is, this result  essentially arises from the idea that the radius of the horizon is uncertain within one Planck length. This is quite true, of course, but
does not lead to large enough fluctuations. 

A more sophisticated way of getting this (wrong) result is to
relate the fluctuations in the cosmological constant
 to that of the volume of the universe is by using a canonical ensemble description for universes
 of proper Euclidean 4-volume \cite{hawk}. 
 Writing $V\equiv {\mathcal V}/8\pi L_P^2$ and treating
 $V$ and $\Lambda$ as the relevant variables, one can write a partition function for the 
 4-volume as 
 \begin{equation}
 Z(V) = \int_0^\infty g(\Lambda) e^{-\Lambda V} d\Lambda
 \end{equation}
 Taking the analogy with standard statistical mechanics (with the correspondence $V \to \beta$ and
 $\Lambda \to E$), we can evaluate the fluctuations in the cosmological constant in exactly the 
 same way as energy fluctuations in canonical ensemble. 
 (This is done in several standard text books; see, for example, \cite{tpvol1} 
 p. 194.)
 This will give
 \begin{equation}
 \left( \Delta \Lambda\right)^2 ={C\over V^2}; \qquad C = {\partial \Lambda\over \partial (1/V)} = - V^2 {\partial \Lambda\over \partial V}
 \label{fluct}
 \end{equation}
 where $C$ is the analogue of the specific heat. 
 Taking the 4-volume of the universe to be ${\mathcal V} = b H^{-4}=9b\Lambda^{-2}$ where $b$ is a numerical
 factor and using 
 $V = ({\mathcal V}/8\pi  L_P^2) $
we get $\Lambda \propto L_P^{-1} V^{-1/2}$.  It follows from (\ref{fluct}) that 
 \begin{equation}
 (\Delta \Lambda)^2 = {C\over V^2}={12\pi\over b} (L_PH^3)^2
 \label{canens}
 \end{equation}
 In other words $\Delta \Lambda\propto H^2(HL_P)$, which is the same result from area quantization and is
  a lot smaller 
 than the cosmologically significant value.

Interestingly enough, one could do slightly better by assuming that
the horizon {\it radius} is quantized in units of  Planck length, so that $r_H=H^{-1}=NL_P$. This will lead to the
fluctuations $\Delta r_H=\sqrt{r_HL_P}$ and using $r_H=H^{-1}\propto \Lambda^{-1/2}$, we get 
$\Delta\Lambda\propto H^2(HL_P)^{1/2}$ --- larger than (\ref{canens}) but still inadequate.  
In summary,  the existence of two length scales $H^{-1}$ and $L_P$ allows different 
 results for $\Delta \Lambda$ depending on how exactly
 the fluctuations are characterized ($ \Delta V \propto \sqrt{N}, \Delta A \propto \sqrt{N} $ or
$ \Delta r_H \propto \sqrt{N}$). Hence the result obtained above in
 (\ref{dellamb}) is non trivial.

These conclusions stress, among other things, the difference between {\it fluctuations} and the {\it mean values}. For, if one assumes that every patch of the universe with size $L_P$  contained an energy $E_P$, then a universe with characteristic size $H^{-1}$ will contain the energy
 $E=(E_P/L_P)H^{-1}$. The corresponding energy {\it density} will be $\rho_\Lambda=(E/H^{-3})=(H/L_P)^2$ which leads to the correct result. But, of course, we do not know why every length scale $L_P$ should contain an energy $E_P$ and --- more importantly --- contribute coherently  to give the total energy.

  \subsection{Anthropic interpretation of the \cc }\label{anthrocc}
  
      The anthropic principle \cite{bartip,hogan99}
      is an interpretational paradigm which argues that, while discussing the origin of physical phenomena and
      the values of 
      constants of nature, we must recognize the fact that only certain combination and range
      of values will lead to the existence of intelligence observers in the universe who could
      ask questions related to these issues. This paradigm has no predictive 
      power in the sense that none of the values of the cosmological parameters were ever predicted
      by this method.\footnote{Some advocates of the anthropic principle cite 
      Fred Hoyle predicting the existence of excited state of carbon nucleus, thereby leading
      to efficient triple alpha reaction in stellar nucleosynthesis,
       as an example  of  a prediction from anthropic principle;
      it is very doubtful whether Hoyle applied anthropic considerations in arriving at
      this conclusion.} In fact some   cosmologists have advocated  the   
       model with $\Omega_{\rm NR} =1, \Omega_\Lambda=0$ strongly
      and later --- when observations indicated  $\Omega_\Lambda \ne 1$ ---
     have  advocated the anthropic interpretation of cosmological constant with
      equal fluency. This is defended by the argument that not all guiding principles
      in science (Darwinian evolution, Plate tectonics, ....) need to be predictive in 
      order to be useful. In this view point, anthropic principle is a back drop for
      discussing admittedly complicated conceptual issues. 
      Within this paradigm there have been many attempts to explain (after the fact)
      the values of several fundamental constants with varying degree of
      success. 
      
      In the context of \cc, the anthropic interpretation works as follows.
      It is assumed that  widely disparate values
      for the constants of nature can occur in an ensemble of universes (or 
      possibly in different regions of the universe causally unconnected with
      each other). Some of these values for constants of nature --- and in particular
       for the \cc\ --- will lead broadly to the kind of universe we seem to live in.
      This is usually characterized by  formation of: (i) structures by
      gravitational instability, (ii) stars which act as gravitationally
      bound nuclear reactors that synthesize the elements and distribute
      them  and (iii) reasonably complex molecular structures which could
      form the basis for some kind of life form.      
      Showing that such a scenario can exist only for a particular
      range of values for the cosmological constant is considered an
      explanation for the value of cosmological constant by the 
      advocates of anthropic principle. (More sophisticated versions
      of this principle exist; see, for example \cite{banks}, 
      and references cited therein.)
      
       The simplest constraint on the \cc\
      is that it should not be so high as to cause rapid expansion of 
      the universe early on preventing the formation of galaxies
      \cite{weinberg87}. 
      If the energy density of the \cc\ has to be less than that of
      energy density of matter at the redshift $z_{\rm gal}(\approx 4)$ at
      which galaxy formation takes place, 
      then we must have 
      \begin{equation}
      \frac{\Omega_\Lambda}{\Omega_{\rm NR}} \lesssim (1+z_{\rm gal})^3 \approx 125
      \end{equation}
      This gives a bound on $\Omega_\Lambda$ which is ``only'' a couple of orders of magnitude
      larger than what is observed. 
      
      More formally, one could ask:
      What is the most  probable value of $\Omega_\Lambda$ if it is interpreted as
      the value that would have been observed by the largest number of 
      observers \cite{vilenkin,efstath}?
      Since a universe with $\Omega_\Lambda \approx \Omega_{\rm NR}$ will have
      more galaxies than one with a universe with $\Omega_\Lambda \approx 10^2 \Omega_{\rm NR}$,
      one could argue that most observers will measure a value $\Omega_\Lambda \approx 
      \Omega_{\rm NR}$. The actual probability $dP$  for measuring a particular value for $\Omega_\Lambda$
       in the range $(\Omega_\Lambda, \Omega_\Lambda + d\Omega_\Lambda)$ is the product
        $(dP/d\Omega_\Lambda)= Q(\Omega_\Lambda) \mathcal{P} (\Omega_\Lambda)$
      where $\mathcal{P}$ is the a priori probability measure for a specific value of 
      $\Omega_\Lambda$ in a member of an ensemble of universes (or in a region of the 
      universe) and $Q(\Omega_\Lambda)$ is the average number of galaxies which 
      form in a universe with a given value of $\Omega_\Lambda$. There has been several attempts
      to estimate these quantities (see, for example, \cite{martel,garriga}) but  all of them 
      are necessarily speculative.   
       The first --- and the most serious --- difficulty with this approach
      is the fact that we simply do not have any reliable way of estimating $\mathcal{P}$; in fact, if we really  had a way of calculating it from a fundamental theory, such a theory probably
      would have provided a deeper insight into the \cc\ problem itself. The second
      issue has to do with the dependence of the results on other parameters
      which describe
      the cosmological structure formation (like for example, the spectrum of initial perturbations).
      To estimate $Q$ one needs to work in a multi parameter space and marginalize
      over other parameters --- which would involve more assumptions regarding
      the priors. And finally,  anthropic paradigm  itself
      is suspect in any scientific discussion, for reasons mentioned earlier. 
  
  \subsection{Probabilistic interpretation of the \cc }\label{probintcc}
  
  It is also  possible to produce more complex scenarios which 
  could justify the small or zero value of \cc. One such idea, which enjoyed 
   popularity for a few years \cite{tpbaum1,tpbaum2,tpbaum3,coleman},
  is based on the conjecture that quantum wormholes can change the
  effective value of the observed constants of nature. The wave function of the 
  universe, obtained by a path integral over all possible spacetime metrics
  with wormholes, will receive dominant contributions from those 
  configurations for which the effective values of the physical constants
  extremize the action. Under some assumptions related to Euclidean quantum
  gravity, one could argue that the configurations with zero \cc\ will occur
  at late times. It is, however, unlikely that the assumptions of Euclidean quantum
  gravity has any real validity and hence this idea must be considered as
   lacking in concrete justification.  
  
  \section{Relaxation mechanisms for the \cc}\label{relaxcc}
  
  One possible way of obtaining a small, non-zero, cosmological constant at the present epoch 
  of the universe is to make the cosmological constant evolve in time due to
  some physical process. At a phenomenological level this can be done
  either by just postulating such a variation and explore its consequences or
  --- in a slightly more  respectable way --- by postulating a scalar
  field potential as described in section \ref{ccmodels}. 
  These models, however, cannot explain why a bare \cc\ [the first term 
  on the right hand side of (\ref{nine})] is zero. To tackle this issue, one can
  invoke some field [usually a scalar field] which  {\em directly couples} to the cosmological constant and decreases its ``effective value". We shall now examine two such models.
  
  The key idea is to introduce a field which couples to the trace $T= T^a_a$ of the energy momentum tensor.
   If $T$ depends on $\phi$ and vanishes at some value $\phi=\phi_0$, then $\phi$ will evolve towards $\phi=\phi_0$ at which $T=0$. This equilibrium solution will have zero \cc  \cite{dolgov,wilzcek,peccei,barr}.
  While this idea sounds attractive, there are general arguments as to why it does \emph{not} work in the simplest context \cite{swlambda}.
  
  A related attempt was made by several authors,    \cite{dolgov,barr87,ford,hebecker},  who coupled the scalar field directly to $R$ which, of course, is proportional to $T$ because of Einstein's equations.
  Generically, these models have the Lagrangian
  \begin{equation}
  L = \left[ \frac{1}{2} \partial_\mu \phi \partial^\mu \phi + \frac{1}{16\pi G} (R - 2\Lambda) - U(\phi) R\right]
  \end{equation}
  The field equations of this model has flat spacetime solutions at $\phi=\phi_0$ provided $U(\phi_0)=\infty$.
  Unfortunately, the effective gravitational constant in this model evolves as
   \begin{equation}
   G_{\rm eff} = \frac{G}{1+ 16\pi G U(\phi_0)}
   \end{equation}
  and vanishes as $U\to\infty$. Hence these models are not viable.
  
  The difficulty in these models arise because they do not 
  explicitly couple the trace of the $T_{ab}$ of  the scalar field itself.  
  Handling this consistently \cite{tpstp88}
  leads to a somewhat different model which we will briefly describe because of its conceptual interest.
  
  Consider a system consisting of the gravitational fields $g_{ab}$, radiation fields, and a  scalar field 
  $\phi$ which couples to the trace of the energy-momentum tensor of all fields, including its own. 
  The {\em zeroth order} action for this system is given by
  \begin{equation}
  A^{(0)} = A_{\rm grav} + A_\phi^{(0)} + A_{\rm int}^{(0)} + A_{\rm radn}
  \end{equation}
  where 
  \begin{equation}
  A_{\rm grav} = (16 \pi G )^{-1} \int R \sqrt{-g} \, d^4x - \int \Lambda \sqrt{-g} \, d^4x,
  \end{equation}
  \begin{equation}
  A_\phi^{(0)} = \frac{1}{2} \int \phi^i\phi_i\sqrt{-g} \, d^4x; \quad A_{\rm int}^{(0)} = \eta \int Tf(\phi/\phi_0) 
  \sqrt{-g} \, d^4x
  \end{equation}
 Here, we have explicitly included the cosmological constant term and $\eta $ is a dimensionless number which 
 `switches on' the interaction. In the zeroth order action, $T$ represents the trace of all fields other 
 than $\phi$. Since the radiation field is traceless, the only zeroth-order contribution to $T$ comes from the $\Lambda$ term, so that we have $T= 4 \Lambda$. The coupling to the trace is through a function $f$ of the scalar
 field, and one can consider various possibilities for this function. The constant $\phi_0$ converts $\phi$
 to a dimensionless variable, and is introduced for dimensional convenience.

 To take into account the back-reaction of the scalar field on itself, we must add to $T$ the contribution
  $T_\phi = - \phi^l\phi_l$  of the scalar field. 
  If we now add $T_\phi$ to $T$ in the interaction term $A_{\rm int}^{(0)}$ further modifies $T^{ik}_\phi$.
  This again changes $T_\phi$. Thus to arrive at the correct action an infinite iteration will have to be performed and the complete action  can be obtained by summing up all the terms. (For a demonstration of this iteration procedure,  see \cite{jvnrawaltp1,jvnrawaltp2}.) The full action  can be found more simply by a  consistency
  argument.
  
  Since the effect of the iteration is to modify the expression for $A_\phi$ and $A_\Lambda$,  we consider
  the following ansatz for the full action:
  \begin{eqnarray}
  A &=& \frac{1}{16 \pi G} \int  R \sqrt{-g} \, d^4x - \int \alpha(\phi) \Lambda  \sqrt{-g} \, d^4x
  \nonumber \\
  && \quad +\frac{1}{2}
  \int \beta(\phi) \phi^i\phi_i  \sqrt{-g} \, d^4x  + A_{\rm rad}
  \end{eqnarray}
  Here $\alpha(\phi)$ and $\beta(\phi)$ are functions of $\phi$ to be determined by the 
  consistency requirement  that they represent the 
  effect of the 
  iteration of the interaction term. (Since radiation makes no contribution to $T$, we expect  $A_{\rm rad}$ to remain unchanged.)
  The energy-momentum tensor for $\phi$ and $\Lambda$ is now given by
  \begin{equation}
  T^{ik} = \alpha(\phi) \Lambda g^{ik} + \beta(\phi) \left[ \phi^i\phi_k - \frac{1}{2} g^{ik} \phi^\alpha\phi_\alpha\right]
  \end{equation}
  so that the total trace is $T_{\rm tot}=4\alpha(\phi)\Lambda - \beta(\phi) \phi^i\phi_i$. The functions
  $\alpha(\phi) $ and $\beta(\phi)$ can now be  determined by the consistency requirement
  \begin{eqnarray}
  && -\int \alpha(\phi) \Lambda  \sqrt{-g} \, d^4x  +\frac{1}{2}
  \int \beta(\phi) \phi^i\phi_i  \sqrt{-g} \, d^4x \nonumber \\
  && \quad = -\int \Lambda \sqrt{-g} \, d^4x +  \frac{1}{2}
  \int \phi^i\phi_i  \sqrt{-g} \, d^4x +\eta \int T_{\rm tot} f(\phi/\phi_0) \sqrt{-g} \, d^4x
  \end{eqnarray}
  Using $T_{\rm tot}$ and comparing terms in the above equation we find that
  \begin{equation}
  \alpha(\phi) = [ 1+ 4 \eta f]^{-1}, \quad \beta(\phi) = [ 1+ 2 \eta f]^{-1}
  \end{equation}
  Thus the complete action can be written as
  \begin{eqnarray}
  A &=& \frac{1}{16 \pi G} \int  R \sqrt{-g} \, d^4x - \int \frac{ \Lambda}{1+4nf}  \sqrt{-g} \, d^4x
  \nonumber \\
  && \quad+ \frac{1}{2}
  \int \frac{ \phi^i\phi_i }{1+2nf} \sqrt{-g} \, d^4x  + A_{\rm rad}
  \label{completeaction}
  \end{eqnarray}
  (The same action would have been obtained if one uses the iteration procedure.) 
  The action in (\ref{completeaction}) leads to the following field equations,
  \begin{eqnarray}
  R_{ik} - \frac{1}{2} g_{ik} R &=& -8 \pi G \left[ \beta(\phi) \left( \phi^i\phi^k - \frac{1}{2} g^{ik} \phi^\alpha\phi_\alpha \right) \right. \nonumber\\
  && \hskip8em \left. + \frac{\Lambda}{8\pi G} \alpha(\phi) g_{ik} +T_{ik}^{\rm traceless}\right]
  \end{eqnarray}
  \begin{equation}
  \square \phi +\frac{1}{2} \frac{\beta'(\phi)}{\beta(\phi)} \phi^i\phi_i + \frac{\Lambda}{8\pi G} \frac{\alpha'(\phi)}{\beta(\phi)} =0
  \end{equation} 
  Here, $\square$ stands for a covariant d'Lambertian, $T_{ik}^{\rm traceless}$ is the 
  stress tensor of all fields with traceless stress tensor and a prime denotes differentiation with respect 
  to $\phi$.
  
  In the cosmological context, this reduces to
  \begin{equation}
  \ddot \phi +\frac{3\dot a}{a} \dot \phi = \eta \dot \phi^2 \frac{f'}{1+2\eta f} + \eta \frac{\Lambda}{2\pi G} \frac{f'(1+2\eta f)}{(1+4\eta f)^2}
  \end{equation}
  \begin{equation}
  \frac{\dot a^2 +k}{a^2} = \frac{8\pi G}{3} \left[ \frac{1}{2} \frac{\dot \phi^2}{1+2\eta f} + \frac{\Lambda}{8\pi G}
  \frac{1}{(1+4\eta f)} + \frac{\rho_0}{a^4}\right]
  \end{equation}

  It is obvious that the effective cosmological constant can decrease if $f$ increases in an expanding universe.The result can be easily generalized for a scalar field with a potential by
  replacing $\Lambda$ by $V(\phi)$.
  This model is conceptually attractive since it correctly accounts for the coupling of the scalar field with 
  the trace of the stress tensor. 
  
  The trouble with this model is two fold: (a) If one uses natural initial conditions and do {\it not} fine tune the parameters, then one does not get a viable model. (b) Since the scalar field couples to the trace of all sources, it also couples to dust-like matter and ``kills" it, making the universe radiation dominated at 
  present. This reduces the age of the universe and could also create difficulties for structure formation.
  These problems can be  circumvented by invoking a suitable potential $V(\phi)$ within this
  model \cite{samitp}. However, such an approach takes away the naturalness of the model
  to certain extent.

   \section{Geometrical structure of the de Sitter spacetime}\label{desittergeom}
   
   The most symmetric {\em vacuum} solution to Einstein's equation, of course, is the flat spacetime. If we now add
   the cosmological constant as the only source of curvature in Einstein's equation, the resulting spacetime 
    is also highly symmetric and has an interesting geometrical structure. In the case of a positive cosmological constant, this is the {\em de Sitter manifold} and in the case of negative cosmological constant, it is known as {\em anti-de Sitter manifold}. We shall now discuss some features of the former, corresponding to the positive \cc . (For a nice, detailed, review of the classical geometry of de Sitter spacetime, see \cite{yoonbai}.)
     
 To understand the geometrical structure of the de Sitter spacetime, 
 let us begin by noting that a spacetime
 with the source $T^a_b = \rho_\Lambda \delta^a_b$ must have
  3-dimensional section which are   homogeneous and isotropic. This will lead us to
  the Einstein's equations for a
 FRW universe with  \cc\ as source
   \begin{equation}
   \frac{\dot a^2}{a^2} +\frac{k}{a^2} = \frac{8\pi G }{3} \rho_\Lambda \equiv H^2
   \end{equation}
   This equation can be solved with any of the following three forms of $(k, a(t))$ pair. The first
   pair is the spatially flat universe with $(k=0, a= e^{Ht})$. The second corresponds to
   spatially open universe with $(k=-1, a=H^{-1} \sinh Ht)$ and the  third
   will be $(k=+1, a=H^{-1} \cosh Ht)$.  Of these, the last pair gives a coordinate system
   which covers the full de Sitter manifold. In fact, this is the metric on a 4-dimensional
   hyperboloid, embedded in a 5 dimensional Minkowski space with the metric 
\begin{equation}
ds^2 = dt^2 - dx^2 - dy^2 - dz^2 - dv^2, 
\label{eq:5dmetric}
\end{equation}
The equation of the hyperboloid in 5-D space is 
\begin{equation}
t^2 - x^2 - y^2 - z^2 - v^2 = - H^{-2}.
\label{eq:5dhyper}
\end{equation}
We can introduce a parametric representation of the hyperbola with
 the four  variables $(\tau, \chi, \theta, \phi)$     where
\begin{eqnarray}
&&x = H^{-1} \cosh(H\tau) \sin \chi \sin\theta \cos\phi;~~ 
y = H^{-1} \cosh(H\tau) \sin \chi \sin\theta \sin\phi;
\nonumber \\
&&z = H^{-1} \cosh(H\tau) \sin \chi \cos\theta;
\nonumber \\
&&v = H^{-1} \cosh(H\tau) \cos \chi;
\nonumber \\
&& t = H^{-1} \sinh(H\tau).
\label{eq:clos_tran}
\end{eqnarray}
This set, of course, satisfies (\ref{eq:5dhyper}). Using (\ref{eq:5dmetric}), 
we can compute the metric induced on the
hyperboloid  which --- when expressed in terms 
of the four coordinates $(\tau, \chi, \theta, \phi)$ ---  is given by 
\begin{equation}
ds^2 = d\tau^2 - H^{-2} \cosh^2(H\tau) \left[d\chi^2 + \sin^2 \chi (d\theta^2 + \sin^2 \theta d\phi^2)\right],
\label{eq:5dclosed}
\end{equation}
This  is precisely the de Sitter manifold with closed spatial sections.

All the three forms of FRW universes with $k=0, \pm 1$ arise by taking
different cuts in this 4-dimensional hyperboloid embedded in the  5-dimensional spacetime.
Since two of these dimensions (corresponding to the polar angles $\theta$ and $\phi$) merely
go for  a ride, it is more convenient (for visualization) to work with a 3-dimensional spacetime
having  the metric 
\begin{equation}
ds^2 = dt^2 - dx^2 - dv^2.
\label{eq:3dmetric}
\end{equation}
instead of the 5-dimensional metric (\ref{eq:5dmetric}). Every point in this 3-dimensional
space corresponds to a 2-sphere whose coordinates $\theta$ and $\phi$ are suppressed for
simplicity. The $(1+1)$ de Sitter spacetime is the 2-dimensional hyperboloid 
[instead of the four dimensional hyperboloid of (\ref{eq:5dhyper})]
with the equation 
\begin{equation}
t^2 - x^2 - v^2 = - H^{-2}
\label{eq:3dhyper}
\end{equation}
embedded in the 3-dimensional space with metric (\ref{eq:3dmetric}).
  The three different coordinate systems which are natural on this hyperboloid are the 
  following:
  \begin{itemize}
  \item  {\em Closed Spatial Sections}: This is obtained by introducing the coordinates 
  $t = H^{-1} \sinh(H\tau); x = H^{-1} \cosh(H\tau)\sin \chi;
v =H^{-1} \cosh(H\tau) \cos \chi $ on the hyperboloid, in terms of which the induced 
metric on the hyperboloid has the
form 
\begin{equation}
ds^2 = d\tau^2 - H^{-2} \cosh^2(H\tau) d\chi^2
\label{eq:3dclosed}
\end{equation}
This is the two dimensional de Sitter space which is analogous to the 4-dimensional
case described by (\ref{eq:5dclosed}). 
\item  {\em Open Spatial Sections}: These are obtained by using the coordinates
$ t = H^{-1} \sinh(H\tau) \cosh \xi;~~x = H^{-1} \sinh(H\tau)\sinh \xi;~~
v = H^{-1} \cosh(H\tau)$
  on the hyperboloid in terms of which the induced 
metric on the hyperboloid has the
form 
\begin{equation}
ds^2 = d\tau^2 - H^{-2} \sinh^2(H\tau) d\xi^2
\label{eq:3dopen}
\end{equation}
\item  {\em Flat Spatial Sections}: This corresponds to the choice
 $t=H^{-1} \sinh(H\tau) + (H^{-1}/2) \xi^2\exp(H\tau)$;
$x=H^{-1} \cosh(H\tau) - (H^{-1}/2)\xi^2 \exp(H\tau)$;
$v= \xi \exp(H\tau)$ leading to the metric 
\begin{equation}
ds^2 = d\tau^2 - \exp(2H\tau) d\xi^2
\label{eq:3dflat}
\end{equation}
This covers one half of the de Sitter hyperboloid bounded by the null rays
$t+x=0$. 
\end{itemize}

All these metrics have an apparent time dependence. But, in 
the absence of any source other than \cc, there is no preferred notion of
time and the spacetime manifold cannot have any intrinsic time dependence.
This is indeed true, in spite of the expansion factor $a(t)$ ostensibly
depending on time. The translation along the time direction merely slides the point 
on the surface of the hyperboloid. 
[This is obvious in the coordinates ($k=0, a\propto e^{Ht}$) in which the time 
translation $t \to t+\epsilon$ merely rescales the coordinates by $(\exp H\epsilon)$.]

The time independence of the metric
can be made explicit in another set of coordinates called `static coordinates'.
To motivate these coordinates, let us note that a spacetime with only \cc\ 
as the source is  certainly static and possesses spherical symmetry. Hence we can 
also express the metric in the form
\begin{equation}
ds^2 = e^\nu dt^2 - e^\lambda dr^2 - r^2 ( d\theta^2 + \sin^2\theta d\phi^2)
\end{equation}
where $\nu$ and $\lambda$ are functions of $r$. The Einstein's equations for this metric
has the solution $e^\nu = e^{-\lambda} = (1- H^2 r^2)$ leading to 
\begin{equation}
ds^2 = (1-H^2r^2) dt^2 - \frac{dr^2}{(1-H^2r^2)} - r^2 (d\theta^2 + \sin^2\theta d\phi^2)
\label{desitterstatic}
\end{equation}
  This form of the metric makes the static nature apparent.  
  This metric also describes a hyperboloid embedded in a higher  dimensional flat space. For example,
  in the $(1+1)$ case (with $\theta, \phi$ suppressed) this metric  
    can be obtained by the following
  parameterization of the hyperboloid in equation (\ref{eq:3dhyper}):
  \begin{equation}
t= (H^{-2} - r^2)^{1/2} \sinh(H\tau); \quad 
v= (H^{-2} - r^2)^{1/2} \cosh(H\tau); \quad x=r
\label{eq:3d_sta_tr}
\end{equation}
The key feature of the manifold, revealed by 
   equation (\ref{desitterstatic})  is the existence of a horizon at $r=H^{-1}$. It also
   shows that
  $t$ is a time-like coordinate only in the region $r< H^{-1}$. 
  
  The structure of the metric
  is very similar to the Schwarzschild metric: 
  \begin{equation}
  ds^2 = \left(1-\frac{2M}{r}\right) dt^2 - \frac{dr^2}{ \left(1-\frac{2M}{r}\right)} -
   r^2 (d\theta^2 + \sin^2\theta d\phi^2)
   \label{schwarzmetric}
  \end{equation}
  Both the metrics (\ref{schwarzmetric}) and  (\ref{desitterstatic}) are spherically symmetric
  with $g_{00} = - (1/g_{11})$. 
  Just as the Schwarzschild metric
  has a horizon at $r=2M$ (indicated by $g_{00}\to 0, g_{11}\to \infty$),
   the de Sitter metric also has a horizon at $r=H^{-1}$.
  From the slope of the light cones $(dt/dr) = \pm (1-H^2 r^2)^{-1}$ [corresponding to $ds=0=d\theta
  =d\phi$ in (\ref{eq:3d_sta_tr})] it is clear that
  signals sent from the region $r<H^{-1}$  cannot go beyond the surface $ r= H^{-1}$. 
  
  This feature, of course, is independent of the coordinate system used.
  To see how the horizon in de Sitter universe arises in the
  FRW coordinates, let us recall the equation governing the propagation of 
  light signals between the events ($t_1,r_1$) and ($t,r$):
  \begin{equation}
  \int_{r_1}^r \frac{dx}{\sqrt{1-k x^2}} = \int_{t_1}^{t}\frac{dt'}{a(t')}.
\label{lightprop}
  \end{equation}
  Consider a photon emitted by an observer at the origin  at the present epoch $(r_1=0, t_1=t_0)$.
  The maximum {\em coordinate} distance  $x_H$ reached by this photon as $t\to \infty$ is determined
  by the equation 
  \begin{equation}
  \int_0^{x_H} \frac{dx}{\sqrt{1-k x^2}} = \int_{t_0}^\infty \frac{dt'}{a(t')}.
\label{lightprop1}
  \end{equation}
   If the integral on the right hand side
  diverges as $t\to \infty$, then, in the same limit, $x_H \to \infty$ and 
  an observer can send signals to any event provided (s)he waits
  for a sufficiently long time. But if the integral on the right hand side
  converges to a finite value as $t\to \infty$, then
  there is a finite horizon radius   beyond  which
  the observer's signals will not reach  even if (s)he waits
  for infinite time. In the de Sitter universe with $k=0$ and $a(t) = e^{Ht}$,
  $x_H=H^{-1} e^{-H t_0}$; the corresponding maximum proper distance up to which
  the signals can reach is $r_H= a(t_0) x_H = H^{-1}$.
  Thus we get the same result in any other coordinate system.
  
  Since the result depends essentially on the behaviour of $a(t)$ as 
  $t\to \infty$, it will persist even in the case of a universe containing
  {\it both} non relativistic matter and cosmological constant. For example,
  in our universe, we can ask what is the highest redshift
  source from 
    which we can ever receive  a light signal, if the signal
  was sent today.
  To compute  this explicitly, consider a model
  with $\Omega_{\rm NR} +\Omega_\Lambda =1$. Let us assume
  that light from an event at $(r_H, z_H)$ reaches $r=0$ at 
  $z=0$ giving 
  \begin{equation}
r_H = \int_{t_H}^{t_0}{dt\over a(t)}=\int_0^{z_H} \frac{dz}{H_0\left[ 1-\Omega_{\rm NR} + \Omega_{\rm NR} (1+z)^3\right]^{1/2}}.
\label{lighttrav}
\end{equation}
  If we take $r_H$ to be the size of the horizon, then it also follows
  that the light emitted today from this event will just reach us at 
  $t=\infty$. This gives 
  \begin{equation}
 r_H = \int_{t_0}^{\infty}{dt\over a(t)}= \int_{-1}^{0} \frac{dz}{H_0\left[ 1-\Omega_{\rm NR} + \Omega_{\rm NR} (1+z)^3\right]^{1/2} }.
\label{lighthorizon}
  \end{equation}
  Equating the two expressions, we get an implicit expression for $z_H$. If $\Omega_{\rm NR}
  =0.3$, the limiting redshift is quite small: $z_H \approx 1.8$. This implies that
  sources with $z>z_H$ can never be influenced by light signals from us
  in a model with \cc\ \cite{loeb98,starkman99}.

  \section{Horizons, temperature and entropy}\label{horizons}
   
   In the description of standard cosmology
    $\Omega_\Lambda$ appears as a parameter like, say, the Hubble
   constant $H_0$. There is, however, a significant difference between these two  parameters
   as far as fundamental physics is concerned. The exact numerical value of $h$ is not of 
   major concern to fundamental physics. But, the non-zero value for $\Omega_\Lambda$
   signifies the existence of an exotic form of energy density with negative pressure which is a 
   result of deep significance to the whole of physics.
   We shall now take up an  important aspect of the \cc\ which is somewhat 
   different in spirit compared to the results covered  so far. 
   
It turns out that the universe with a non-zero value for \cc\ behaves in many ways
   in a manner similar to a black hole. Just as the black hole has close links with thermodynamics
   (like having a finite temperature, entropy etc.) the de Sitter universe also possesses thermodynamic
   features which makes it peculiar and important in understanding the \cc.  This thermodynamic
relationship of the \cc has not been adequetely explored or integrated into the standard cosmological
description so far. But since it is likely to have a important implications for the eventual resolution of
the \cc problem, 
 we shall provide
   a fairly self contained description of the same. 
   
   One of the remarkable features of classical gravity is that it can wrap up regions of spacetime thereby producing surfaces which act as one way membranes. The classic example is that  of  Schwarzschild black hole    of mass $M$ which has a   compact spherical surface of radius $r = 2M$ that
act as  a horizon. Since the horizon can hide information --- and information is deeply connected with entropy --- one would expect a fundamental relationship between gravity and thermodynamics.
[There is extensive literature in this subject and our citation will be representative rather than exhaustive;
for a text book  discussion and earlier references, see
 \cite{birrel}; for a recent review, see \cite{tprecent1}.]  
As we saw in the last section, 
the de Sitter universe also has a horizon which 
  suggests that de Sitter spacetime will have non trivial thermodynamic features \cite{GH}.

   This result can be demonstrated mathematically in many different ways of which
   the simplest procedure is based on the relationship between temperature and the Euclidean
   extension of the spacetime. To see this connection, let us recall that the mean value
   of some dynamical variable $f(q)$ in quantum statistical mechanics can be expressed
   in the form 
   \begin{equation}
   <f> = \frac{1}{Z}\sum_E \int \phi^*_E(q) f(q) \phi_E(q)e^{-\beta E} \, dq
   \label{avef}
   \end{equation}
   where $\phi_E(q)$ is the stationary state eigen function of the Hamiltonian
   with $H\phi_E =E\phi_E, 
    \beta = (1/T)  $  is the inverse temperature  and $Z(\beta)$ is the partition function.
   This expression calculates the mean value $<E|f|E>$ in a given energy state and then 
   averages over a Boltzmann distribution of energy states with the weightage 
   $Z^{-1} \exp(-\beta E)$.  On the other hand, the quantum mechanical kernel
   giving the probability amplitude for the system to go from  the state $q$ at time $t=0$ 
   to the state $q' $ at time $t$ is given by
   \begin{equation}
   K(q',t;q,0) = \sum_E\phi^*_E(q')  \phi_E(q) e^{-itE}
   \label{defkernel}
   \end{equation}
  Comparing (\ref{avef}) and (\ref{defkernel}) we find
   that the thermal average in (\ref{avef}) can be obtained by
   \begin{equation}
   <f> = \frac{1}{Z} \int dq \, K(q, -i\beta; q,0) f(q)
   \label{newthermave}
   \end{equation}
   in which we have done the following: (i) The time coordinate has been analytically continued 
   to imaginary values with $it =\tau$. (ii) The system is assumed to exhibit periodicity
    in the imaginary time $\tau$  with period $\beta$ in the sense that the state variable $q$ has the same 
    values at    $\tau =0 $ and at $\tau = \beta$. These considerations continue to hold even 
    for a field theory with $q$ denoting the field configuration at a given time. If the system, in 
    particular the Greens functions describing the dynamics, are periodic with a period $p$  in imaginary time,
    then  one can attribute a temperature $T = (1/p)$ to the system.
    It may be noted that the partition function $Z(\beta)$ can also be expressed in the 
    form 
    \begin{equation}
    Z(\beta)= \sum_E e^{-\beta E} = \int dq \, K(q, -i\beta;q,0) = \int {\mathcal D}q \exp[-A_E(q,\beta; q,0)]
    \label{piz}
    \end{equation}
    The first equality is the standard definition for $Z(\beta)$; the second equality
    follows from (\ref{defkernel}) and the normalization of $\phi_E(q)$; the last equality
    arises from the standard path integral expression for the kernel in the Euclidean sector
    (with $A_E$ being the Euclidean action) and imposing the periodic boundary conditions.
    (It is assumed that the path integral measure ${\mathcal D}q$ includes an integration
    over $q$.) We shall have occasion to use this result later. Equations (\ref{newthermave}) and 
    (\ref{piz}) represent the relation between the periodicity in Euclidean time and 
    temperature.

    Spacetimes with horizons possess a natural analytic continuation from Minkowski
    signature to the Euclidean signature with $t\to \tau = it$. If the metric is periodic
    in $\tau$, then one can associate a natural notion of  a temperature  to such spacetimes.
    For example, the de Sitter manifold with the metric  
   (\ref{eq:5dclosed})
   can be continued to imaginary time arriving at the metric 
   \begin{equation}
   - ds^2 = d\tau^2 + H^{-2} \cos^2 H\tau \left[ d\chi^2 + \sin^2 \chi ( d\theta^2 + \sin^2 \theta d\phi^2)\right]
   \label{imagds} 
   \end{equation}
   which is clearly periodic in $\tau$ with the period $(2\pi/H)$. [The original metric was a 
   4-hyperboloid in the 5-dimensional space while equation (\ref{imagds})
   represents a 4-sphere in the 5-dimensional space.]
   It follows that de Sitter spacetime has a natural notion of temperature
   $T= (H/2\pi)$ associated with it.

  It is instructive to see how this periodicity arises in the static form of the metric in
  (\ref{desitterstatic}). Consider a metric of the form 
   \begin{equation}
ds^2 = f(r)dt^2 -  \frac{dr^2}{f(r)} - dL^2_\perp 
\label{generic}
\end{equation}
  where $dL_\perp^2$ denotes the transverse 2-dimensional metric and $f(r)$
  has a simple zero at $r=r_H$. Near $r=r_H$, we can expand $f(r)$ in a 
  Taylor series and obtain $f(r)\approx B(r-r_H)$ where $B\equiv f'(r_H)$.
  The structure of the metric in (\ref{generic}) shows that there is a 
  horizon at $r=r_H$. Further, since the general relativistic metric
  reduces to $g_{00} \approx (1+2\phi_N)$ in the Newtonian limit,
  where $\phi_N$ is the Newtonian gravitational potential, the quantity
  \begin{equation}
  \kappa = |\phi_N' (r_H)| =\frac{1}{2}| g_{00}'(r_H)| = \frac{1}{2}| f'(r_H)| =\frac{1}{2}|B| 
  \end{equation} 
   can be interpreted as the gravitational attraction on the surface of the 
   horizon --- usually called the surface gravity.  Using the form $f(r)\approx 2\kappa (r-r_H)$
   near the horizon and shifting to the coordinate $\xi \equiv [2\kappa^{-1}(r-r_H)]^{1/2}$
   the metric near the horizon becomes
   \begin{equation}
   ds^2\approx \kappa^2 \xi^2 dt^2 - d\xi^2 - dL^2_\perp
   \end{equation}
   The Euclidean continuation $t\to \tau = it$ now leads to the metric 
   \begin{equation}
   -ds^2\approx  \xi^2 d(\kappa\tau)^2 + d\xi^2 + dL^2_\perp
   \end{equation}
   which is essentially the metric in the polar coordinates in 
   the $\tau-\xi$ plane. For this metric to be well defined near the origin, $\kappa\tau$
   should behave like an angular coordinate $\theta$ with periodicity $2\pi$. Therefore,
   we require all well defined physical quantities defined in this spacetime
   to have a periodicity in $\tau$ with the period $(2\pi/|\kappa |)$. Thus, all metrics of 
   the form in (\ref{generic}) with a simple zero for $f(r)$ leads to a horizon
   with temperature $T=|\kappa|/2\pi = |f'(r_H)|/4\pi$. In the case of de Sitter spacetime,
   this gives $T= (H/2\pi)$; for the Schwarzschild metric, the corresponding analysis 
   gives the  well known 
    temperature $T= (1/8\pi M)$ where $M$ is the mass of the black-hole.

   \subsection{The connection between thermodynamics and spacetime geometry}\label{thermgeo}

The existence of one-way membranes, however,  is not necessarily a feature of gravity or curved spacetime
and can be induced even in flat Minkowski spacetime. It is possible to introduce coordinate charts in Minkowski spacetime such that regions are separated by horizons, a familiar example being the
coordinate system used by a uniformly accelerated frame
 (Rindler frame) which has a non-compact  horizon.
 The natural coordinate system $(t,x,y,z)$ used by an observer moving with a uniform acceleration 
 $g$ along the x-axis is related to the inertial coordinates $(T,X,Y,Z)$ by 
 \begin{equation}
 gT=\sqrt{1+2gx}\sinh (gt);\quad (1+gX)=\sqrt{1+2gx}\cosh (gt); 
 \end{equation}  
 and $ Y=y;  Z=z.$ The metric in the accelerated frame will be
 \begin{equation}
 ds^2 =(1+2gx)dt^2 -\frac{dx^2}{(1+2gx)} -dy^2 - dz^2
 \end{equation}
 which has the same form as the metric in (\ref{generic}) with $f(x) = (1+2gx)$. This
 has a horizon at $x=-1/2g$ with the surface gravity $\kappa =g$ and 
 temperature $T=(g/2\pi)$.
   All the horizons are implicitly defined with respect to certain class of observers; for example,
   a suicidal observer plunging into the Schwarzschild black hole will describe the physics very 
   differently from an observer at infinity. From this point of view, which we shall adopt, there is no 
   need to distinguish between observer dependent and observer independent horizons.
   This allows a powerful way of describing the thermodynamical behaviour of all these spacetimes (Schwarzschild, de Sitter, Rindler ....) at one go.
   
   The Schwarzschild, de Sitter and Rindler metrics are symmetric under time reversal and there exists a `natural' definition of  a time symmetric
vacuum state in all these cases.  Such a vacuum state will appear to be 
described a thermal density matrix in a subregion ${\mathcal R}$ of spacetime with the horizon as a boundary. The QFT based on such a state will be manifestedly time symmetric and will describe an isolated system in thermal equilibrium in the subregion ${\mathcal R}$. No time asymmetric phenomena like evaporation, outgoing radiation, irreversible changes etc can take place in this situation. 
We shall now describe how this arises.

    Consider a (D+1) dimensional flat Lorentzian manifold ${\mathcal S}$ with the signature (+, $ -,-,  ...)$
and Cartesian coordinates $Z^A$ where $A=(0,1,2,..., D)$. 
A four dimensional sub-manifold ${\mathcal D}$ in this (D+1) dimensional space can be defined
through a mapping $Z^A=Z^A(x^a)$ where $x^a$ with $a=(0,1,2,3)$ are the four
dimensional coordinates on the surface. The flat Lorentzian metric in the (D+1)
dimensional space induces a metric $g_{ab}(x^a)$ on the four dimensional space
which --- for a wide variety of the mappings $Z^A=Z^A(x^a)$ --- will have the signature
$(+,-,-,-)$ and  will represent, in general, a curved four geometry. The quantum theory of a free scalar field in ${\mathcal S}$ 
is well defined in terms of the, say, plane wave modes which satisfy the wave equation in ${\mathcal S}$. A subset of these
modes, which does not depend on the `transverse' directions, will satisfy the corresponding wave equation
in  ${\mathcal D}$ and will depend only on $x^a$. These modes induce a natural QFT in ${\mathcal D}$. We are interested
in the mappings $Z^A=Z^A(x^a)$ which leads to a horizon in ${\mathcal D}$ so that we can investigate the QFT
in spacetimes with horizons using the free, flat spacetime, QFT in ${\mathcal S}$ (\cite{deserlevin}  \cite{tprecent1}).

 For this purpose, let us restrict attention to a class of surfaces defined by the mappings $Z^A=Z^A(x^a)$ 
which ensures  the following properties for ${\mathcal D}$: (i) The induced metric $g_{ab}$ has the signature
$(+,-,-,-)$. 
(ii) The induced metric is static in the sense
that $g_{0\alpha}=0$ and all $g_{ab}$s are independent of $x^0$.
[The Greek indices run over 1,2,3.]
  (iii) Under the transformation $x^0 \to x^0\pm i(\pi/g)$, where
$g$ is a non zero, positive constant, the mapping of the coordinates changes
as $Z^0\to -Z^0, Z^1\to -Z^1$ and $Z^A\to Z^A$ for $A= 2,...,D$. 
It will turn out
that the four dimensional manifolds defined by such mappings possess 
a horizon and most of the interesting features of the thermodynamics related
to the horizon can be obtained from the above characterization. 
Let us first determine the nature of the mapping $Z^A = Z^A(x^a) = Z^A(t,{\bf x})$
such that the above conditions are satisfied. 

The condition (iii) above singles out the spatial coordinate
$Z^1$ from the others. To satisfy this condition we can take the mapping
$Z^A = Z^A(t,r,\theta,\phi)$   to be of the form
$Z^0 =Z^0(t,r), Z^1 =Z^1(t,r), Z^\perp =Z^\perp(r,\theta,\phi)$ where $Z^\perp$
denotes the transverse coordinates $Z^A$ with $A=(2,...,D)$. To impose the condition
(ii) above, one can make use of the fact that ${\mathcal S}$
 possesses invariance under translations, rotations and Lorentz boosts,
which are characterized by the existence of a set of $N = (1/2)(D+1)(D+2)$
Killing vector fields $\xi^A(Z^A)$. Consider any linear combination $V^A$
of these Killing vector fields which is time like in a region of ${\mathcal S}$.
 The integral curves to this vector field $V^A$ will
define   time like curves in ${\mathcal S}$. If one treats these 
curves as the trajectories of a hypothetical observer, then one can
set up the proper Fermi-Walker transported coordinate system
for this observer. Since the four velocity of the observer is along the
Killing vector field, it is obvious that the metric
in this coordinate system will be static \cite{sriramtp}.  In particular,  there
exists a Killing vector which  corresponds to Lorentz boosts along the $Z^1$ direction
that can be interpreted as rotation in imaginary time coordinate allowing a natural realization of (iii) above. 
Using the property of Lorentz boosts, it is easy to see that the transformations
of the form $Z^0 =  lf(r)^{1/2} \sinh gt;  Z^1 =  \pm lf(r)^{1/2} \cosh gt$
will satisfy both conditions (ii) and (iii) where $(l,g)$ are constants introduced for dimensional reasons 
and $f(r)$ is a given
 function. This map covers only the two quadrants
with $|Z^1| > |Z^0|$ with positive sign for the right quadrant and negative sign
for the left. To cover the entire $(Z^0,Z^1)$ plane, we will use the full set
\begin{eqnarray}
Z^0&=& lf(r)^{1/2} \sinh gt; \ Z^1 =  \pm lf(r)^{1/2} \cosh gt \quad ({\rm for}\  |Z^1| >|Z^0|) \label{eqn:two}
\\
Z^0&=&\pm  l [- f(r)]^{1/2} \cosh gt; \ Z^1 =   l[- f(r)]^{1/2} \sinh gt \quad ({\rm for}\  |Z^1| <|Z^0|) \nonumber
\end{eqnarray}
 The inverse
transformations corresponding to (\ref{eqn:two}) are
\begin{equation}
l^2f(r)=(Z^1)^2 -(Z^0)^2;\quad gt=\tanh ^{-1}(Z^0/Z^1) \label{eqn:three}
\end{equation}
Clearly, 
 to cover the entire two dimensional plane of $ - \infty < (Z^0,Z^1) < + \infty $, it is necessary
to have both $f(r)>0$ and $f(r)<0$.  
The pair of points $(Z^0,Z^1)$ and $(-Z^0,-Z^1)$ are mapped to the same
$(t,r)$ making this a 2-to-1 mapping.
The null surface $Z^0=\pm Z^1$ is mapped to the surface $f(r) =0$. 

The transformations given above with any arbitrary mapping for the transverse
coordinate $Z^\perp =Z^\perp(r,\theta,\phi)$ will give rise to an induced 
metric on ${\mathcal D}$ of the form
\begin{equation}
ds^2 = f(r)(lg)^2dt^2 - {l^2 \over 4} \left({ f^{\prime 2} \over f}\right) dr^2 - dL^2_\perp \label{eqn:four}
\end{equation}
where $dL_\perp^2$ depends on the form of the mapping  $Z^\perp =Z^\perp(r,\theta,\phi)$.
This form of the  metric is valid in all the quadrants even though
we will continue to work in the right quadrant and will comment
on the behaviour in other quadrants only when necessary. It is 
obvious that the ${\mathcal D}$, in general, is curved and has a horizon
at $f(r)=0$.

As a specific example, let us consider the case of (D+1)=6 with the coordinates
$(Z^0,Z^1,Z^2,Z^3,Z^4,Z^5)=(Z^0,Z^1,Z^2,R,\Theta,\Phi)$ and consider a mapping to
4-dimensional subspace in which: (i) The  $(Z^0,Z^1)$ are mapped to $(t,r)$ as before; (ii) the 
spherical coordinates $(R,\Theta,\Phi)$ in ${\mathcal S}$ are mapped to standard spherical polar coordinates in
${\mathcal D}$: $(r, \theta, \varphi)$ {\it and} (iii) we take $Z^2$ to be an arbitrary function of $r$: $Z^2=q(r)$. This leads to the metric
\begin{equation}
ds^2 = A(r) dt^2 - B(r) dr^2 - r^2 d \Omega^2_{\rm 2-sphere}; \label{eqn:tena}
\end{equation}
with
\begin{equation}
A(r) = (lg)^2f; \qquad B(r) = 1 + q^{\prime 2} + {l^2 \over 4} {f^{\prime 2 }\over f} \label{eqn:eleven}
\end{equation}
Equation (\ref{eqn:tena}) is the form of a general, spherically symmetric, static metric in 4-dimension with two arbitrary functions $f(r),q(r)$. Given any specific metric
with $A(r)$ and $B(r)$, equations (\ref{eqn:eleven}) can be solved to determine
$f(r),q(r)$. As an example, let us consider the Schwarzschild solution for which we will take $f=4 \left[ 1 - (l/ r) \right]$; the condition $g_{00}=(1/g_{11})$ now determines $q(r)$ through the equation
\begin{equation}
(q^{\prime})^ 2 = \left( 1 + {l^2 \over r^2} \right) \left( 1 + {l \over r} \right) - 1 = \left( {l \over r}\right)^3 + \left({l  \over r}\right)^2 + {l \over r}\label{eqn:twelve}
\end{equation}
That is
\begin{equation}
q(r) = \int^r\left[\left( {l \over r}\right)^3 + \left({l  \over r}\right)^2 + {l \over r}\right]^{1/2} dr \label{eqn:thirteen}
\end{equation}
Though the integral cannot be expressed in terms of elementary functions, it
is obvious that $q(r)$ is well behaved everywhere including at $r=l$. The transformations 
$(Z^0,Z^1) \to (t,r); Z^2\to q(r); (Z^3,Z^4,Z^5)\to(r, \theta, \varphi)$
thus provide the embedding of Schwarzschild metric in a 6-dimensional space.
[This result was originally obtained by Frondsal \cite{fronsdal};  but the derivation in that paper is somewhat obscure and does not bring out the generality of the situation]. As a corollary, we may note that this procedure leads to a spherically symmetric Schwarzschild-like metric in arbitrary dimension, with the
2-sphere in (\ref{eqn:tena}) replaced any  $N$-sphere. 

The choice $lg=1,f(r)=[1-(r/l)^2]$ will provide an embedding
of the de Sitter spacetime in 6-dimensional space with $Z^2=r, (Z^3,Z^4,Z^5)\to (r,\theta,\phi)$. Of course, in this case, one of the coordinates is actually redundant and --- as  we have seen earlier --- one can
achieve the embedding in a 5-dimensional space. 
    A still more trivial case is that of Rindler metric  which can be obtained 
    with D=3, $lg=1, f(r)=1+2gr$; in this case, the ``embedding'' is just a reparametrization within four
    dimensional spacetime and --- in this case --- $r$ runs in the range $(-\infty, \infty)$.
    The key point is that the metric in  (\ref{eqn:four}) is fairly generic and can describe
    a host of spacetimes with horizons located at $f=0$.     
     We shall discuss several
   features related to the thermodynamics of the horizon in the next few 
   sections.
    
    \subsection{Temperature of horizons}\label{horizontemp}
    
    There exists a natural definition of QFT in the original (D + 1)-dimensional space; in particular,
we can define a vacuum state for the quantum field on the $Z^0=0$ surface,
which coincides with the $t=0$ surface. By restricting the field modes (or the field configurations
in the Schrodinger picture) to depend only on the coordinates in ${\mathcal D}$, we will obtain a quantum
field theory in ${\mathcal D}$ in the sense that these modes will satisfy the relevant field equation
defined in ${\mathcal D}$. 
In general, this is a complicated problem and it is not easy to have a 
choice of modes in ${\mathcal S}$ which will lead to a natural set of modes in ${\mathcal D}$.
We can, however, take advantage of the arguments given in the last section --- that
all the interesting physics arises from the $(Z^0,Z^1)$ plane and the other 
transverse dimensions are irrelevant near the horizon.
In particular, solutions to the wave equation in ${\mathcal S}$ which depends only on  the 
coordinates $Z^0$ and $Z^1$ will satisfy the wave equation in ${\mathcal D}$ and will depend
only on $(t,r)$. Such modes will define a natural $s-$wave
QFT in ${\mathcal D}$.
  The positive frequency modes of the above kind 
  (varying as $\exp(-i\Omega Z^0)$ with $\omega >0$.)
  will be a specific superposition of
negative  (varying as $e^{i\omega t} $) and positive  (varying as $e^{-i\omega t} $) frequency modes in ${\mathcal D}$ leading to
  a  temperature $T=(g/2\pi)$ in the 4-dimensional subspace on
 one side of the horizon. 
There are several ways of proving this result, all of which depend essentially on the property
  that under the transformation $t\to t\pm (i\pi/g)$ the two coordinates $Z^0$ and $Z^1$ reverses
  sign.

Consider a positive frequency mode of the form
$ F _\Omega(Z^0,  Z^1) \propto
\exp[-i\Omega Z^0+ iPZ^1]$ with $\Omega > 0$.
These set of modes can be used to expand the quantum field thereby defining
the creation and annihilation operators $A_\Omega,A_\Omega^\dagger$:
\begin{equation}
\phi(Z^0,Z^1)=\sum_\Omega [A_\Omega F_\Omega(Z^0,Z^1) +A_\Omega^\dagger F_\Omega^*(Z^0,Z^1)]
\end{equation}
The vacuum state
defined by $A_\Omega|{\rm vac}>=0$ corresponds to a globally time symmetric
state which will be interpreted as a no particle state by observers using
$Z^0$ as the time coordinate. Let us now consider the same mode which can 
be described in terms of the $(t,r)$ coordinates. Being a scalar, this mode
 can be expressed in the 4-dimensional sector in the form $F_\Omega(t,r)= F_\Omega [Z^0(t,r),  Z^1(t,r)]$.
  The Fourier transform of $F_\Omega(t,r)$ with respect to $t$ will be:,
\begin{equation}
 K_\Omega(\omega,r) = \int_{-\infty}^\infty dt\,  e^{+i\omega t} F_\Omega[Z^0(t,r), Z^1(t,r)]; \qquad (-\infty<\omega<\infty)
 \label{ftfoft}
 \end{equation}
Thus a positive frequency mode in the higher dimension  can only be
expressed as an integral over $\omega$ with $\omega $ ranging over both positive and
negative values. However, using the fact that $t\to t-(i\pi/g)$ leads to $Z^0\to -Z^0, Z^1\to -Z^1$, 
it is easy to show  that 
\begin{equation}
  K_\Omega(-\omega, r) = e^{-(\pi \omega/g)}K_\Omega^*(\omega,r) 
  \label{analyticity}
\end{equation}
This allows us to write the inverse relation to (\ref{ftfoft}) as
\begin{eqnarray}
F_\Omega(t,r) &=&\int^\infty_{-\infty} {d\omega\over 2\pi} K_\Omega(\omega,r) e^{-i\omega t}
\nonumber\\
&=&\int_0^\infty {d\omega\over 2\pi}\left[K_\Omega(\omega,r)e^{-i\omega t}+
e^{-\pi \omega/g} K_\Omega^*(\omega,r)e^{i\omega t}\right]
\end{eqnarray}
The  term with $K_\Omega^*$ represents the contribution of negative frequency modes in the 
the 4-D spacetime to the pure positive frequency mode in the embedding spacetime.
A field mode of the embedding spacetime containing creation and annihilation operators $(A_\Omega,A_\Omega^\dagger)$ can now be
represented in terms of the creation and annihilation operators $(a_\omega,a_\omega^\dagger)$ appropriate to the
  $(t,r)$ coordinates as 
  \begin{eqnarray}
  A_\Omega F_\Omega + A_\Omega^\dagger F_\Omega^* &=& \int_0^\infty {d\omega\over 2\pi} \left[ \left( A_\Omega + A_\Omega^\dagger e^{-\pi \omega/g}\right) K_\Omega e^{-i\omega t} + {\rm h.c.}\right] \nonumber \\
  &=& \int_0^\infty {d\omega\over 2\pi}{1\over N_\omega} \left[ a_\omega K_\Omega e^{-i\omega t} + {\rm h.c.}\right] 
  \end{eqnarray}
  where $N_\omega$ is a normalization constant. Identifying $a_\omega = N_\omega(A_\Omega+e^{-\pi \omega/g} A_\Omega^\dagger)$ and using
  the conditions $[a_\omega,a_\omega^\dagger] = 1, [A_\Omega,A_\Omega^\dagger] = 1$ etc., we get $N_\omega=[1-\exp(-2\pi \omega/g)]^{-1/2}$. It follows that
  the number of $a-$particles in the vacuum defined by $A_\Omega|{\rm vac}>=0$ is given by 
  \begin{equation}
  <{\rm vac} | a_\omega^\dagger a_\omega|{\rm vac}> = N_\omega^2 e^{-2\pi \omega/g} = (e^{2\pi \omega/g}-1)^{-1}
  \end{equation}
  This is a Planckian spectrum with temperature $T=g/2\pi$.
  The key role in the derivation is played by equation (\ref{analyticity}) which, in turn,
  arises from the analytic properties of the spacetime under Euclidean continuation.

   \subsection{Entropy and energy of  de Sitter spacetime}\label{horizonent}
   
    The best studied spacetimes with horizons are the black hole spacetimes. (For a sample of references, see
    \cite{BHLS1,BHLS2,bhentropy1,bhentropy2,bhentropy3,bhentropy4,bhentropy5,bhentropy6,bhentropy7,bhentropy8}). In the simplest context of a Schwarzschild black hole of mass $M$,
    one can attribute an energy  $E=M$,  temperature    $T=(8\pi M)^{-1}$  and
    entropy $S=(1/4)(A_H/L_P^2)$ where
   $A_H$ is the area of the horizon and $L_P = (G\hbar /c^3)^{1/2}$ is the Planck length. (Hereafter, we will
   use units with $G=\hbar=c=1$.)  These are clearly related by the thermodynamic identity
   $TdS = dE$, usually called the first law of black hole dynamics. This result has been obtained
   in much more general contexts and  has been investigated from many different points of 
   view in the literature. The simplicity of the result depends on the following features: 
   (a)  The Schwarzschild metric  is a vacuum solution
    with no pressure so that there is no $PdV$ term in the first law of thermodynamics.
    (b) The metric has only one parameter $M$ so that changes in all physical parameters can be related
    to $dM$. (c) Most importantly, there exists a well defined notion of energy $E$ to the 
    spacetime and the changes in the energy $dE$ can be interpreted in terms of the 
    physical process of the black hole evaporation. 
The idea can be generalized to other 
    black hole spacetimes in a rather simple manner only because of well defined notions of
    energy, angular momentum etc.
    
    Can one generalize the thermodynamics of horizons to cases other than black holes
    in a straight forward  way ? In spite of years of research in this field, 
 this generalization remains non trivial and challenging when the conditions
    listed above are not satisfied. 
    To see the importance of the above conditions, we only need to contrast the situation in Schwarzschild 
    spacetime with that of de Sitter spacetime:
    
   \begin{itemize}
   \item As we saw in section \ref{horizontemp}, the
      notion of temperature is  well defined in the case of de Sitter spacetime and  we have $T= H/2\pi$
    where $H^{-1}$ is the radius of the de Sitter horizon.  But the correspondence probably
    ends there. 
    A study of literature shows that there exist very  few concrete calculations of energy, entropy
   and laws of horizon dynamics in the case of de Sitter spacetimes, in sharp contrast to
   BH space times. 
   
   \item There have been several attempts in the literature to define the concept of
   energy using local or quasi-local concepts (for a sample of references,
   see  \cite{energydef1,energydef2,energydef3,energydef4,energydef5,energydef6,energydef7,energydef8,energydef9}). The problem is that not all definitions of energy  
   agree with each other
   and not all of them can be applied to de Sitter type universes.  
   
   \item Even when a notion of 
   energy can be defined, it is not clear  how to
    write and interpret an equation analogous to $dS = (dE/T)$ in this spacetime,
    especially since the physical basis for $dE$ would require
    a notion of evaporation of the de Sitter universe.
    
    \item Further, we know that de Sitter spacetime is a solution to Einstein's equations with a 
    source having non zero pressure. Hence one would very much doubt whether $TdS $ is 
    indeed equal to $dE$. It would be necessary to add a $PdV$ term for consistency.
    \end{itemize}

   All these suggest that to make any progress, one might require a local approach
   by which one can define the notion of entropy {\it and} energy for spacetimes with 
   horizons. This conclusion is strengthened further by the following argument:
   Consider a class of spherically symmetric spacetimes of the form
    \begin{equation}
     ds^2=f(r)dt^2-f(r)^{-1}dr^2 - r^2(d\theta^2+\sin^2\theta
     d\phi^2)
     \label{ssmetric}
     \end{equation}
   If $f(r)$ has a simple zero at $r=a$ with $f'(a)\equiv B$ remaining finite, then
   this spacetime has a horizon at $r=a$. 
   Spacetimes like Schwarzschild or de Sitter have only one free parameter
    in the metric (like $M$ or $H^{-1}$) and hence the scaling of all other thermodynamical 
    parameters is uniquely fixed by purely dimensional considerations.  But, for a general  metric
    of the form in (\ref{ssmetric}), with an arbitrary $f(r)$, the 
    area of the horizon  (and hence the entropy) is determined by the location of the zero of the function $f(r)$
    while the temperature --- obtained from the periodicity considerations ---
    is determined by the value of $f'(r)$ at the zero. For a general function, of course,
    there will be no relation between the location of the zero and the slope of the 
    function at that point. It will, therefore, be incredible if there exists any a priori relationship 
     between the temperature (determined by $f'$ ) and the entropy
    (determined by the zero of $f$) even in the context of horizons in spherically symmetric spacetimes. 
    If we take the entropy to be $S=\pi a^2$ (where $f(a)=0$ determines the radius of the horizon) and
    the temperature to be $T=|f'(a)|/4\pi $ (determined by the periodicity of Euclidean time),
    the quantity $TdS  = (1/2) |f'(a)| a da$ will depend both on the slope $f'(a)$ as well as
    the radius of the horizon. This implies that any local interpretation of thermodynamics
    will be quite non trivial.

     Finally, the need for local description of thermodynamics of horizons becomes crucial
     in the case of spacetimes with multiple horizons. 
     The strongest 
   and the most robust result we have, regarding spacetimes with a horizon,
   is the notion of temperature associated with them. This, in turn, depends
    on the study of the periodicity of the Euclidean time coordinate.  
   This approach does not work very well if the spacetime has more than one horizon like, for example,
  in  the Schwarzschild-de Sitter metric which has the form in (\ref{ssmetric}) with
     \begin{equation}
     f(r) = \left(1-{2M\over r} - H^2 r^2\right)
    \end{equation}
    This spacetime has  two horizons at $r_\pm$ with 
    \begin{equation}
    r_+ =\sqrt{4\over 3} H^{-1} \cos {x+4\pi\over 3};\quad
     r_- =\sqrt{4\over 3} H^{-1}  \cos {x \over 3}
     \end{equation}
     where
     $\cos x =-3\sqrt{3} MH^{-1}$.
     (The parameter $x$ is in the range ($\pi, (3/2)\pi]$ and 
     we assume that $0\le 27M^2H^{-2} <1$.)
     Close to either horizon the spacetime can be approximated as Rindler.
      Since the surface gravities on the two
     horizons are different, we get two different   temperatures $T_\pm = |f' (r_\pm)|/ 4\pi$.
     To maintain invariance under $it\to it+\beta$ (with some finite $\beta$) it is necessary
     that $\beta$ is an integer multiple of both $4\pi/   |f' (r_+)| $
     and $4\pi/   |f' (r_-)| $ 
     so that $\beta = (4 \pi n_\pm/ |f' (r_\pm)| )$ where $n_\pm$ are integers.
     Hence the ratio of surface gravities $ |f' (r_+)|/ |f' (r_-)| = (n_+/n_-)$
     must be a rational number. 
      Though irrationals can be approximated by rationals,
     such a condition definitely excludes a class of values for 
     $M$ if $H$ is specified and vice versa.
     It is not clear why the existence of a cosmological constant 
     should imply something for the masses of black holes (or vice versa).
    Since there is no physical basis for such a condition,
    it seems reasonable to conclude that these difficulties arise
    because of our demanding the existence of a finite periodicity $\beta$ in the
    Euclidean time coordinate. This demand is related to
    an expectation of thermal equilibrium which is violated in spacetimes with 
    multiple horizons having different temperatures. 
    
 If even the simple notion of temperature falls apart in the presence
     of multiple horizons, it is not likely that the notion of energy or entropy can be 
     defined by {\it global} considerations. 
    On the other hand, it will be equally  strange  if  we cannot attribute a temperature to a black
     hole formed in some region of the universe just because the universe
     at the largest scales is described by a de Sitter spacetime, say.
     One is again led to searching for a {\it local} description of the thermodynamics of 
     {\it all types of}  horizons. 
     We shall now see how this can be done.
     
     Given the notion of temperature,  there are two  very different ways of defining the entropy:  (1)  In statistical mechanics, the   partition function $Z(\beta)$
of the canonical ensemble of systems with constant temperature $\beta^{-1}$  is related to the entropy $S$ and energy $E$ by
$Z(\beta)\propto \exp(S-\beta E)$. (2) In classical thermodynamics, on the other hand, it is the {\it change in} the entropy, which
can be operationally defined via $dS=dE/T(E)$. Integrating this equation will
lead to the function $S(E)$ except for an additive constant which needs to be
determined from additional considerations. Proving the equality of these two concepts was nontrivial and --- historically --- led to the unification of thermodynamics with mechanics. 

In the case of time symmetric  state,  there will be no change of entropy $dS$ and the thermodynamic
route is blocked. It is, however, possible to construct a canonical
ensemble of a class of  spacetimes and evaluate the partition function $Z(\beta)$. For  spherically symmetric spacetimes with a horizon at $r=l$,  the  partition function has the generic form
$Z\propto \exp[S-\beta E]$, where $S= (1/4) 4\pi l^2$  and $|E|=(l/2)$. This analysis reproduces the conventional result for the black hole spacetimes and provides a simple and consistent interpretation of entropy and energy for de Sitter spacetime, with the latter being given by $E=-(1/2) H^{-1}$. 
In fact, it is possible to write Einstein's equations for a spherically symmetric spacetime as a thermodynamic
identity $TdS -dE = P dV$ with $T,S$ and $E$ determined as above and the $PdV$ term arising from the
source \cite{tplongpap}. We shall now discuss some of these issues.
   
    Consider a class of spacetimes with the metric
    \begin{equation}
     ds^2=f(r)dt^2-f(r)^{-1}dr^2 -dL_\perp^2
     \label{basemetric}
     \end{equation}
  where $f(r)$ vanishes at some surface $r=l$, say, with $f'(l)\equiv B$ remaining finite. When $dL_\perp^2=r^2dS_2^2$
 with $[0\leq r\leq \infty]$, equation (\ref{basemetric}) covers a variety of spherically symmetric spacetimes  with a compact horizon at $r=l$. 
 Since the metric is static, Euclidean continuation is trivially effected by $t\to
  \tau=it$ and an examination of the conical singularity near $r=a$ [where $f(r) \approx B(r-a)$] shows that $\tau$ should be interpreted as periodic with period $\beta=4\pi/|B|$ corresponding to the temperature $T=|B|/4\pi$. 
 Let us consider a set ${\mathcal S}$ of such metrics
  in
  (\ref{basemetric}) with the restriction that $[f(a)=0,f'(a)=B]$ but $f(r)$ is otherwise arbitrary and has no
  zeros.  The partition function for this  set of metrics ${\mathcal S}$  is given  by the path integral sum
   \begin{equation}
    Z(\beta)=\sum_{g\epsilon {\mathcal S}}\exp (-A_E(g))  
 =\sum_{g\epsilon {\mathcal S}}\exp \left(-{1\over 16\pi}\int_0^\beta  d\tau \int d^3x \sqrt{g_E}R_E[f(r)]\right)
     \label{zdef}
     \end{equation}
  where   Einstein action  has been continued in the Euclidean sector 
  and we have  imposed the periodicity in $\tau$
  with period $\beta=4\pi/|B|$.
  The sum is restricted to the set ${\mathcal S}$ of all metrics of the form in 
  (\ref{basemetric}) with the behaviour $[f(a)=0,f'(a)=B]$ and the Euclidean Lagrangian is a functional of $f(r)$.
   The spatial integration will be restricted to a region bounded by the 2-spheres $r=a$ and $r=b$, where
  the choice of $b$ is arbitrary except for the requirement that  within the region of integration the Lorentzian
  metric must have the proper signature with $t$ being a time coordinate. 
  The remarkable feature is the form of the
  Euclidean action for this class of spacetimes.  Using the result
  $ R=\nabla_r^2 f -(2/r^2)(d/ dr)\left[r(1-f)\right]$
  valid for metrics of the form in (\ref{basemetric}), a
   straight forward calculation shows that
   \begin{equation}
  - A_E={\beta\over 4}\int_a^b dr\left[-[r^2f']'+2[r(1-f)]'\right] 
={\beta\over 4}[a^2B -2a]+Q[f(b),f'(b)]
     \label{zres}
     \end{equation}
where $Q$ depends on the behaviour of the metric near $r=b$ and we have used the 
conditions $[f(a)=0,f'(a)=B]$. The sum in (\ref{zdef}) now reduces to summing over the values of $[f(b),f'(b)]$
with a suitable (but unknown) measure. 
This sum, however, will only lead to a factor which we can ignore in deciding about the dependence of $Z(\beta)$ on the form of the metric near $r=a$. Using $\beta=4\pi/B$
(and taking $B>0$, for the moment)  the final result can be written in a very suggestive form:
 \begin{equation}
  Z(\beta)=Z_0\exp \left[{1\over 4}(4\pi a^2) -\beta({a\over 2})  \right]\propto 
  \exp \left[S(a) -\beta E(a)  \right]
     \label{zresone}
     \end{equation}
with the identifications for the entropy and energy being given by:
\begin{equation}
S={1\over 4} (4\pi a^2) = {1\over 4} A_{\rm horizon}; \quad E = {1\over 2} a = \left( {A_{\rm horizon}\over 16 \pi}\right)^{1/2}
\label{mainresult}
\end{equation}

     In the case of the Schwarzschild black hole with $a=2M$, the energy turns out to
    be $E=(a/2) = M$ which is as expected. (More generally,
	$E=(A_{\rm horizon}/16\pi)^{1/2}$ corresponds to the so called `irreducible mass' in 
       BH spacetimes \cite{irrmass}.) Of course, the identifications $S=(4\pi M^2)$,
    $E=M$, $T=(1/8\pi M)$ are consistent with the result $dE = TdS$ in this particular case. 
    
   The above  analysis also provides an interpretation of entropy and energy in the case
    of de Sitter universe. In this case, $f(r) = (1-H^2r^2)$, $a=H^{-1}, B=-2H$.
Since the region where $t$ is time like is ``inside'' the horizon, the integral for $A_E$ in (\ref{zres}) should be taken from some arbitrary value $r=b$ to $r=a$ with $a>b$. So the horizon contributes in the upper limit of the integral
introducing a change of sign in (\ref{zres}). Further, since $B<0$, there is another negative sign in the area term
from $\beta B\propto B/|B|$. Taking all these into account we get, in this case, 
\begin{equation}
  Z(\beta)=Z_0\exp \left[{1\over 4}(4\pi a^2) +\beta({a\over 2})  \right]\propto 
  \exp \left[S(a) -\beta E(a)  \right]
     \label{zrestwo}
     \end{equation}
giving
$S=(1/ 4) (4\pi a^2) = (1/ 4) A_{\rm horizon}$ and  $E=-(1/2)H^{-1}$. 
These definitions do satisfy the relation $TdS -PdV =dE$ when it is noted that the de Sitter universe has 
    a non zero pressure $P=-\rho_\Lambda=-E/V$ associated with the cosmological constant.  
    In fact,
if we use the ``reasonable" assumptions $S=(1/4)(4\pi H^{-2}), V\propto H^{-3}$ and $E=-PV$ in the equation
$TdS -PdV =dE$ and treat $E$ as an unknown function of $H$, we get the equation 
$H^2(dE/dH)=-(3EH+1)$
which integrates to give precisely $E=-(1/2)H^{-1}. $ 
(Note that we only needed the proportionality, $V \propto H^{-3}$ in this argument since $PdV \propto 
(dV/V)$. The ambiguity between the coordinate and proper volume is  irrelevant.)

 A peculiar feature of the metrics in (\ref{basemetric}) is worth stressing.  
 This metric  will satisfy Einstein's equations  provided the  the source stress 
     tensor has the form 
     $T^t_t = T^r_r \equiv (\epsilon(r)/ 8\pi );  T^\theta_\theta = T^\phi_\phi\equiv (\mu(r)/ 8\pi )$.
    The Einstein's equations now reduce to: 
\begin{equation}
    {1\over r^2} (1-f) - {f'\over r} = \epsilon; \quad \nabla^2 f = -2 \mu
    \label{albert}
    \end{equation}
    The remarkable feature about the metric in (\ref{basemetric}) is that the Einstein's 
    equations become linear in $f(r)$ so that solutions for different $\epsilon(r)$
    can be superposed.  Given any $\epsilon (r)$ the solution becomes
    \begin{equation}
    f(r) = 1-{a\over r} -{1\over r}\int_{a}^r \epsilon(r)  r^2\, dr
    \label{epsol}
    \end{equation}
    with $a$ being an integration constant and $\mu(r)$ is fixed by $\epsilon(r)$ through:
   $ \mu(r) = \epsilon + (1/ 2)r\epsilon' (r)$.
   The integration constant $a$ in (\ref{epsol})  is chosen such that $f(r)=0$ at $r=a$ so that this surface
    is a horizon. 
    Let us now assume that the solution  (\ref{epsol}) is such that $f(r) =0$ at $r=a$ with
    $f'(a)=B$ finite leading to  leading to a  notion of temperature with $\beta =(4\pi/|B|)$. 
     From  the first of the equations (\ref{albert}) evaluated at $r=a$, we get     
         \begin{equation}
    {1\over 2} Ba - {1\over 2} =- {1\over 2}\epsilon(a) a^2
    \label{albertmod}
    \end{equation}
    It is possible to provide an interesting interpretation of this equation which 
    throws light on the notion of entropy and energy.
   Multiplying the above equation by
    $da$ and using $\epsilon = 8\pi T^r_r$, it is trivial to rewrite  equation (\ref{albertmod})
    in the form    
    \begin{equation}
    {B\over 4\pi} d\left( {1\over 4} 4\pi a^2 \right) - {1\over 2} da=- T^r_r(a) d \left( {4\pi \over 3}  a^3 \right)
     =- T^r_r(a) [4\pi a^2]da 
    \label{keyeqn}
    \end{equation}
    Let us first consider the case in which a particular horizon has
    $f'(a)= B >0$ so that the temperature is $T=B/4\pi$. Since $f(a)=0,f'(a)>0$, it follows that $f>0$ for
    $r>a$ and $f<0$ for $r<a$; that is, the ``normal region'' in which $t$ is time like is outside
    the horizon as in the case of, for example, the Schwarzschild metric.  
    The first term in the left hand side of (\ref{keyeqn}) clearly has the form of $TdS$ since we have an independent identification of temperature from the periodicity argument in the local Rindler coordinates. 
    Since the pressure is $P=-T^r_r$, 
     the right hand side has the structure of $PdV$ or --- more
    relevantly --- is the product of the radial pressure times the transverse area times the 
    radial displacement. This is important because, for the metrics in the form (\ref{basemetric}), the proper transverse area is just that of a 2-sphere though
    the proper volumes and coordinate volumes differ. 
    In the case of horizons with $B=f'(a)>0$ which we are considering (with $da>0$), the volume of the region where $f<0$ will increase and the volume of the region
    where $f>0$ will decrease. 
    Since the entropy is due to the existence of an inaccessible region, $dV$ must refer to the change 
    in the volume of the inaccessible region where $f<0$.
    We can now identify
    $T$ in $TdS$ and $P$ in $PdV$ without any difficulty and  interpret the remaining term (second term in the left hand side) as $dE=da/2$.
    We thus get the expressions for the entropy $S$ and energy $E$ (when $B>0$) to be
   the same as in (\ref{mainresult}). 
   
   Using  (\ref{keyeqn}), we can again provide an interpretation of entropy and energy in the case
    of de Sitter universe. In this case, $f(r) = (1-H^2r^2)$, $a=H^{-1}, B=-2H<0$
    so that the temperature --- which should be positive --- is $T=|f'(a)| / (4\pi) =(-B)/4\pi$.
   For horizons with $B=f'(a)<0$ (like the de Sitter horizon) which we are now considering, 
   $f(a)=0, f'(a)<0$, and it follows that $f>0$ for
    $r<a$ and $f<0$ for $r>a$; that is, the ``normal region'' in which $t$ is time like is inside
    the horizon as in the case of, for example, the de Sitter metric.  
    Multiplying equation (\ref{keyeqn}) by $(-1)$,  we get
     \begin{equation}
    {-B\over 4\pi} d\left( {1\over 4} 4\pi a^2 \right) + {1\over 2} da= T^r_r(a) d \left( {4\pi \over 3}  a^3 \right)
    =P(-dV)
    \label{tdseqn}
    \end{equation}
    The first term on the left hand side is again of the form $TdS$ (with positive temperature and entropy).
    The term on the right hand side has the correct sign since the
     inaccessible region (where $f<0$) is now outside the horizon and the volume of this region 
    changes by $(-dV)$.
   Once again, we can use (\ref{tdseqn}) to identify \cite{tplongpap} the entropy and the energy: 
 $S=(1/  4) (4\pi a^2) = (1/  4) A_{\rm horizon};  E=-(1/ 2)H^{-1}$.
    These results agree with the previous analysis. 
  
  \subsection{Conceptual issues in de Sitter thermodynamics}\label{conceptds}
  
    The analysis in the last few sections 
     was based on a strictly {\it static} 4-dimensional spacetime. The black hole metric, for example, corresponds to an eternal black hole and the vacuum state  which we constructed in section \ref{horizontemp} corresponds to the Hartle-Hawking vacuum  \cite{hh} of the Schwarzschild spacetime,  describing a black hole in thermal equilibrium. There is no net radiation flowing to infinity and the entropy and temperature obtained in the previous sections were based on equilibrium considerations.
  
   As we said before, there are two   different  ways of defining the entropy. In statistical mechanics, the entropy $S(E)$ is related to the 
degrees of freedom [or phase volume] $g(E)$ by $S(E)=\ln g(E)$. Maximization of the phase volume for systems which can exchange energy
will then lead to equality of the quantity $T(E)\equiv (\partial S/\partial E)^{-1}$ for the systems. It is usual to identify this variable as the thermodynamic temperature. 
The analysis of BH temperature based on Hartle-Hawking state is analogous to this approach.

In classical thermodynamics, on the other hand, it is the {\it change in} the entropy which
can be operationally defined via $dS=dE/T(E)$. Integrating this equation will
lead to the function $S(E)$ except for an additive constant which needs to be
determined from additional considerations. 
This suggests an alternative point of view  regarding thermodynamics of horizons. The Schwarzschild metric, for example, can be thought of as an {\em asymptotic limit} of a  metric arising from the collapse of a body forming a black-hole. While developing the QFT in such a spacetime containing
a collapsing black-hole,
 we need not maintain time reversal invariance for the vacuum state and --- in fact --- it is more natural to choose a state with purely in-going modes at early times like the Unruh vacuum state  \cite{unruh}. The  study of QFT in such a spacetime shows that,
at late times, there will exist an outgoing thermal  radiation of particles
which is totally independent of the details of the collapse. The temperature in this case will be $T(M)=1/8\pi M$, which is the same as the one found in the case of the state of thermal
equilibrium around an ``eternal" black-hole.  In the Schwarzschild spacetime, which is asymptotically flat, it is also possible to associate an energy $E=M$ with the black-hole.
 Though the  calculation  was done in a metric with a fixed value of energy $E=M$, it seems reasonable to assume that ---
as the energy flows to infinity at late times ---  the mass of the black hole
will decrease.  {\it If} we make this assumption --- that the evaporation of black hole will lead to a decrease of $M$ --- {\it then} one can integrate the equation $dS=dM/T(M)$ to obtain the
entropy of the black-hole to be $S=4\pi M^2=(1/4)(A/L_P^2)$ where $A=4\pi (2M)^2$
is the area of the event horizon and $L_P=(G\hbar/c^3)^{1/2}$ is the Planck length.\footnote{This integration  can determine the entropy only
up to an additive constant. To fix this constant, one can make the additional assumption that $S$ should vanish when $M=0$. One may think that this assumption is eminently reasonable since the Schwarzschild metric reduces to the Lorentzian
metric when $M\to 0$.  But note that in the same limit of $M\to 0$, the temperature of the black-hole diverges !. Treated as a limit of Schwarzschild spacetime, normal flat spacetime has infinite --- rather than zero --- temperature.}
The procedure outlined above is similar in spirit to the approach of classical
thermodynamics rather than statistical mechanics.   

Once  it is realized that only the asymptotic form of the metric matters, we can simplify the above analysis by
just choosing a time {\it asymmetric} vacuum and working with the asymptotic
form of the metric  with the understanding that the asymptotic form arose due to a time asymmetric process (like gravitational collapse).
In the case of black hole spacetimes this is accomplished --- for example --- by choosing the Unruh vacuum \cite{unruh}.
 The question arises as to how our unified approach fares in handling such a situation which is not time symmetric and the horizon
forms only asymptotically as $t\to\infty$.

There exist analogues for the collapsing black-hole in the case of de Sitter (and even Rindler)  \cite{tplongpap} . The analogue in the case of de Sitter
spacetime will be an FRW universe which behaves like a de Sitter universe only at late times
[like in equation (\ref{univevol}); this is indeed the metric describing our universe if $\Omega_\Lambda =0.7,
\Omega_{\rm NR}=0.3$].  Mathematically,
we only need to take $a(t)$ to be a function which has the asymptotic form
$\exp(Ht)$ at late times. Such a spacetime is, in general, time asymmetric and one can choose a vacuum state at early times in such a way that a thermal spectrum of particles exists at late times.
Emboldened by the analogy with black-hole spacetimes, one can also directly construct quantum states (similar to Unruh vacuum of black-hole spacetimes) which are time asymmetric, even in the exact de Sitter spacetime, with the understanding that the de Sitter universe came about at late times through a time asymmetric evolution.  

The analogy also works for Rindler spacetime.  The coordinate system for an observer with {\it time dependent} acceleration will generalize the  standard Rindler spacetime
in a  time dependent manner.  In particular, one can have an observer who was inertial (or at rest) at early times and is uniformly accelerating at late times. In this case an event horizon forms at late times exactly in analogy with a collapsing black-hole. It is now possible to choose quantum states which are analogous to
the Unruh vacuum - which will correspond to an inertial vacuum state at early times and will appear as a thermal state at late times. The study of different `vacuum' states  shows \cite{tplongpap} that radiative flux exists in the quantum states which are time asymmetric analogues of the Unruh vacuum state.

A formal analysis of this problem will involve setting up the in and out 
vacua of the theory,  evolving the modes from $t=-\infty$ to $t=+\infty$,
and computing the Bogoliubov coefficients. It is, however, not necessary to perform the
details of such an analysis because
all the three spacetimes (Schwarzschild, de Sitter and Rindler)
have virtually identical kinematical structure.
In the case of Schwarzschild metric, it is well known that 
the thermal spectrum at late times arises because
the modes
which reach spatial infinity at late times propagate from near the event
horizon at early times and undergo exponential redshift.
The corresponding result occurs in all the 
three spacetimes (and a host of other spacetimes).

Consider the propagation of a wave packet centered around a  radial null ray in a  spherically symmetric
 (or Rindler) spacetime which  has the form in equation
(\ref{eqn:tena}) or (\ref{basemetric}).
The trajectory of the null ray which goes from the initial position $r_{in}$
at $t_{in}$ to a final position $r$ at $t$ is determined by the equation
\begin{equation} 
t - t_{in} =  \pm\left({1\over 2g}\right)\int_{r_{in}}^r\left({f'\over f}\right)\left (1+\cdots\right)^{1/2}dr\label{eqn:thirtynine}
\end{equation} 
where the $\cdots$ denotes terms arising from the transverse part
containing $dr^2$ (if any).
Consider now a ray which was close to the horizon initially so that $(r_{in} - l) \ll l$ and propagates to a region far away from the horizon at late times. (In
a black hole metric $r\gg r_{in}$ and the propagation will be outward directed;
in the de Sitter metric we will have $r\ll r_{in}$ with rays propagating towards the origin. )
Since we have $f(r) \to 0$ as $r\to l$, the integral will be dominated
by a logarithmic singularity near the horizon and the regular term denoted
by $\cdots$ will not contribute. 
[This can be verified directly from (\ref{eqn:tena}) or (\ref{basemetric}).]
Then we get
\begin{equation} 
t - t_{in} =  \pm\left({1\over 2g}\right)\int_{r_{in}}^r\left({f'\over f}\right) \left(1+\cdots\right)^{1/2}dr\approx\pm \left({1\over 2g}\right)\ln |f(r_{in})|+{\rm const.}
\end{equation} 
As the wave propagates away from the horizon its frequency will be red-shifted by the factor $\omega \propto (1/ \sqrt{g_{00}})$ so that
\begin{equation} 
{\omega(t) \over \omega(t_{in})} = \left( {g_{00}(r_{in}) \over g_{00} (r) } \right)^{1/2}
=\left[ {f(r_{\rm in})\over f(r)}\right]^{1/2} \approx K e^{\pm gt}
\label{eqn:fortytwo}
\end{equation} 
where $K$ is an unimportant constant. It is obvious that the dominant behaviour
of $\omega(t)$ will be exponential for any null geodesic starting 
near the horizon and proceeding away since all the transverse
factors will be sub-dominant to the diverging logarithmic 
singularity arising from the integral of $(1/f(r))$ near the horizon. Since
$\omega(t) \propto 
  \exp [\pm gt ]$ 
and the phase $\theta(t)$ of the wave will be vary with time as
$\theta(t) = \int \omega(t)dt  \propto   \exp [\pm gt ]$,
the time dependence of the wave at late times will be
\begin{equation}
\psi(t) \propto \exp[i\theta(t)] \propto \exp i\int w(t) dt \propto \exp i Q e^{\pm gt}
\end{equation}
where $Q$ is some constant.
An observer at a fixed  $r$ will see the wave to have the time dependence $\exp  [i \theta (t)]$ which, of course, is not monochromatic. If this wave is decomposed into different Fourier components with respect to $t$, then the amplitude at frequency $\nu$ is given by the Fourier transform
\begin{equation} 
f(\nu)= \int_{-\infty}^\infty dt\, \psi(t) \,e^{-i \nu t}  \propto \int e^{i\theta(t) - i \nu t} dt \propto \int\limits^{\infty}_{-\infty} dt e^{-i(\nu t - Q\exp[\pm gt])}\label{eqn:fortysixa}
\end{equation} 
 Changing the variables from $t$ to $\tau$ by  
$ Qe^{\pm gt} = \tau$,
evaluating the integral by analytic continuation to Im $\tau$ and taking the modulus one finds that the result is  a thermal spectrum:
\begin{equation} 
\vert f (\nu)\vert^2 \propto {1 \over e^{\beta \nu} - 1 } ; \quad \beta = {2\pi \over g}\label{eqn:fortyeight}
\end{equation} 
The standard expressions for the temperature are reproduced for Schwarzschild ($g= (4M)^{-1}$), 
de Sitter ($g=H$) and Rindler spacetimes. 
This analysis  stresses the fact that the origin of thermal spectrum lies in the Fourier transforming of an exponentially red-shifted spectrum. 

But in de Sitter or Rindler spacetimes there is {\it no}
natural notion of  ``energy source" analogous to the mass of the black-hole. 
The conventional view is to assume that:  (1) In the case of black-holes, one considers the collapse scenario as ``physical" and the natural quantum state is the Unruh vacuum. The notions of evaporation, entropy etc. then follow in a concrete manner. The eternal black-hole (and the Hartle-Hawking vacuum state) is taken to be just a mathematical construct not realized in nature.  (2) In the case of Rindler, one may like to think of a time-symmetric vacuum state as natural and treat the situation as one of thermal equilibrium. This forbids using quantum states with outgoing radiation which could make the  Minkowski spacetime radiate energy -- which seems unlikely.  

The real trouble  arises for spacetimes which are asymptotically de Sitter.  Does such a spacetime have temperature and  entropy like a collapsing black-hole? Does it ``evaporate" ? Everyone is comfortable with the idea of associating temperature with the de Sitter spacetime and most people seem to be willing to associate even an entropy. However, the idea of the cosmological constant changing due to evaporation of the de Sitter spacetime seems too radical.
  Unfortunately, there is no clear mathematical reason for  a dichotomous approach as regards a collapsing black-hole and an asymptotically de Sitter spacetime, since:
  (i) The temperature and entropy 
  for  these spacetimes arise in identical manner due to identical mathematical
  formalism.  It will be surprising if one has
  entropy while the other does not. 
  (ii) Just as collapsing black hole leads to an asymptotic event horizon, a universe
  which is dominated by cosmological constant at late times will also lead to a horizon.
  Just as we can mimic the time dependent effects in a collapsing black hole by
  a time asymmetric quantum state (say, Unruh vacuum), we can mimic the late time behaviour of 
  an asymptotically de Sitter universe by a corresponding time asymmetric quantum state.
  Both these states will lead to stress tensor expectation values in which there will be a flux
  of radiation. 
(iii) The energy source for expansion at early times (say, matter or radiation) is irrelevant just as the collapse details are irrelevant in the case of a black-hole. 

If one treats the de Sitter horizon as a  `photosphere' with temperature $T=(H/2\pi)$ and area $A_H=4\pi H^{-2}$,
then the radiative luminosity will be $(dE/dt)\propto T^4A_H\propto H^2$. If we take $E=(1/2)H^{-1}$, this will
lead to a decay law \cite{tpccdecay} for the cosmological constant of the form:
\begin{equation}
\Lambda(t)=\Lambda_i\left[1+k (L_P^2\Lambda_i)(\sqrt{\Lambda_i}(t-t_i))  \right]^{-2/3}\propto (L_P^2t)^{-2/3}
\end{equation}
where $k$ is a numerical constant and the second proportionality is for $t\to \infty$. It is interesting that  this naive model leads to a late time cosmological constant which is independent of the initial value ($\Lambda_i$). Unfortunately, its value is still far too large.
These issues are not analyzed in adequate detail in the literature and might have important implications for the \cc\  problem.

   \section{Cosmological constant and the string theory}\label{ccstring}

   A relativistic point particle is a zero dimensional object; the world line of such
   a particle, describing its time evolution, will be one dimensional and the standard
   quantum field theory (like QED) uses real and virtual world lines of particles in
   its description. In contrast, a string (at a given moment of time) 
   will be described by an one dimensional
   entity and its time evolution will be a two dimensional world surface  called the {\it world
   sheet}. 
   The basic formalism of string theory --- considered to be a possible candidate for
   a model for quantum gravity --- uses a two dimensional world sheet rather than the one 
   dimensional world line of a  particle to describe fundamental physics. 
   Since the point particle has been replaced by a more extended structure, string
   theory can be made into a finite theory and, in general, the excitations of the 
   string can manifest as low energy particles. This provides a hope for describing
   both gauge theories and gravity in a unified manner. (For a text book description of 
   string theory, see \cite{stringtext1,stringtext2}; 
   for a more popular description, see \cite{stringpop1b,stringpop1h,stringpop2}).  
   
   It was realized fairly early on that string theory can be consistently formulated
   only in 10 dimension and it is necessary to arrange matters so that six of these
   dimensions are compact (and very small) while the other four --- which
   represents the spacetime --- are presumably large and non compact.
   There is no fundamental understanding of how this comes about; but the 
   details of the four dimensional theory depends on the way in which six extra
   dimensions are compactified. The simplest example corresponds to a situation
   in which the six dimensional geometry is what is known as {\it calabi-yau manifold}
   \cite{calabiyau1,calabiyau2,calabiyau3} 
   and the four dimensions exhibit $N=1$ supersymmetry. The current paradigm, however,
   considers different ten dimensional theories as weakly coupled limits of a single theory and 
   not as inequivalent theories. Depending on the choice of parameters in the 
   description, one can move from one theory to other. In particular, as the parameters
   are changed, one can make a transition from weakly coupled limit of  one theory
   to the strongly coupled limit of another. These strong-weak coupling dualities
   play an important role in the current paradigm of string theories though explicit
   demonstration of dualities exists only for limited number of cases \cite{font1,font2,font3,font4}.
   
   The role of cosmological constant in string theories came into the forefront when 
   it was realized that there exists a peculiar equivalence between a class of theories
   containing gravity and pure gauge theories. One example of such a duality \cite{malda}
    arises as follows: A particular kind of string theory in ten dimension (called {\it type II B} string theory)
    can  be compactified with five of the dimensions wrapped up as 5-sphere ($S^5$) and the 
    other five dimensions taken to describe a 5-dimensional anti de Sitter spacetime with negative
    cosmological constant ($AdS_5$). The whole manifold will then be $S^5 \times AdS_5$ with the 
    metric on the $AdS$ sector given by
    \begin{equation}
    ds^2 = dr^2 +e^{2r} (\eta_{\mu\nu} dx^\mu dx^\nu) \qquad \mu, \nu =1,2,3,4.
    \end{equation}
    This string theory has an exact equivalence with the 4-dimensional $N=4$ supersymmetric
    Yang-Mills theory. It was known for a long time that the latter theory is conformally invariant;
    the large symmetry group of the $AdS_5$ matches precisely with the invariance group of Yang-Mills
    theory. The limit $r\to \infty$ is considered to be the boundary of $AdS$ space on which
    the dual field theory is defined. This allows one to obtain a map from the string theory states
    to the field which lives on the boundary. It must be stressed that it is hard to prove directly
    the equivalence between type II B $AdS_5 \times S^5$ string theory and the four dimensional
    Yang-Mills theory especially since we do not have a non perturbative description of the former.
    In this sense the Yang-Mills theory actually provides a definition of the non perturbative
    type II B $AdS_5 \times S^5$ string theory. It is, however, possible to verify the correspondence
    by restricting to low energies on the string theory side. 
    
    If gravity behaves as a local field theory, then the entropy in 
    a compact region of volume $R^3$ will scale as $S\propto R^3$ while indications from
    the physics of the horizons is that it should scale as $S\propto R^2$. One  can provide a consistent
    picture if gravity in $D-$dimensions is equivalent to a field theory in $D-1$ dimension with the 
    entropy of the field theory scaling as the volume of the $D-1$ dimensional space which, of course,
    is the same as the area in the original $D-$dimensional space. This is achieved in a limited sense
    in the above model. 
    
    The $AdS$ spacetime has a negative cosmological constant while the standard de Sitter
    spacetime has a positive cosmological constant. This change of sign is crucial and the 
    asymptotic structure of these theories are quite different. We do not, however, know of
    any solution to string theory which contains de Sitter spacetime or even any solution to 
    standard Einstein's equation with a positive cosmological constant.
    There are, in fact, some no-go theorems which state that such solutions cannot exist
    \cite{nogo1,nogo2,nogo3}.
    This, however, is not a serious concern since the no-go theorems assume 
    certain positive energy conditions which are indeed violated in string theory.

    If de Sitter solutions of the string theory exists, then it would be interesting to 
    ask whether they would admit a dual field theory description as in the case
    of anti de Sitter space. Some preliminary results indicate that if such a duality
    exists, then it would be with respect to a rather peculiar type of conformal
    field theories \cite{vbs1,vbs2,vbs3}. 
     The situation
    at present is reasonably open. 
    
   There is another indirect implication of the string theory paradigm for the cosmological
   constant problem. The detailed  vacuum structure in string theory is at present quite unknown
   and the preliminary indications are that it can be fairly complicated. Many
     believe that the ultimate theory may not lead to a unique vacuum state
   but instead could lead to a set of degenerate vacua. The properties of physical
   theories built out of these  vacua could be different and it may be necessary to invoke some 
   \emph{additional} criterion to select one vacuum out of many as {\em the} ground state of 
   the observed universe. Very little is known about this issue \cite{tptirthmpl}
   but advocates of anthropic principle sometimes use the possibility multiple degenerate
   vacua as a justification for anthropic paradigm. 
   While this is not the only possibility, it must be stressed that the existence of degenerate
   vacua introduces an additional feature as regards the cosmological constant  \cite{degeneratevac}.
   The problem arises from the fact that quantum theory allows tunneling between
   the degenerate vacua and makes the actual ground state a superposition of
   the degenerate vacua. There will be an energy difference between: (i) the degenerate
   vacua and (ii)  the vacuum state obtained by including the effects of tunneling.
   While the fundamental theory may provide some handle on the \cc\
   corresponding to the degenerate vacua, the {\em observed } vacuum energy
   could correspond to the real vacuum which incorporates the effect of 
   tunneling. In that case it is the dynamics of tunneling which will determine the 
   ground state energy and the \cc.

  \section*{Acknowledgement}
  
  I thank J.S. Bagla, J.V. Narlikar, T. Roy Choudhury and K. Subramanian for comments on the earlier
  version of the draft.

  \end{document}